\documentclass[twocolumn]{aastex631}

\newcommand{\eazy}{{\tt{EAZY-py}}}
\newcommand{\bagpipes}{{\tt{Bagpipes}}}

\newcommand{\prospector}{{\tt{Prospector}}}
\newcommand{\sextractor}{{\tt{SExtractor}}}

\newcommand{\MUV}{$M_{\mathrm{UV}}$}

\newcommand{\jaguar}{{\tt{JAGUAR}}}
\newcommand{\beagle}{{\tt{BEAGLE}}}
\newcommand{\cloudy}{{\tt{CLOUDY}}}

\def\casgm20{CAS-G-M$_{20}\,$}
\def\m20{M$_{20}\,$}

\def\Spitzer{\textit{Spitzer}}
\def\HST{\textit{Hubble Space Telescope}}

\def\ACSWFC{Advanced Camera for Surveys Wide Field Channel}
\newcommand{\hbeta}{H$\beta$}
\usepackage{amsmath}
\usepackage{savesym}
\savesymbol{tablewidth}
\usepackage{tabularray}
\restoresymbol{x}{tablewidth}
\usepackage{multirow}
\usepackage{subfigure}
\usepackage[nameinlink]{cleveref}

\makeatletter
\DeclareRobustCommand{\oiii}{%
  [\mbox{O\check@mathfonts\fontsize\sf@size\z@\selectfont III}]%
}
\makeatother

\begin{document}

\title{EPOCHS IV: SED Modelling Assumptions and their impact on the Stellar Mass Function at $6.5~\leq z \leq~13.5$ using PEARLS and public JWST observations}

\author[0000-0002-4130-636X]{Thomas Harvey}
\affiliation{Jodrell Bank Centre for Astrophysics, University of Manchester, Oxford Road, Manchester M13 9PL, UK}

\author[0000-0003-1949-7638]{Christopher J. Conselice}
\affiliation{Jodrell Bank Centre for Astrophysics, University of Manchester, Oxford Road, Manchester M13 9PL, UK}

\author[0000-0003-4875-6272]{Nathan J. Adams}
\affiliation{Jodrell Bank Centre for Astrophysics, University of Manchester, Oxford Road, Manchester M13 9PL, UK}

\author[0000-0003-0519-9445]{Duncan Austin}
\affiliation{Jodrell Bank Centre for Astrophysics, University of Manchester, Oxford Road, Manchester M13 9PL, UK}

\author[0009-0003-7423-8660]{Ignas Juodžbalis}
\affiliation{Jodrell Bank Centre for Astrophysics, University of Manchester, Oxford Road, Manchester M13 9PL, UK}
\affiliation{Kavli Institute for Cosmology, University of Cambridge, Cambridge, Madingley Road, Cambridge, CB3 0HA}

\author[0000-0002-9081-2111]{James Trussler}
\affiliation{Jodrell Bank Centre for Astrophysics, University of Manchester, Oxford Road, Manchester M13 9PL, UK}

\author[0000-0002-3119-9003]{Qiong Li}
\affiliation{Jodrell Bank Centre for Astrophysics, University of Manchester, Oxford Road, Manchester M13 9PL, UK}

\author[0000-0003-2000-3420]{Katherine Ormerod}
\affiliation{Jodrell Bank Centre for Astrophysics, University of Manchester, Oxford Road, Manchester M13 9PL, UK}
\affiliation{Astrophysics Research Institute, Liverpool John Moores University, 146 Brownlow Hill, Liverpool, L3 5RF}

\author[0000-0002-8919-079X]{Leonardo Ferreira}
\affiliation{Department of Physics \& Astronomy, University of Victoria, Finnerty Road, Victoria, British Columbia, V8P 1A1, Canada}

\author[0000-0001-7964-5933]{Christopher C. Lovell}
\affiliation{Institute of Cosmology and Gravitation, University of Portsmouth, Burnaby Road, Portsmouth, PO1 3FX, UK}

\author[0009-0009-8105-4564]{Qiao Duan}
\affiliation{Jodrell Bank Centre for Astrophysics, University of Manchester, Oxford Road, Manchester M13 9PL, UK}

\author[0009-0008-8642-5275]{Lewi Westcott}
\affiliation{Jodrell Bank Centre for Astrophysics, University of Manchester, Oxford Road, Manchester M13 9PL, UK}

\author[0009-0005-0817-6419]{Honor Harris}
\affiliation{Jodrell Bank Centre for Astrophysics, University of Manchester, Oxford Road, Manchester M13 9PL, UK}

\author[0000-0003-0883-2226]{Rachana Bhatawdekar}
\affiliation{European Space Agency (ESA), European Space Astronomy Centre (ESAC), Camino Bajo del Castillo s/n, 28692 Villanueva de la Cañada, Madrid, Spain}

\author[0000-0001-7410-7669]{Dan Coe} 
\affiliation{AURA for the European Space Agency (ESA), Space Telescope Science
Institute, 3700 San Martin Drive, Baltimore, MD 21218, USA}

\author[0000-0003-3329-1337]{Seth H. Cohen} 
\affiliation{School of Earth and Space Exploration, Arizona State University,
Tempe, AZ 85287-1404}

\author[0000-0002-6089-0768]{Joseph Caruana}
\affiliation{Department of Physics, University of Malta, Msida MSD 2080, Malta; \& Institute of Space Sciences \& Astronomy, University of Malta, Msida MSD 2080, Malta}

\author[0000-0003-0202-0534]{Cheng Cheng}
\affiliation{Chinese Academy of Sciences South America Center for Astronomy, National Astronomical Observatories, CAS, Beijing 100101, People's Republic of China}
\affiliation{2 CAS Key Laboratory of Optical Astronomy, National Astronomical Observatories, Chinese Academy of Sciences, Beijing 100101, People's Republic of China}

\author[0000-0001-9491-7327]{Simon P. Driver} 
\affiliation{International Centre for Radio Astronomy Research (ICRAR) and the
International Space Centre (ISC), The University of Western Australia, M468,
35 Stirling Highway, Crawley, WA 6009, Australia}

\author[0000-0003-1625-8009]{Brenda Frye} 
\affiliation{University of Arizona, Department of Astronomy/Steward
Observatory, 933 N Cherry Ave, Tucson, AZ85721}

\author[0000-0001-6278-032X]{Lukas J. Furtak}
\affiliation{Physics Department, Ben-Gurion University of the Negev, P. O. Box 653, Be’er-Sheva, 8410501, Israel}

\author[0000-0001-9440-8872]{Norman A. Grogin} 
\affiliation{Space Telescope Science Institute, 3700 San Martin Drive, Baltimore, MD 21218, USA}

\author[0000-0001-6145-5090]{Nimish P. Hathi}
\affiliation{Space Telescope Science Institute, 3700 San Martin Drive, Baltimore, MD 21218, USA}

\author[0000-0002-4884-6756]{Benne W. Holwerda}
\affiliation{Department of Physics and Astronomy, University of Louisville, 102 Natural Sciences Building, Louisville, KY 40292, USA}

\author[0000-0003-1268-5230]{Rolf A. Jansen} 
\affiliation{School of Earth and Space Exploration, Arizona State University,
Tempe, AZ 85287-1404}

\author[0000-0002-6610-2048]{Anton M. Koekemoer} 
\affiliation{Space Telescope Science Institute, 3700 San Martin Drive, Baltimore, MD 21218, USA}

\author[0000-0001-6434-7845]{Madeline A. Marshall} 
\affiliation{National Research Council of Canada, Herzberg Astronomy \&
Astrophysics Research Centre, 5071 West Saanich Road, Victoria, BC V9E 2E7,
Canada; \& ARC Centre of Excellence for All Sky Astrophysics in 3
Dimensions (ASTRO 3D), Australia}

\author[0000-0001-6342-9662]{Mario Nonino}
\affiliation{INAF-Osservatorio Astronomico di Trieste, Via Bazzoni 2, I-34124 Trieste, Italy}


\author[0000-0002-1905-4194]{Aswin P. Vijayan}
\affiliation{Cosmic Dawn Center (DAWN)}
\affiliation{Astronomy Centre, Department of Physics and Astronomy, University of Sussex, Brighton, BN1 9QH, UK}
\affiliation{DTU-Space, Technical University of Denmark, Elektrovej 327, DK-2800 Kgs. Lyngby, Denmark}

\author[0000-0003-3903-6935]{Stephen M. Wilkins}
\affiliation{Astronomy Centre, Department of Physics and Astronomy, University of Sussex, Brighton, BN1 9QH, UK}

\author[0000-0001-8156-6281]{Rogier Windhorst}
\affiliation{School of Earth and Space Exploration, Arizona State University, Tempe, AZ 85287-1404}

\author[0000-0001-9262-9997]{Christopher N. A. Willmer} 
\affiliation{Steward Observatory, University of Arizona, 933 N Cherry Ave, Tucson, AZ, 85721-0009}

\author[0000-0001-7592-7714]{Haojing Yan} 
\affiliation{Department of Physics and Astronomy, University of Missouri,
Columbia, MO 65211}

\author[0000-0002-0350-4488]{Adi Zitrin}
\affiliation{Department of Physics, Ben-Gurion University of the Negev, P.O. Box 653, Be'er-Sheva 84105, Israel}



\begin{abstract}

We utilize deep JWST NIRCam observations for the first direct constraints on the Galaxy Stellar Mass Function (GSMF) at $z>10$. Our EPOCHS v1 sample includes 1120 galaxy candidates at $6.5<z<13.5$ taken from a consistent reduction and analysis of publicly available deep JWST NIRCam data covering the PEARLS, CEERS, GLASS, JADES GOOD-S, NGDEEP, and SMACS0723 surveys, totalling 187 arcmin$^2$. We investigate the impact of SED fitting methods, assumed star formation histories (SFH), dust laws, and priors on galaxy masses and the resultant GSMF. Whilst our fiducial GSMF agrees with the literature at $z<13.5$, we find that the assumed SFH model has a large impact on the GSMF and stellar mass density (SMD), finding a 0.75~dex increase in the SMD at $z=10.5$ between a flexible non-parametric and standard parametric SFH. Overall, we find a flatter SMD evolution at $z \geq 9$ than  some studies predict, suggesting a rapid buildup of stellar mass in the early Universe. We find no incompatibility between our results and those of standard cosmological models, as suggested previously, although the most massive galaxies may require a high star formation efficiency. We find that the ‘Little Red Dot’ galaxies dominate the $z=7$ GSMF at high-masses, necessitating a better understanding of the relative contributions of AGN and stellar emission. We show that assuming a theoretically motivated top-heavy IMF reduces stellar mass by 0.5~dex without affecting fit quality, but our results remain consistent with existing cosmological models with a standard IMF.

\end{abstract}



\section{Introduction} \label{sec:intro}

The James Webb Space Telescope (JWST) has pushed backed the redshift frontier when searching for the earliest galaxies. The highly sensitive Near Infrared Camera (NIRCam) on JWST has led to an influx of high redshift galaxy candidates through photometry, in surveys such as CEERS, GLASS, PEARLS, NGDEEP \& JADES \citep{Adams2023, Castellano2022, Windhorst2023, austin2023first, Hainline2023,Bagley2023ngdeep}. The wavelength coverage from 0.6 to 5$\mu$m, enables identification of Lyman-break galaxies (LBGs) at redshifts $z\geq 6.5$ by their photometry. In the first 18 months of science operations, tens of candidates above  $z \geq 10$ have been identified \citep{Adams2023, Donnan2022, Atek2022, Naidu2022a, Naidu2022b, Harikane2023, Castellano2022, Finkelstein2022c,austin2023first, Hainline2023, Furtak2022, 2023Natur.616..266L, perez2023life,Finkelstein2023,Leung2023,McLeod2023,Willott2023}, including a spectroscopically confirmed galaxy at $z=13.27$ \citep{2023NatAs.tmp...66C} and candidates at $z \geq 16$ \citep[e.g.][]{Atek2022,Yan2022a,Furtak2022,Hainline2023}. 

An immediate result was a potential overabundance of high-redshift galaxies compared to theoretical predictions and extrapolations of \HST{} (HST)/\Spitzer{} results \citep{Naidu2022a,haslbauer2022,mauerhofer2023dust,mason2023brightest}. The mere existence of some of the galaxies at the inferred redshifts and stellar masses, ($\geq 10^{10.5}$ M$_{\odot}$ at $z \geq 7.5$) seem to be in tension with standard $\Lambda$CDM cosmology given the small areas currently probed with JWST \citep{2023Natur.616..266L,lovell2023extreme, boylan2023stress, desprez2024lambdacdm}. However, spectroscopic observations of some of the highest-mass candidates in \cite{2023Natur.616..266L} have reduced their redshifts and stellar masses, or hinted at the presence of an Active Galactic Nuclei (AGN), and hence reduced possible $\Lambda$CDM tension \citep{kocevski2023hidden,fujimoto2022alma}. Massive ($\geq 10^{11} \rm{M}_\odot$) quiescent galaxies at $ z\approx 4$ also appear to challenge theories of galaxy evolution given the old stellar populations implied by their star formation histories \citep{2024Natur.628..277G,2024ApJ...965...98C, carnall2024jwst}.

Initial overlap of galaxy candidates between independent studies was poor, but has since improved due to agreement on photometric calibration and reduction techniques \citep{rieke2022nircam,adams2023epochs}. Spectroscopic confirmation exists only for a fraction of potential candidates, but most spectroscopic programs have had a high success rate, along with a few notable low-$z$ interlopers \citep{Robertson2022,2023NatAs.tmp...66C,Haro2023b,Tang2023,Laseter2023,2023arXiv230109482F,Bunker2023,wang2023uncover}.

The combination of NIRCam's high sensitivity and infrared (IR) wavelength coverage allows characterisation of the rest-frame optical emission of $0.5 \leq z \leq 10$ galaxies, which is crucial for accurate stellar mass estimates. A more complete census of the high-redshift Universe is also possible, as galaxies without a strong Lyman-break (so called `HST-dark' galaxies) were often missed in the Ultraviolet (UV) selected samples of pre-JWST studies \citep{perezgonzalez2022}. Observations with Spitzer IRAC were available only for the brightest sources due to low sensitivity and angular resolution. Intrinsically UV-faint galaxies are often found to be dusty or evolved systems, and accurate characterisation of this population is essential when measuring the total buildup of stellar mass in the Universe, which is typically done by measuring the Galaxy Stellar Mass Function (GSMF). 

Stellar masses are typically estimated from fitting spectral energy distributions to broadband photometry, and inferring a star formation history along with other physical parameters such as metallicity, dust obscuration and ionization state of the gas \cite[e.g.,][]{Brinchmann2000, Bundy2006, Mortlock2011, duncan2014mass, Mortlock2015,  grazian2015galaxy, Song2016, bhatawdekar2019evolution}. This approach has generally been found to be reliable in the local Universe. 
At high-redshift there are a number of complicating factors that must be considered, which have different effects but overall act to increase the uncertainty in stellar masses beyond the statistical uncertainty from the fitting \citep{lower2020well,wang2023quantifying}. There is a growing consensus that the star formation histories of many high-$z$ galaxies are stochastic and characterised by repeated cycles of a short burst of rapid star formation followed by a temporary period of quiescence \citep{faucher2018model, Looser2023, asada2023bursty, Dressler2023, dome2024mini}. Young, bright stars created in the most recent bursts of star formation can dominate the spectral energy distributions of galaxies and obscure older populations, known as 'outshining' and lead to the stellar mass being underestimated by up to 1 dex \citep{gimenez-artega2022, endsley2021iii, perezgonzalez2022, papovich2022ceers, jain2024motivation,giménezarteaga2024outshining}. For extremely stochastic SFHs, information loss of the first periods of star formation may occur, leading to large uncertainties in stellar mass \citep{narayanan2023outshining, shen2023impact,van2023no}. Flexible `non-parametric' SFHs, such as those presented in \cite{leja2019measure, tacchella2022stellar, Robertson2022}, are able to reproduce these bursty SFHs more accurately at high-$z$, typically finding systematically larger stellar masses, \citep{gimenez-artega2022,jain2024motivation}, and may produce more reliable stellar mass estimates when compared to traditional parametric SFHs, which typically vary smoothly \citep[e.g. delayed exponential]{carnall2019measure}. 

Another assumption is that of a possible universal initial mass function (IMF), which predicts the number of stars as a function of stellar mass \citep{salpeter1955luminosity}. The presence of low-mass stars, which dominate the stellar mass, is inferred entirely from the shape of the assumed IMF in most galaxies.
A universal IMF \citep[e.g.][]{salpeter1955luminosity,Chabrier2000,kroupa2001variation}, which is constant across time and space, has long been assumed in the majority of galaxy studies. 
Models of the the physics of high-$z$ star formation suggest that it may have deviated from the universal IMF above redshift $z=8$ \citep{Hopkins2005, Jermyn2018,Steinhardt2021, Steinhardt2022}. Low metallicity, high star formation rates, an increasing CMB temperature and high cosmic ray density could all contribute to heating of star formation regions and lead to a top-heavy IMF at high redshift \citep{gunawardhana2011galaxy,clauwens2016implications,papadopoulos2011extreme, 2024A&A...686A.138C}. Observations of local elliptical galaxies have also found them to be inconsistent with the universal IMF, and instead find evidence for an IMF gradient, with evidence for a different IMF between  younger and older stellar populations \citep[e.g.][\& references therein]{weidner2013galaxy, 2021A&A...655A..19Y}.  Studies such as \cite{2018A&A...620A..39J,Steinhardt2022,Sneppen2022,woodrum2023jades,2024A&A...686A.138C} have introduced temperature/time-dependent IMF models for use in SED-fitting, which can decrease stellar mass estimates by up to 1~dex at constant redshift. Constraining the IMF is extremely difficult, but recent studies are beginning to find possible evidence for a top-heavy IMF in the early Universe \citep[e.g.][]{katz2022nature,cameron2023nebular,2024arXiv240208696M}.  Ultimate conclusions on this are however far from certain. 

In the EPOCHS paper series we have presented an independent and consistent reduction of deep JWST observations from available GTO, GO and ERS data including the CEERS, GLASS, SMACS-0723, JADES and PEARLS fields. We found 1165 robust galaxy candidates above redshift $z=6.5$, with a total area of 187 arcmin$^2$ \citep[][]{2024arXiv240714973C, adams2023epochs}.

In this paper we present a detailed examination of the inferred physical parameters of our high-redshift sample, with a particular focus on the stellar mass of our galaxy candidates. We derive a galaxy stellar mass function at $6.5 \leq z \leq$13.5, and estimate the stellar mass density in order to trace the buildup of stellar mass in the early Universe. Importantly, we explore the impact of some of the key assumptions used in deriving stellar masses at high redshifts, such as star formation histories and the IMF.

In \autoref{sec:data} we present a brief overview of the data products used in this work and detail our data reduction procedure. \autoref{sec:cat} describes our process for catalogue creation and robust sample selection of high-$z$ galaxy candidates. We detail our SED-fitting procedure and the impact of different assumptions on the inferred properties of high-$z$ galaxies in \autoref{sec:gal_prop}. In \autoref{sec:gsmf} we use our stellar mass estimates to build on the UV luminosity function presented in \cite{adams2023epochs} and construct a galaxy stellar mass function at redshifts from 6.5 $\leq z \leq$ 13.5. We discuss our findings and make comparisons to the literature in \autoref{sec:discuss}. Finally \autoref{sec:conclusions} summarises the findings of this work and looks at possibilities for future studies.

We assume a standard $\Lambda\mathrm{CDM}$ cosmology with $H_0=70$\,km\,s$^{-1}$\,Mpc$^{-1}$, $\Omega_{\rm M}=0.3$ and $\Omega_{\Lambda} = 0.7$. All magnitudes listed follow the AB magnitude system \citep{Oke1974,Oke1983}. All stellar masses measured in this work use a \cite{Kroupa_IMF_2002} unless otherwise indicated.

\section{Products and Data Reduction} 

 \label{sec:data}
This section briefly details the JWST programs and data products used in this analysis.
For further details on the fields used please see \cite{austin2023first}, \cite{adams2023epochs} and \cite{2024arXiv240714973C}. \autoref{tab:areas_depths} shows the available unmasked areas, JWST NIRCam filters and 5$\sigma$ depths of each dataset used.

\begin{table*}[]
\centering
\caption{Table showing the unmasked areas and depths of the observations used in this work. Depths are given at 5$\sigma$ in AB magnitudes, measured in 0$\farcs$16 radius apertures. Depths are calculated by placing non-overlapping apertures in empty regions of the image, as measured by the \sextractor{} segmentation maps and our image masks. The nearest 200 apertures are used to calculate the Normalised Mean Absolute Deviation (NMAD) to derive local depths for each individual source. Where depths are tiered across mosaics (e.g. HST \ACSWFC{} (ACS/WFC) observations in the Hubble Ultra Deep Field (HUDF) Parallel 2) we have listed the depths and areas separately. The four spokes of the NEP-TDF and ten CEERS pointings have uniform depths (within 0.1 mags) with the exception of CEERS P9 which we list separately. Areas are given in arcmin$^2$ and measured from the mask to account for the masked areas of the image and unused cluster modules. Fields with a `*' indicate that we have excluded the NIRCam module containing a lensing cluster from our analysis.}
\label{tab:areas_depths}

\setlength{\tabcolsep}{3pt}
\begin{tabular}{|l|l|ll|lllllllll|}
\hline
              & Area      & \multicolumn{2}{l|}{HST/ACS\_WFC} & \multicolumn{9}{c|}{JWST/NIRCam}                                       \\
Field         & (arcmin$^2$)  & F606W           & F814W           & F090W & F115W & F150W & F200W & F277W  & F335M & F356W & F410M & F444W \\ \hline
NEP           & 57.32 & 28.74           & -               & 28.50 & 28.50 & 28.50 & 28.65 & 29.15  & -     & 29.30 & 28.55 & 28.95 \\
El Gordo*      & 3.90  & -               & -               & 28.23 & 28.25 & 28.18 & 28.43 & 28.96  & -     & 29.02 & 28.45 & 28.83 \\
MACS-0416*     & 12.3  & -               & -               & 28.67 & 28.62 & 28.49 & 28.64 & 29.16  & -     & 29.33 & 28.74 & 29.07 \\
CLIO*          & 4.00  & -               & -               & 28.12 & -     & 28.07 & 28.21 & 28.675 & -     & 28.91 & -     & 28.71 \\
CEERS         & 66.40 & 28.6            & 28.30           & -     & 28.70 & 28.60 & 28.89 & 29.20  & -     & 29.30 & 28.50 & 28.85 \\
CEERSP9       & 6.08  & 28.31           & 28.32           & -     & 29.02 & 28.55 & 28.78 & 29.20  & -     & 29.22 & 28.50 & 29.12 \\
SMACS-0723*    & 4.31  & -               & -               & 28.75 & -     & 28.81 & 28.95 & 29.45  & -     & 29.55 & -     & 29.28 \\
GLASS         & 9.76  & -               & -               & 29.14 & 29.11 & 28.86 & 29.03 & 29.55  & -     & 29.61 & -     & 29.84 \\
NGDEEP HST-S  & 1.28  & 29.20           & 28.80           & -     & 29.78 & 29.52 & 29.48 & 30.28  & -     & 30.22 & -     & 30.22 \\
NGDEEP HST-D  & 4.03  & 30.30           & 30.95           & -     & 29.78 & 29.52 & 29.48 & 30.28  & -     & 30.22 & -     & 30.22 \\
JADES Deep GS & 22.98 & 29.07           & -               & 29.58 & 29.78 & 29.68 & 29.72 & 30.21  & 29.58 & 30.17 & 29.64 & 29.99 \\ \hline

\end{tabular}
\end{table*}

\subsection{PEARLS}
We incorporate NIRCam observations from the proprietary GTO survey \textit{Prime Extragalactic Areas for Reionization Science} \citep[PEARLS, PI: R. Windhorst \& H.Hammel, PID: 1176 \& 2738, ][]{Windhorst2023}. We use observations of three fields targeting gravitationally lensed clusters, and one blank field consisting of a mosaic of 8 JWST NIRCam pointings. The gravitationally lensed clusters consist of MACS J0416.1-2403 (hereafter referred to as MACS-0416), El Gordo \citep[$z \sim 0.87$, ACT-CL J0102-4915 in the Atacama Cosmology survey, \citep{2012ApJ...748....7M} and Clio $z \sim 0.42$, Designation GAMA100050 in the GAMA Galaxy Group Catalog v6+,][]{robotham2011galaxy}. El Gordo and Clio have been visited once with JWST/NIRCam, with the cluster centered in one NIRCam module and the other observing a neighbouring blank field $\sim  3\arcmin$ away \citep{griffiths2018muse}. MACS-0416 has been observed 3 times, resulting in 3 separate parallel observations at different position angles. The PEARLS blank field is the North Ecliptic Pole Time Domain Field \citep[(NEP-TDF,]{Jansen2018}. The NEP-TDF is positioned so it can be observed throughout the year, making it ideal for time-domain science and constructing a large deep field. Observations of the NEP-TDF consist of four pairs of overlapping NIRCam pointings (8 pointings total), with each of these four pairs orientated at 90 degree intervals like spokes on a windmill. NIRCam observations of the NEP-TDF, El Gordo and MACS0416 use the standard 8 photometric bands; F090W, F115W, F150W, F200W, F277W, F356W, F410M and F444W. Clio uses 6 of the 8 previous bands, but lacks F115W and F410M. Within the NEP-TDF field we incorporate HST \ACSWFC{} (ACS/WFC) imaging in the F606W filter, collected as part of the GO-15278 (PI: R.~Jansen) and GO-16252/16793 (PIs: R.~Jansen \& N.~Grogin) between October 1 2017 and October 31 2022. Mosaics of these data, astrometrically aligned to Gaia/DR3 and resampled on 0\farcs03 pixels, were made available pre-publication by R.~Jansen \& R.~O'Brien \citep[private comm.][]{o2024treasurehunt}. For full details of the PEARLS program please see \cite{Windhorst2023}. 

\subsection{ERS and GO Data}

We incorporate Early Release Science (ERS) and public General Observer (GO) data from SMACS-0723 \citep[PID: 2736, PI: K. Pontoppidan,][]{pontoppidan2022jwst}, the \textit{Cosmic Evolution Early Release Science Survey} \citep[CEERS, PID: 1345, PI: S. Finkelstein, see also][]{Bagley2023ceers}, the \textit{Grism Lens Amplified Survey from Space} survey \citep[GLASS, PID: 1324, PI: T. Treu,][]{Treu2022} and the \textit{Next Generation Deep Extragalactic Exploratory Public Survey} \citep[NGDEEP, PID: 2079, PIs: S.\@ Finkelstein, Papovich and Pirzkal,][]{Bagley2023ngdeep}. We incorporate HST ACS/WFC observations of the Extended Groth Strip \citep[EGS][]{davis2007all} into our CEERS dataset in the F606W and F814W filters. This was obtained as part of the \textit{Cosmic Assembly Near-infrared Deep Extragalactic Legacy Survey} \citep[CANDELS,][]{Grogin2011,Koekemoer2011}, with updated astrometric alignment to Gaia EDR3 \citep{brown2021gaia} by the CEERS team\footnote{\url{https://ceers.github.io/hdr1.html}} and released as \textit{Hubble Data Release 1}. The addition of these observations compensates for the lack of F090W observations in the CEERS survey. 

We also incorporate NIRCam imaging of the \textit{Great Observatories Origins Deep Survey South} (GOODS-South) field collected as part of \textit{JWST Advanced Deep Extragalactic Survey} \citep[JADES, PID:1180, PI: D. Eisenstein, ][]{Eisenstein2023} and released publicly as JADES DR1 \citep{Rieke2023}. In the JADES and NGDEEP fields, which lie on the GOODS-South footprint, we add in existing HST data from F606W and F814W from the most recent mosaic (v2.5) from the Hubble Legacy Fields team \citep{Illingworth2016,Whitaker2019}.

\subsection{JWST NIRCam Data Reduction}

We have uniformly reprocessed all lower-level JWST data products following our modified version of the official JWST pipeline. This is a similar process to that used in \citet{Leo2022}, \citet{Adams2023}, \citet{austin2023first}, and in particular \citet{adams2023epochs} but with updates based on new flat-fielding and techniques for dealing with NIRCam imaging artefacts.

We use version 1.8.2 of the official STScI JWST Pipeline\footnote{\url{https://github.com/spacetelescope/jwst}} \citep{Bushouse2022} and Calibration Reference Data System (CRDS) v1084, which contains the most up-to-date NIRCam calibrations at the time of writing and includes updated flat-field templates for the LW detectors, resulting in improved average depths across a single pointing of $\sim$ 0.2 dex in F444W. Next we subtract templates of wisps, artefacts present in the F150W and F200W imaging, between stage 1 and stage 2 of the pipeline. After stage 2 of the pipeline, we apply the $1/f$ noise correction derived by Chris Willott\footnote{\url{https://github.com/chriswillott/jwst}}, which removes linear features caused by read-noise from the images. We do not use the sky subtraction step included in stage 3 of the pipeline, instead performing background subtraction on individual NIRCam frames between stage 2 and stage 3 (`cal.fits' files), consisting of an initial uniform background subtraction followed by a 2-dimensional background subtraction using {\tt photutils} \citep{larry_bradley_2022_6825092}. This allows for quicker assessment of the background subtraction performance and immediate fine-tuning of configuration parameters. After stage 3 of the pipeline, we align the final F444W image onto a Gaia-derived World Coordinate System (WCS) \citep{GAIADR2,GAIADR3} using {\tt tweakreg}, part of the DrizzlePac python package\footnote{\url{https://github.com/spacetelescope/drizzlepac}}, and then match all remaining filters to this derived WCS, ensuring the individual images are aligned to one another. In some cases (NEP, CEERS), we match to a WCS-frame derived from other space or ground based imaging with a larger FOV, given the low number of Gaia stars in some individual NIRCam pointings. We then pixel-match the images to the F444W image with the use of {\tt astropy, reproject} \citep{Astropy2013,astropy:2022,Hoffmann2021}.\footnote{\url{https://reproject.readthedocs.io/en/stable/}} The final resolution of the drizzled images is 0\farcs03/pixel. Comparison of our reduction to the official PEARLS reduction pipeline \citep{Windhorst2023} is given in \cite{adams2023epochs}, finding excellent agreement in both observed fluxes (within 0.03 (0.01) magnitudes in the blue (red) NIRCam photometric filters) and astrometry (within 2 pixels (0\farcs07)).

\section{Catalogue Creation and Sample Selection}
\label{sec:cat}
Full details of our catalogue creation and sample selection pipeline, called {\tt GALFIND}, is available in \cite{2024arXiv240714973C}. We briefly summarise the procedure used here. 

\subsection{Catalogue Creation}
We use the code {\tt SExtractor} \citep{Bertin1996} for source identification and photometric measurements. We use an inverse-variance weighted stack of the NIRCam F277W, F356W and F444W images for source detection in order to reliably identify faint sources and then carry out forced aperture photometry in all photometric bands. This photometry is calculated within 0.32 arcsecond diameter circular apertures, correcting for the aperture size with an aperture correction derived from simulated \texttt{WebbPSF} point spread functions (PSFs) for each band used \citep{Perrin2012,Perrin2014}. This diameter was chosen to enclose the central/brightest $70-80$ per cent of the flux of a point source without a large amount of contamination from neighbouring sources. This reduces the reliance on a potentially uncertain PSF model whilst still using the brightest pixels when calculating fluxes.

\textbf{We choose not to PSF homogenize our imaging, and instead rely on a band dependent aperture correction to account for the changing PSF. Other studies have shown that the WebbPSF can underestimate the broadening of the PSF at small scales \citep[e.g.][]{2024ApJ...963....9M, 2024ApJS..270....7W}, and measuring an empirical PSF is extremely difficult in some of our fields given both the limited number of isolated stars in single pointing fields (e.g. NGDEEP, El Gordo or CLIO), or more complex position dependent PSFs due to the stacking of exposures taken at different times and position angles (e.g. NEP and MACS-0416). In the JADES-Deep-GS field we tested whether not PSF-matching had a systematic impact on the derived fluxes, or the galaxy properties derived from SED fitting. We repeated our full analysis of this field, using PSF matched imaging, using PSF models and kernels derived following the methodology of \cite{2024ApJS..270....7W, suess2024medium}, finding very little impact ($<$ 0.05 magnitudes) on the resultant aperture fluxes across all bands, which is well within the 10\% flux uncertainty floor used when SED fitting.}

As {\tt SExtractor} requires all images to be on the same pixel grid, for aligned HST imaging on a different pixel grid we use {\tt photutils} to perform forced aperture photometry in the same diameter apertures \citep{larry_bradley_2022_6825092}.

We next produce masks for our images by eye to cover diffraction spikes, any remaining snowballs, the cross pattern between SW detectors, image edges (including a $\sim50-100$ pixel border around detector edges) and any large foreground galaxies. The total amount of unmasked area used in this study is listed alongside the average 5$\sigma$ depths of each field in \autoref{tab:areas_depths}.

Following the generation of source catalogues and segmentation maps for each image, we calculate local depths for each source in each filter. This accounts for variation in background and noise across the image. Apertures of 0$\farcs$32 diameter are placed in empty regions of the image, as calculated from the segmentation map to be $\geq 1\arcsec$ from the nearest source. For each source the nearest 200 apertures are used to calculate the Normalised Mean Absolute Deviation (NMAD) of the fluxes measured in the apertures, which corresponds to the 1$\sigma$ flux uncertainty. We convert this to a 5$\sigma$ depth, displaying the average depth in AB magnitudes for each field in \autoref{tab:areas_depths}. Where fields consist of mosaics of multiple pointings we display the average depth across the entire field, but we note that 9/10 CEERS pointings and 4/4 NEP-TDF have depths consistent within 0.1 magnitudes. The exception in CEERS is pointing 9 (P9), which has an additional exposure in F115W and F444W resulting in increased depths of $\sim 0.2$ mags \citep{adams2023epochs}. Due to correlated noise,  the flux uncertainties calculated by {\tt SExtractor} are underestimated \citep{2020AJ....160..231S}. We replace these uncertainty estimates with the local-depth derived flux errors \citep{adams2023epochs}. 

\subsection{Sample Selection}
\label{sec:sample_select}
To select a sample of high-redshift galaxies we introduce selection criteria based primarily on photometric SED-fitting with \eazy{} \citep{brammer2008eazy}. We aim to select a robust sample of galaxies above z$ \geq 6.5$ , where the Lyman break is within the NIRCam F090W filter.

We use the default EAZY templates (tweak\_fsps\_QSF\_12\_v3), along with Set 1 and Set 4 of the SED templates generated by \cite{Larson2022}. These additional templates were developed to have bluer rest-frame UV colors than the default templates, as well as stronger emission lines, both of which have been observed in high-redshift galaxies \citep{Finkelstein2022a,Nanayakkara2022,cullen2023ultraviolet,withers2023spectroscopy}. These templates have young stellar populations, low metallicities and active star formation. \cite{Larson2022} has shown that they improve the accuracy of photo-$z$ estimates at high redshift. 

We run \eazy{} initially with a uniform redshift prior of $0 \leq z \leq 25$, but then repeat the fitting with a reduced upper redshift limit of $z \leq 6$. This allows us to compare the goodness of fit of both a high and low-redshift solution for all galaxies in our sample. We use a minimum flux uncertainty of 10\% to account for uncertainties in flux calibrations and aperture corrections \citep{rieke2022nircam}, as well as potential differences between the synthetic templates and our galaxies.

For reproducibility our selection criteria are designed to be based as much as possible on specific cuts in computed quantities, rather than individual inspection of candidates which can introduce hard to measure bias and incompleteness. 
To ensure robustness in our sample, our final selection criteria includes a visual review of the cutouts and SED-fitting solutions for all sources by authors TH, DA, NA, \& QL but we reject less than 5\% of our original sample by eye at this stage, which is much lower than comparable studies \citep[reaching $\geq$50\% in some cases, e.g. ][]{Hainline2023}.  

Our selection criteria for robust high redshift galaxies are as follows. 
\begin{enumerate}
    \item We require that the bandpass of the lowest wavelength photometric band must be entirely below the Lyman break, given the primary photo-$z$ solution. This sets a lower limit of $z \approx6.5$ in most of our fields. 
    \item We require a $\leq 3 \sigma$ detection in band(s) blueward of the Lyman break. 
    \item We require a $\geq 5\sigma$ detection in the 2 bands directly redward of the Lyman break, and $\geq 2\sigma$ detection in all other redward bands, excluding observations in NIRCam medium-bands (e.g. F335M, F410M). If the galaxy appears only in the long wavelength NIRCam photometry (i.e. a F200W or higher dropout), we increase the requirement on the 1st band to 7$\sigma$ detection.
    \item The integral of the photo-$z$ PDF is required to satisfy $\int^{1.10\times z_{\textrm phot}}_{0.90\times z_{\textrm phot}} \ P(z) \ dz \ \geq \ 0.6 $ to ensure the majority of the redshift PDF is located within the primary peak, and that the peak is sufficiently narrow to provide a strong constraint on the redshift. $z_{\textrm phot}$ refers to the redshift with maximum likelihood from the \eazy{} redshift posterior. 
    \item We require the best-fitting \eazy{} SED to satisfy $\chi^2_{red} < 3 (6)$ to be classed as a robust (good) fit.
    \item We require a difference of $\Delta \chi^2 \geq$ 4 between the high-$z$ and low-$z$ \eazy{} runs. This ensures that the high-$z$ solution is much more statistically probable. 
    \item If the half-light radius ({\tt FLUX\_RADIUS} parameter in {\tt SExtractor}) is smaller than the FWHM of the PSF in the F444W band, then we require that $\Delta \chi^2 \geq$ 4 between the best-fitting high-$z$ galaxy solution and the best-fitting brown dwarf template. This requirement is designed to remove Milky Way brown dwarf contaminants and is discussed further in \autoref{sec:brown_dwarfs}.
    \item We require the 50\% encircled flux radius from {\tt SExtractor} to be $\geq$1.5 pixels in the long-wavelength wideband NIRCam photometry (F277W, F356W, F444W). This avoids the selection of oversampled hot pixels in the LW detectors as F200W dropouts. 
    
\end{enumerate}
Given our requirement to observe the Lyman break, the lowest redshift at which we select `robust' galaxies with NIRCam photometry only is $\sim$6.5, where the break falls within the NIRCam F090W filter. In the fields where HST ACS/WFC imaging is available, we can robustly identify the Lyman break at lower redshifts.

Our selection criteria are similar to other high-$z$ galaxy studies, such as \cite{Hainline2023,Finkelstein2022c,Finkelstein2023,Naidu2022a,Castellano2022}, who also fit galaxies using \eazy{}, and select their samples from the rest-UV SNR of their candidates and the resultant redshift PDF and $\Delta \chi^2$ from the SED fitting. The $\Delta \chi^2 \geq 4$ requirement was chosen to ensure that the high-z solution is a better fit than any possible low-z explanation, and is consistent with the criteria used in other studies, such as \cite{Finkelstein2023}. SNR requirements vary between studies, and are also somewhat dependent on the size of the extraction aperture used. Our 0\farcs16 radius apertures are larger than the 0\farcs1 radius apertures used in \cite{Finkelstein2023}, resulting in a number of their candidates being removed from our sample because they do not meet our 5$\sigma$ threshold, despite agreeing on the photo-$z$ solution. Other studies, such as as \cite{Harikane2023,Yan2022b}, use a combination of color-color cuts and SED-fitting. A comprehensive comparison of the EPOCHS v1 galaxy sample to other studies is given in Appendix A of \cite{adams2023epochs}. 

Spectroscopy of high redshift galaxies with NIRSpec \citep[e.g. ][]{wang2023uncover,2023arXiv230109482F,Bunker2023,2023NatAs.tmp...66C} has shown that many high redshift galaxies have strong emission lines, with \hbeta+\oiii \ reaching observed equivalent widths of up to 2000-3000\AA{} \citep{withers2023spectroscopy}. For our photometric observations this can result in an excess in the band covering these emission lines of up to $\sim$ 1 dex. The emission line modelling in the SED fitting codes used span only a limited parameter space in equivalent width and line ratios, so this can result in high $\chi^2$ values even if the model is a good fit to the rest of the photometry. To avoid removing these galaxies from our sample we introduce a secondary group, referred to as `good' galaxies, which have $3< \chi_{r}^2 < 6$ but meet the rest of our criteria. This applies to a very small fraction of our total sample, with only 23 galaxies meeting our other selection criteria but falling within this $\chi^2$ range.

\section{Galaxy Properties from SED Fitting}
\label{sec:gal_prop}

The basic properties, redshift distribution, number counts and UVJ colors of the EPOCHS v1 sample are described in detail in \cite{2024arXiv240714973C}. The UV properties (\MUV, $\beta$ slopes) of this sample are explored in \cite{2024arXiv240410751A}. The UV luminosity function is presented in \cite{adams2023epochs}, Here we briefly summarise the basic statistics of the EPOCHS v1 sample, following the selection criteria described in \autoref{sec:sample_select}. 

For full transparency, we list the number of candidate objects removed by each stage of our selection criteria as follows. We note that as we apply all our selection criteria to all galaxies, each object can fail multiple selection criteria. Across all the survey fields used, the parent sample consists of 211,469 objects which are unmasked in all available photometric bands. Of these, only 1545 have photometric redshifts between $6.5 \leq z \leq 13.5$ and also pass the SNR criteria (criteria 1 - 3, above). 32 are removed because they fail criteria 4, as the photo-z is not well constrained. 9 are removed by criteria 5, as the $\chi^2$ suggests that the best fitting SED does not reproduce the observed photometry. 333 are removed by criteria 6, which requires that the high-$z$ solution has a robust statistical improvement when compared with the low-$z$ ($z < 6.5$) solution. 63 brown dwarfs are identified and removed from the sample in criteria 7, which is discussed further in \autoref{sec:brown_dwarfs}. Finally 5 hot-pixels are removed by criteria 8, which have unphysically small sizes. Accounting for the 49 galaxies which fail multiple of the preceding criteria, this leaves 1214 galaxies  which pass our initial selection criteria with $z \geq 6.5$ over a total unmasked area of 187.27 arcmin$^2$, comprising one of the largest samples of high-$z$ JWST-selected galaxies.

Our requirement constraining the redshift PDF is important for confident photo-$z$'s and does not significantly bias our sample as only 7 galaxies of the full parent sample are rejected for this reason alone, and visual inspection of these 7 objects show they have SNR $<$ 10 and physically implausible SEDs.

Our visual inspection removes 49 completely, leaving 1165 galaxies, which we refer to as the EPOCHS v1 sample. Of these 1165, 1054 are classed as ``certain'' and 111 ``uncertain'' by our visual inspection. In this work we choose to include the visually ``uncertain'' candidates, in order to avoid potentially underestimating the stellar mass function. Filtering the sample to the $6.5 < z < 13.5$ redshift range used in this work, our fiducial sample consists of 1120 galaxies. 

The next sections detail the \bagpipes{} and \prospector{} SED fitting performed for all galaxies in the EPOCHS v1 sample. We perform this SED fitting only for our \eazy{}-selected sample. The purpose of this SED-fitting is to analyse the properties of the stellar and nebular components of these high-$z$ galaxies. We note that whilst the majority of galaxies in the EPOCHS v1 sample do not appear significantly extended beyond our \sextractor{} extraction aperture, all stellar masses quoted in the following results have been corrected by the ratio between the \sextractor{} \textit{FLUX\_AUTO\_F444W} and the aperture corrected \textit{FLUX\_APER\_F444W} fluxes, if it exceeded unity, to account for any residual flux outside the aperture. We use the longest wavelength band to correct stellar masses as this is most representative of the rest-frame optical emission. \textbf{We choose to perform all SED fitting using aperture fluxes, which ensures flux is measured from the same aperture in all filters, and avoids catastrophic issues with the \sextractor{} Kron aperture estimates, which sometimes become unphysically small or large. Some studies \cite[e.g.][]{suess2024medium,2024ApJS..270....7W, 2024arXiv240308872W} calculate total fluxes by scaling all the overall normalisation of all fluxes by the ratio between a Kron and aperture flux in a detection or reference band, which has no impact on the inferred galaxy colors. As SED fitting depends on the measured colors to infer the star formation history and other parameters, and the total stellar mass simply scales linearly with the normalization. Our tests using the JADES-Deep-GS field, in which we repeat our full catalogue creation, selection and SED fitting procedure on PSF-matched imaging with 'TOTAL' fluxes, finds resultant stellar masses consistent to within a mean (median) offset of 0.03 (0.01) dex. Given that this is well within the statistical uncertainty of our stellar mass estimates, we find that our methodology of using a per band aperture correction and aperture corrected stellar masses is completely consistent with the results of comparable studies.}
\subsection{Bagpipes}
\label{sec:bagpipes}
\bagpipes{} \citep[Bayesian Analysis of Galaxies for Physical Inference and Parameter EStimation,][]{bagpipes2018, carnall2019measure} is a Python package which uses Bayesian methods to fit galaxy SEDs to photometry. Our fiducial \bagpipes{} run uses the \cite{bruzual_charlot_2003} 2016 stellar population synthesis models, and a \cite{kroupa2001variation} IMF.

\bagpipes{} can construct and fit SEDs with a variety of SFH models, dust models and priors. We perform multiple fits for each galaxy in order to test the consistencies in derived galaxy parameters. For simplicity our approach is to define an initial `fiducial' model, and then swap out individual model components or priors for other choices. Our model parameters, priors and hyper-parameters are detailed in \autoref{tab:bagpipes_table}. This is similar to the approach of \cite{2021MNRAS.501.1568F}, who define a reference model which is then compared to alternative models.  In the rest of this section we further explain our choices of models and priors. 

For the star formation history we test both parametric and non-parametric models, which have been shown to impact the stellar mass estimates \citep{leja2019measure,tacchella2022stellar}. For our fiducial model, we use a parametric lognormal SFH which allows us to recreate the rising SFHs we expect in the early Universe. 
We compare this SFH with another commonly used parametrization, the delayed exponential. Details of the implementation of these parametric SFHs are available in \cite{carnall2019measure}. We also test a non-parametric ``continuity" SFH model similar to the model added to \prospector{} in \cite{leja2019measure}. This SFH model fits the star formation rates in fixed time bins, with $\Delta \log$SFR between bins linked by a Student's t-distribution. We recreate the methodology of \cite{tacchella2022stellar}, by fitting both a ``continuity" model, where the Student's t-distribution has hyper-parameters $\sigma = 0.3$ and $\nu  = 2$, which weights against rapid changes in star formation rate, as well as a ``continuity bursty" model, with $\sigma = 1$ and $\nu = 2$, which allows more stochastic star formation, with rapid bursts and quenching more similar to the SFH inferred at high-$z$. As is done in \cite{tacchella2022stellar} we fit 6 SFH bins for both models, with the first bin fixed to a lookback time of 0-10 Myr, the last bin ending at z = 20, and the other 4 bins are equally log-spaced in lookback time. \cite{leja2019measure} showed that this model is relatively insensitive to the number of bins used, as long as it is more than four. Like \cite{tacchella2022stellar}, we assume there is no star formation at $z > 20$ \citep{tacchella2022stellar, bowman2018absorption, jaacks2019legacy}. It is important to note that this model allows but does not require `bursty' SFHs, and smooth or quenched SFHs are also possible if favoured during the fitting. 

We include emission lines and nebular continuum based on \cloudy{} v17.03 \citep{2017RMxAA..53..385F}. We regenerate \cloudy{} models in order to probe a wider range of the ionisation parameter U between -3 $\leq \log_{10} U \leq $ -1 using a \cloudy{} configuration file distributed with \bagpipes. 

We use the \cite{calzetti2000dust} prescription for dust. \cite{2023arXiv230917386B} finds that UV-selected high-$z$ galaxies in the ALMA REBELS survey follow the local Calzetti-like IRX-$\beta$ relation, so we do not fit a more complex dust law in our fiducial model. The allowed stellar metallicity ranges from -2.3 $\leq \log_{10} Z_\star/Z_\odot \leq$ 0.70 with a logarithmic prior, as these galaxies are expected  to have low metallicity, but theoretically could enrich their local environments quickly \citep{langeroodi2023evolution,curti2023chemical}. 

In order to constrain the redshift parameter space we fix the redshift prior to the PDF from our \eazy{} SED-fitting, which we approximate as a Gaussian. The redshift prior draws are capped at $\pm 3 \sigma$. We use the default sampling method, using the Python package {\tt PyMultiNest} \citep{feroz2009multinest, buchner2016pymultinest}.

\begin{table*}
\centering
\caption{Summary of parameters, hyper-parameters and priors for our \bagpipes{} SED-fitting. The text color corresponds to a specific \bagpipes{} run, where black corresponds to our default `fiducial' setting for each parameter. Parameters and priors for other iterations can be assumed to be the same as given for the `fiducial' bagpipes run unless otherwise specified. The top section of the table lists parameters that are common to all of our \bagpipes{} models, whereas the lower section gives the model-specific parameters for each of our chosen configurations. \label{tab:bagpipes_table}}
\begin{tblr}{
cell{1}{1} = {c=2}{},
}

\textbf{Common Parameters} & &  \\ \hline \hline
Parameter                             & Prior/Value (Min, Max)                               & Description                         \\ \hline
$z_{\textrm{phot}}$                   & \eazy{} Posterior PDF ($\pm 3 \sigma$)                      & Redshift                            \\
SPS Model                             & \cite{bruzual_charlot_2003}; {\color{red} BPASS v2.2.1}                                     & Stellar population synthesis model  \\
IMF                                   & \cite{kroupa2001variation}; {\color{red} default BPASS IMF}                 & Stellar Initial Mass Function       \\
Dust Law parametrization                            & \cite{calzetti2000dust}; \hypersetup{citecolor=violet}\cite{2018ApJ...859...11S}\hypersetup{citecolor=blue}  & Dust law                            \\
& \hypersetup{citecolor=brown}\cite{charlot2000simple}\hypersetup{citecolor=blue} & \\
A$_\textrm{V}$ & Log-uniform: ($10^{-3}$, 10); {\color{orange}uniform (0, 6)} & V-band attenuation (all stars) \\

SFH                                   & lognormal; {\color{cyan}``continuity bursty"};  \color{blue}delayed-$\tau$                                           & Star formation history              \\
$\log_{10}(M_\star/M_{\odot})$            & uniform: (5, 12)                   & Surviving stellar mass              \\
Z$_\star/\textrm{Z}_{\odot}$              & log-uniform: (0.005, 5); {\color{magenta}uniform (0, 3)}                                    & stellar metallicity                 \\
Z$_{\textrm{gas}}/\textrm{Z}_{\odot}$ &  Fixed to Z$_\star$                          & gas-phase metallicity               \\
$\log_{10}\textrm{U}$                 & uniform: (-3, -1)               & Ionization Parameter                \\
\hline
\end{tblr}

\begin{tblr}{cell{1}{1} = {c=2}{}}
\textbf{Model Specific Parameters} & & \\  \hline \hline

Model & Parameter                             & Prior/Value (Min, Max)                               & Description                         \\ \hline
Fiducial & {t$_\textrm{max}$} & uniform: (10 Myr, 15 Gyr) & Age of Universe at peak SFR \\
& FWHM & uniform: (10 Myr, 15 Gyr)  & FWHM of SFH \\
{\color{blue}delayed-$\tau$ SFH} & $\tau$ &        uniform: (10 Myr, 15 Gyr) & e-folding timescale  \\
& Age & log-uniform:  (10 Myr, t$_{\rm univ}(z_{\rm phot}))$ & Time since SF began \\

{\color{cyan}```continuity bursty"'}  & N$_{\textrm{bins}}$ & 6 bins (5 fitted parameters) & First bin 0 -- 10 Myr, SF begins at z = 20,  \\
{\color{cyan}non-parametric SFH} & & & others distributed equally in log$_{10}$ lookback time \\
& d$_{\log_{10}\textrm{SFR}}$ & Student's-t: $\nu = 2, \sigma = 1.0 $ & Ratio of $\log_{10}$SFR in adjacent bins, coupled by $\sigma$ \\
\hypersetup{citecolor=brown}\cite{charlot2000simple}\hypersetup{citecolor=blue} & n & clipped normal: $\mu=0.7, \sigma = 0.3 $ & Power-law slope of attenuation curve (A $\propto \lambda^{-n}$) \\
\color{brown} Dust Law & & (0.3, 2.5) & For \cite{calzetti2000dust} n $\approx 0.7$ \\
& $\eta$ & clipped normal: $\mu=2, \sigma=0.3$ & A$_{\rm V, < 10 Myr}$/A$_{\rm V}$ ratio between young and old stars  \\ 
& &  (1, 3) & \\
\hypersetup{citecolor=violet}\cite{2018ApJ...859...11S}\hypersetup{citecolor=blue}{\color{violet} \ Dust Law} & $\delta$ & clipped normal: $\mu=0, \sigma = 0.1 $ & Deviation from \cite{calzetti2000dust} slope  \\
& & (-0.3, 0.3) & \\
 & $\beta$ & uniform (0, 5) & Strength of 2175\AA{} bump \\

{\color{red} BPASS SPS Model} & & & No additional components \\

{\color{orange} Uniform A$_\textrm{V}$  Prior} & & & No additional components \\
{\color{magenta} Uniform Z$_{\star}$ Prior} & & & No additional components \\
\hline
\end{tblr}
\end{table*}

\subsection{Prospector}\label{sec:prospector}

In addition to our \bagpipes{} SED-fitting we also fit our sample using the \prospector{} package \citep{Johnson2021} in order to compare the results of these two commonly used SED fitting tools. \prospector{} allows greater flexibility and control of model parameters than \bagpipes{} at the expense of computational time.
\prospector{} uses Bayesian inference to determine galaxy stellar population properties and star formation histories. \prospector{} is built on Flexible Stellar Population Synthesis \citep[FSPS, ][]{conroy2010fsps}), using {\tt python-fsps} \citep{johnson2023dfm} and the Modules for Experiments in Stellar Astrophysics Isochrones and Stellar Tracks (MIST) stellar isochrones \citep{choi2016mesa, dotter2016mesa}. \prospector{} allows for very flexible star formation histories, including the use of non-parametric SFHs, which have become increasingly popular in the JWST era. 
We generally follow the prescription of \cite{tacchella2022stellar} for our \prospector{} model. We test both a traditional parametric star formation history as well as the ``continuity bursty" non-parametric SFH used in \bagpipes. We allow the stellar mass to vary between 6 $\leq \log_{10} M/M_{\odot} \leq$ 12 with a uniform logarithmic prior. For our parametric star formation history we use the `delayed-$\tau$' model, where  $\textrm{SFR}(t) \propto (t - t_{s}) e^{-(t-t_{s})\tau}$. 

We note that by default \prospector{} provides the total stellar mass formed, rather than the surviving stellar mass. Post-fitting we recalculate the return fraction within \prospector{} for the full posterior in order to derive a surviving stellar mass distribution for each galaxy. We allow the V-band optical attenuation due to dust to vary between 0 and 6 magnitudes, with a uniform prior, assuming a \cite{calzetti2000dust} law also used in our `fiducial' \bagpipes{} run. 

We model IGM attenuation following \cite{madau1995radiative}. Following \cite{tacchella2022stellar} and due to possible line-of sight variations in the optical IGM attenuation, we allow the IGM opacity to vary with a clipped Gaussian prior distribution centered on 1, clipped at 0 and 2, and with a dispersion of 0.3. The stellar metallicity is allowed to vary between -4  $\leq \log_{10} Z/\mathrm{Z}_{\odot} \leq$ 0.16 with a uniform prior. We do not link the gas-phase metallicity, which is also free to vary with the same range and prior. Gas-phase metallicity is expected to differ from stellar metallicity at some stages of galaxy evolution (e.g. following significant gas accretion into a galaxy, lowering gas-phase metallicity) and decoupling them permits more flexibility in the stellar and nebular emission line modelling. 

We use a \cite{kroupa2001variation} IMF for consistency with our \bagpipes{} SED fitting, but as \prospector{} allows tabulated IMFs we also test the impact of a top-heavy IMF on the derived stellar masses. Top-heavy IMFs are predicted at high-$z$, and there is preliminary evidence for a top-heavy IMF in extreme nebular dominated galaxies at high-$z$ \citep{cameron2023nebular}. The results for different choices of IMF are detailed in \autoref{sec:imf_results}.

We run \prospector{} using nested sampling with {\tt dynesty} \citep{speagle2020dynesty}, using the default sampling settings with the exception of switching from uniform sampling to the more robust random walk, to efficiently probe the multi-dimensional parameter space. Since \prospector{} is computationally expensive to run, we fit only a subset of our galaxy sample, prioritising those galaxies which are high-redshift or massive. Specifically we fit all galaxies with $z_{\textrm{phot}}$ (from \eazy) above $z \geq 8.5$, or with a fiducial \bagpipes{} stellar mass of $\log_{10} (M_{\star}/M_{\odot}) \geq 9.0$.

\subsection{Stellar Mass Comparisons} \label{sec:mass_comparisons}

\begin{figure*}
    \includegraphics[width=\textwidth]{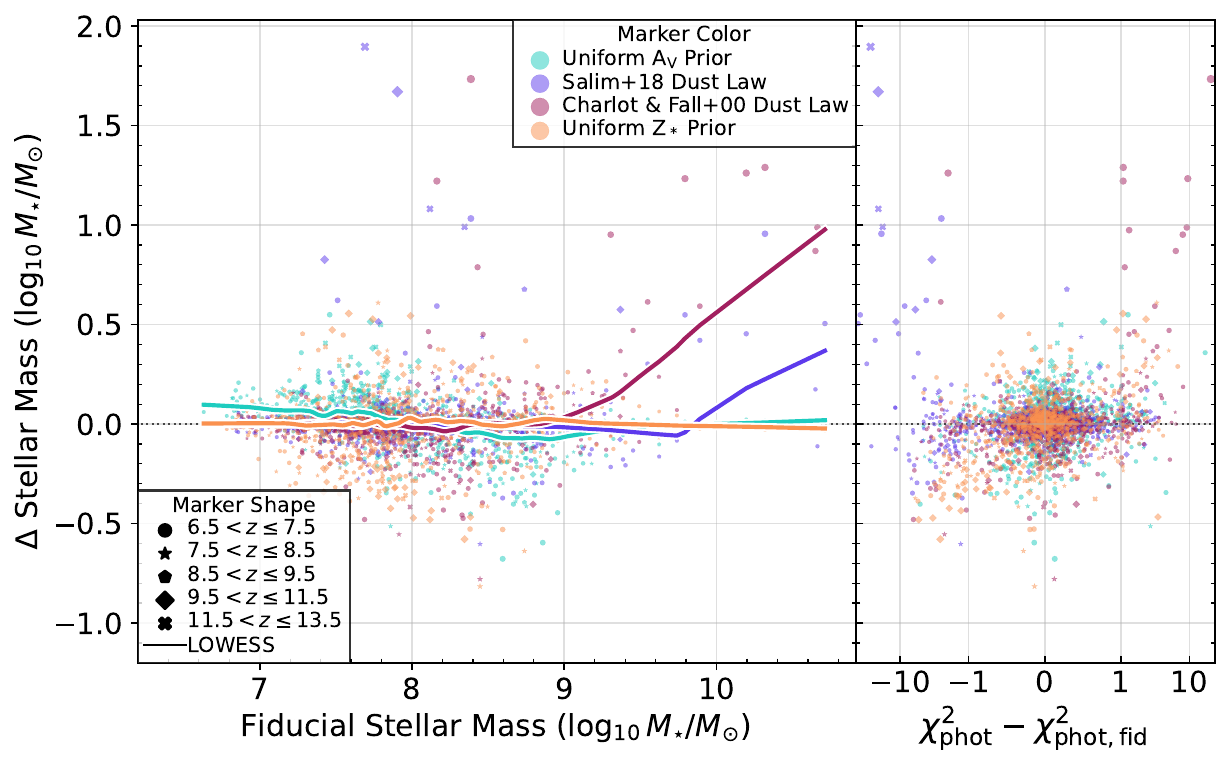}
    \includegraphics[width=\textwidth]{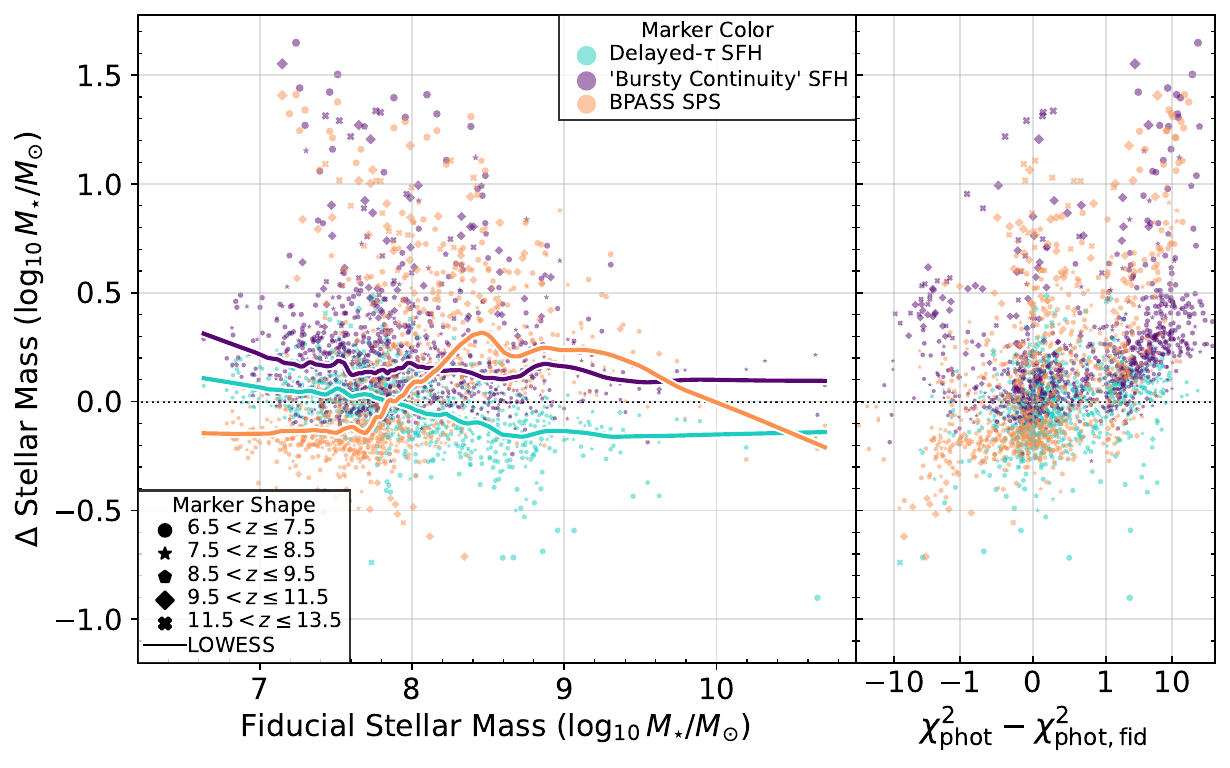}
    \caption{\textbf{Top}) Comparison of the observed stellar mass offset between our fiducial \bagpipes{} model and alternative models for the entire EPOCHS v1 sample, as a function of fiducial stellar mass. The alternative models considered vary priors on the metallicity and dust attenuation, as well as the assumed dust law. Marker shapes show the redshift bin for each galaxy, based on the fiducial redshift, and each \bagpipes{} model considered is shown in a different color. The colored lines show the \textit{Locally Weighted Scatterplot Smoothing} fit \citep[LOWESS,][]{cleveland1979robust}. The right-hand plots show the stellar mass offset as a function of $\Delta \chi^2$, between the two fits. \textbf{Bottom}) Same as the upper plot, but the alternative models vary the assumed star formation history or stellar population synthesis model instead. \label{fig:bagpipes_compar}}
\end{figure*}

In this section we compare the consistency of the derived stellar masses between our fiducial and alternative \bagpipes{} models.
\autoref{fig:bagpipes_compar} shows a comparison of the derived galaxy properties of our fiducial \bagpipes{} run for all sources in our sample to our other \bagpipes{} results. The details of the models compared are shown in \autoref{tab:bagpipes_table}. As discussed in \autoref{sec:bagpipes}, we compare models with different priors, SPS models and SFH parametrizations. The top plot shows the systematic mass offset found when varying the dust law and priors, while the second row shows the same offset for parametric and non-parametric SFHs, as well as a BPASS SPS model \citep{stanway2018re}. For each \bagpipes{} model we show the \textit{Locally Weighted Scatterplot Smoothing} fit \citep[LOWESS,][]{cleveland1979robust}. The LOWESS estimator is a non-parametric fitter for noisy data, and shows the overall trend between the mass discrepancy and fiducial stellar mass. 

\begin{figure*}
    \centering

    \gridline{\fig{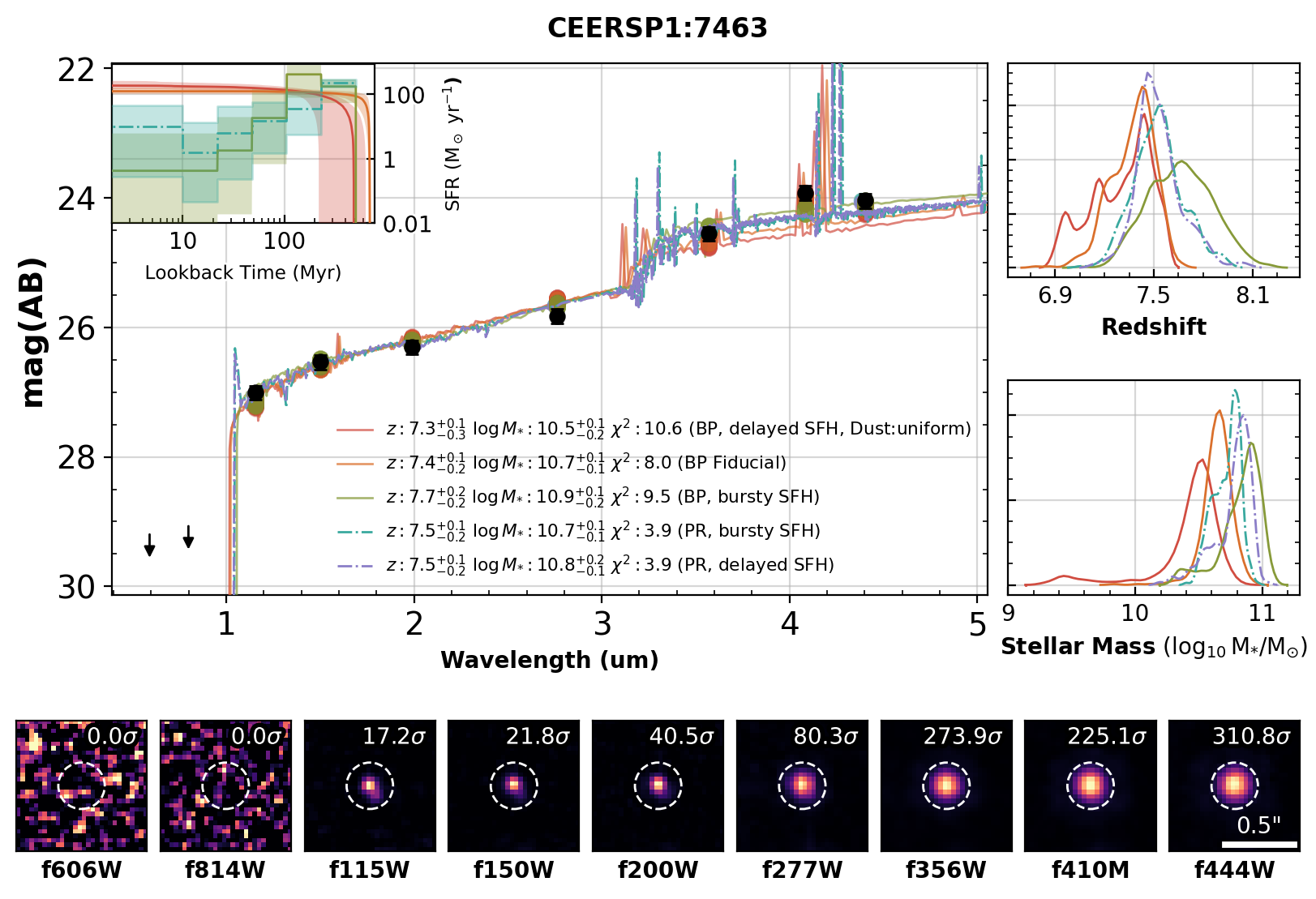}{0.80\textwidth}{}}
    \gridline{\fig{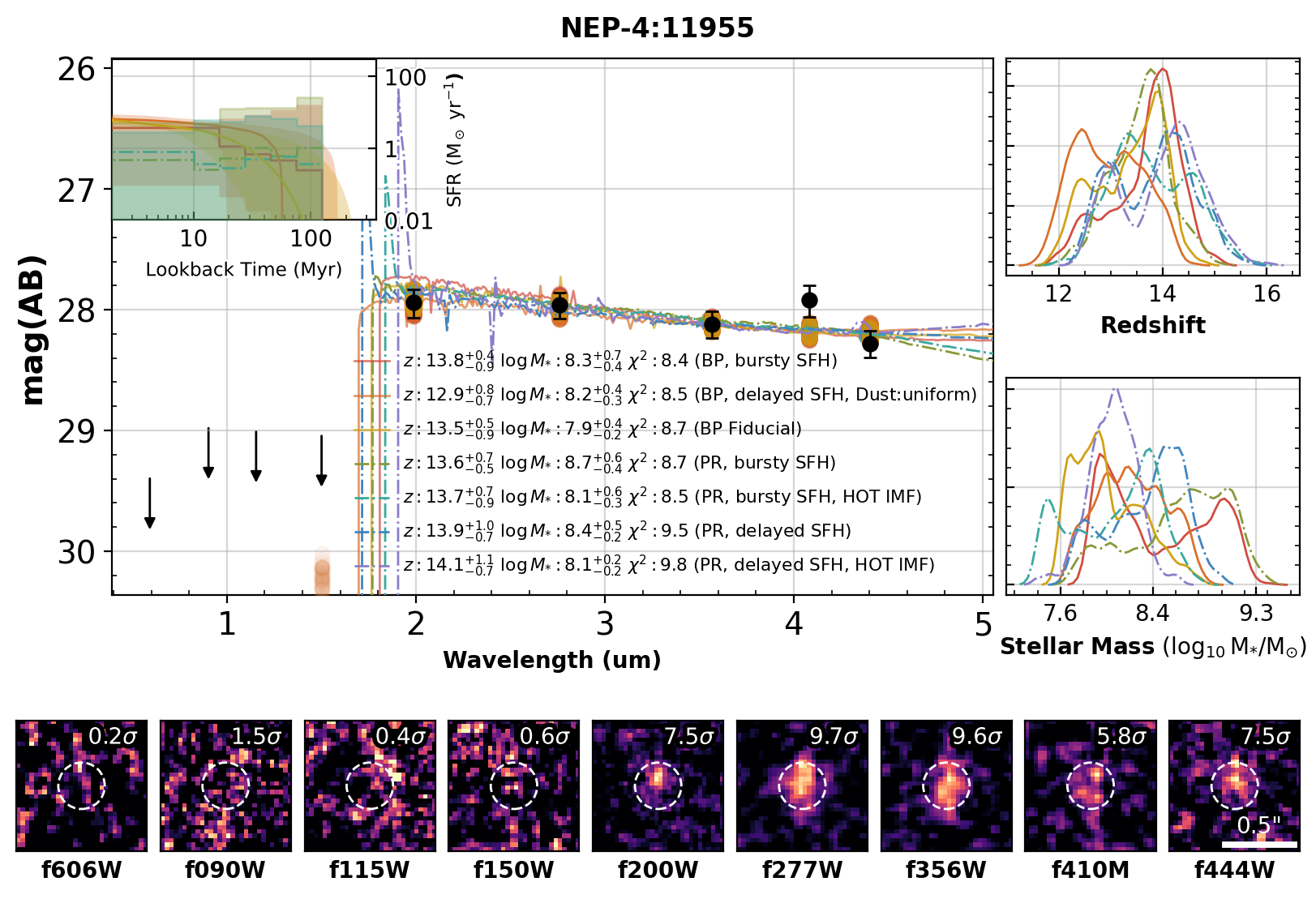}{0.80\textwidth}{}}
          
    \caption{\bagpipes{} (labelled BP) and \prospector{} SEDs (labelled PR) for two of the galaxies in the EPOCHS v1 sample. The top row shows the galaxy labelled \textit{7463\_CEERSP1} with one of the largest fiducial stellar masses in the sample, which is a 'Little Red Dot' also identified by \cite{2023Natur.616..266L,2024arXiv240109981K}. The bottom row shows \textit{11955\_NEP-4}, one of the more massive galaxies found at $z \geq 10$, although the range of stellar mass estimates shows how difficult it is to constrain stellar mass at these redshifts. 
    We label the different SED fits by how they differ to our fiducial \bagpipes{} SED-fitting prescription, with a lognormal SFH, \cite{calzetti2000dust} dust model, logarithmic priors on age, dust and metallicity, \cite{bruzual_charlot_2003} SPS model and a \cite{kroupa2001variation} IMF. For each alternative model we also give the redshift, stellar mass and $\chi^2$ for the fit. On the right of each SED plot we show the posterior redshift and stellar mass PDF distributions and the best-fit SFH history is shown as an inset axes on the upper left. \label{fig:high_mass_SEDs}}
    
\end{figure*}

\begin{figure*}
    \centering

    \gridline{\fig{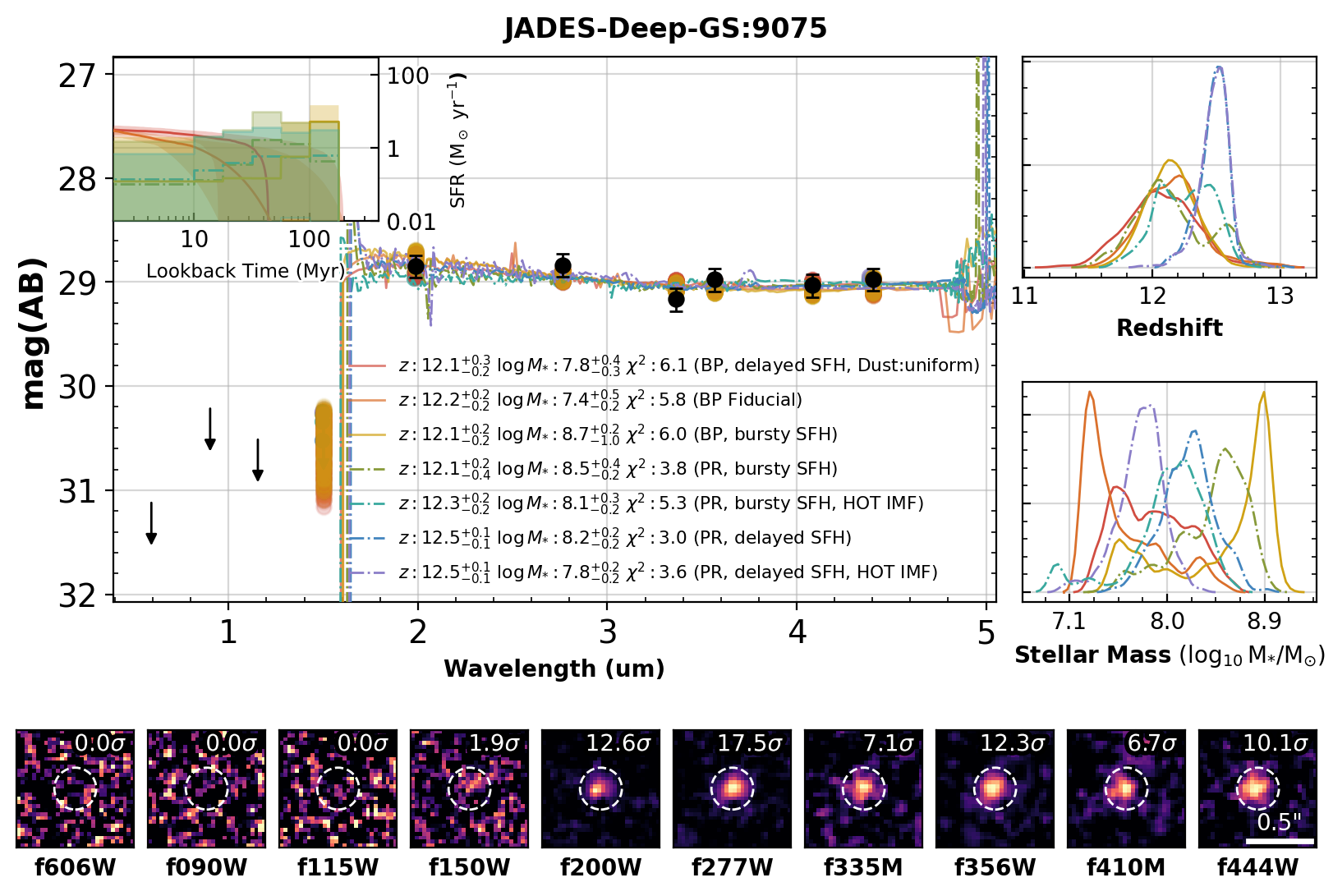}{0.80\textwidth}{}}
    \gridline{\fig{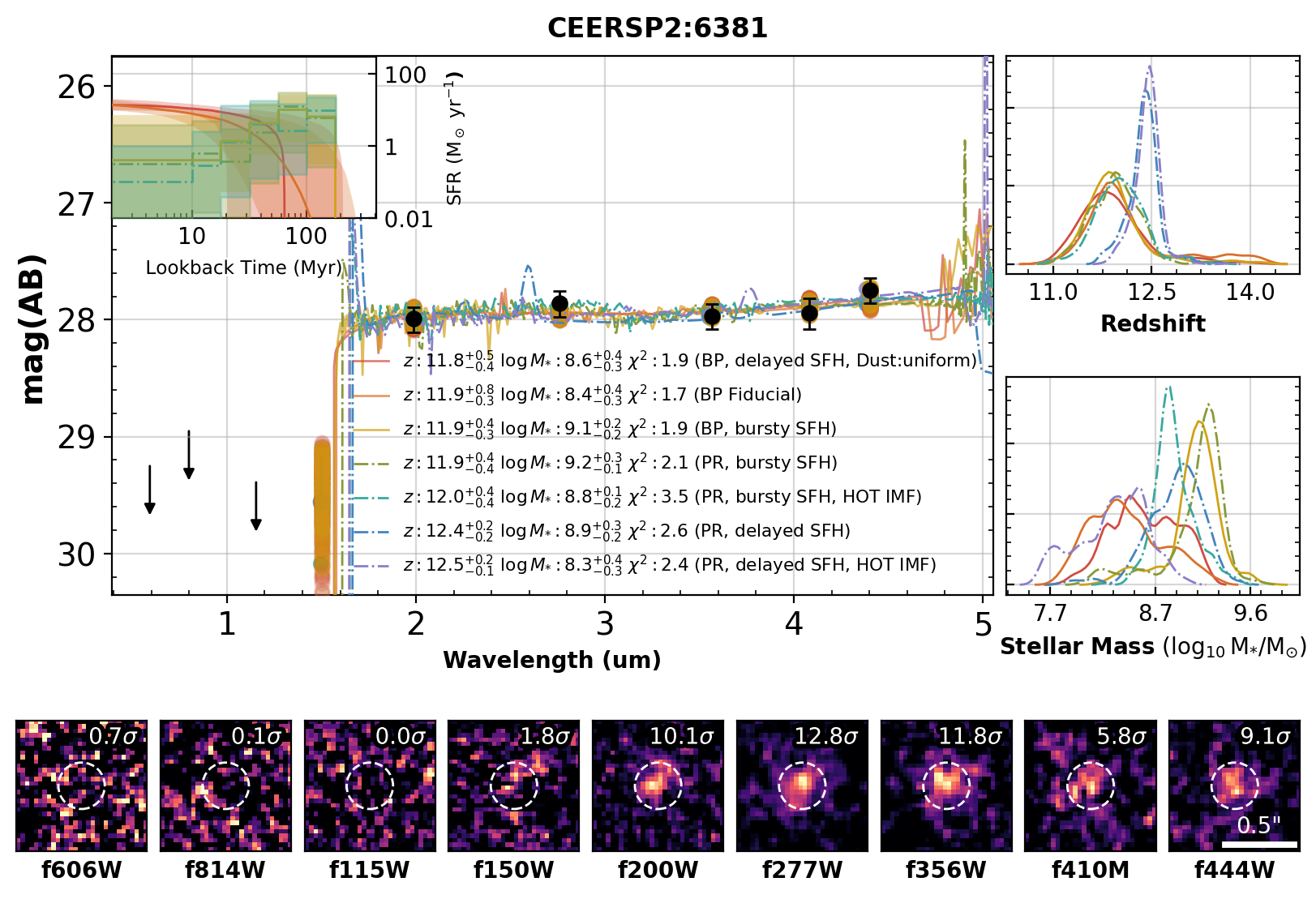}{0.80\textwidth}{}} 
          
    \caption{The same as \autoref{fig:high_mass_SEDs} except these galaxies are shown to demonstrate the  stellar mass discrepancies observed between different \bagpipes{} and \prospector{} SED fits, depending on the chosen star formation history, SED fitting tool, dust law, and IMF. \textit{JADES-Deep-GS:9075} was observed with NIRSpec as part of \cite{Bunker2023} (NIRSpec ID {\tt 00002773}), finding $z_{\rm spec} = 12.63$, meaning that \bagpipes{} and \prospector{} slightly underestimate the redshift in most cases. 
    \label{fig:discrepant_SEDs}}
    
\end{figure*}

In \autoref{fig:high_mass_SEDs} and \autoref{fig:discrepant_SEDs} we show examples of the photometry, best-fitting \bagpipes{} \& \prospector{} SEDs, and posterior redshift and stellar mass estimates for a selection of high-mass and or discrepant galaxies in our sample. We allow the redshift to vary within the \eazy{} posterior PDF for each \bagpipes{} fit, resulting in only small changes in redshift between individual \bagpipes{} results. 99.6\% of photo-$z$ estimates fall within 15\% of our fiducial \bagpipes{} redshift.
The variation in stellar mass and other parameters between models will also depend on the available photometry, and in future surveys with more NIRCam medium bands or deep MIRI data different systematics may be observed. For this reason we place the detailed comparison of each model to the fiducial model, along with further discussion of other galaxy properties in \autoref{sec:appendix_sps} and in this section we summarise the impact each model has on the resultant stellar masses. A comparison between the stellar masses derived with \bagpipes{} and \prospector{} can also be found in the appendix. 

Focusing first on the the upper panel of \autoref{fig:bagpipes_compar}, the dust extinction and metallicity priors appear to have little systematic effect on the stellar masses, although there is significant scatter seen in a few individual galaxies. Replacing the \cite{calzetti2000dust} dust law with \cite{charlot2000simple} or \cite{2018ApJ...859...11S}, we see a significant increase in stellar mass for the most massive galaxies, with the more flexible dust law typically favouring a steeper attenuation power-law than used in \cite{calzetti2000dust}. For these massive galaxies, when comparing to the default \cite{calzetti2000dust} assumption, \cite{2018ApJ...859...11S} typically results in a better fitting SED with a lower $\chi^2$, whereas the opposite is true for the  \cite{charlot2000simple}. 

When we test alternative star formation history models, we see a larger stellar mass discrepancy when comparing to the non-parametric flexible SFH than when we compare to an alternative parametric SFH model (delayed exponential). In particular with the ``continuity bursty" SFH we see offsets of $>1$ dex with the same goodness of fit, as well as systematically larger stellar masses across the galaxy sample. A good example of this is seen in \textit{JADES-Deep-GS:9075} shown in \autoref{fig:discrepant_SEDs}, where the observed flat photometry is consistent with a large range of possible star formation histories, the inferred masses of which can vary by $>1.3$ dex with little variation in the $\chi^2$. Exchanging the default BC03 SPS model for the BPASS model appears to have a complex effect on the derived stellar masses, with the magnitude and direction of the observed scatter dependent on the fiducial stellar mass. 



\subsection{Top-heavy IMF}
\label{sec:imf_results}

As discussed in \autoref{sec:intro}, one possibility to explain the high inferred stellar masses of high-$z$ galaxies is a 'top heavy' IMF, which results in the production of more high mass stars compared to a local IMF, and lowers the inferred mass-to-light ratios in high-$z$ galaxies. We use \prospector{} to investigate the impact of varying the initial mass function on the inferred stellar mass of a subset of the EPOCHS v1 sample. We implement the modified \cite{kroupa2001variation} IMF suggested by \cite{Steinhardt2022}, which assumes a gas temperature evolution T$_{gas} \propto (1+z)^{2}$. We produce two modified IMFs, one with T$_{gas} = 45$K, which we use for  $8 \leq z \leq 12$ , and one with T$_{gas} = 60$K, which we use at  $z \geq 12$. A standard \cite{kroupa2001variation} IMF would have T$_{gas} = 20$K in this parametrization, and given the broken-power-law shape this modification results in an increasingly top-heavy IMF with increasing gas temperature. Whilst it is theoretically possible to produce a unique IMF for each galaxy by assuming a $z-T_{gas}$ relationship, we avoid doing this for simplicity. We otherwise leave unchanged the \prospector{} configuration in order to directly compare the impact of the IMF.

\begin{figure}
    \includegraphics[width=\columnwidth]{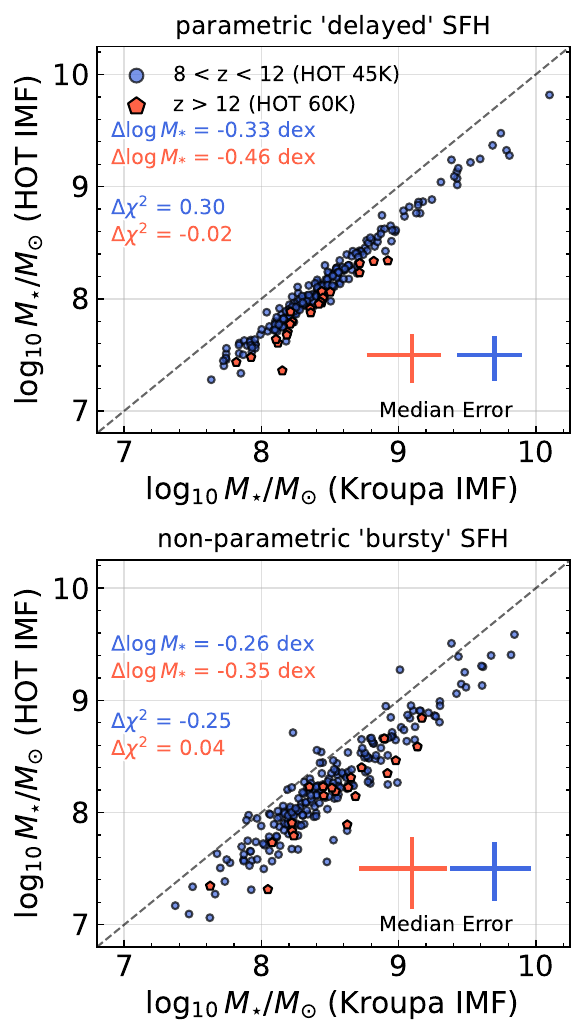}
    
    \caption{(\textbf{Top}) Effect of modified top-heavy IMF on galaxy masses using a parametric delayed-$\tau$ SFH, for both the HOT 45K IMF (at $8< z< 12$) and HOT 60K IMF (at $z >12$) from \cite{Steinhardt2022}. Overlaid on the plot are the median stellar mass and $\chi^2$ offsets, showing that these IMF models systematically reduce the stellar mass with no impact on the goodness of fit. (\textbf{Bottom}) Effect of modified IMF on galaxy masses using a non-parametric ``continuity bursty" SFH for both HOT models. \label{fig:prospect_imf}}
\end{figure}


In \autoref{fig:prospect_imf} we show the results of modifying the IMF on the stellar mass estimates. We compare the masses derived with a standard \cite{kroupa2001variation} IMF to the two modified IMF models, finding significant decreases in stellar mass with very little change in the quality of the fits. Noting that due to the computational intensity of \prospector{} fitting we only fit a subset of our full sample, and calculate the median decrease in mass for both SFH models and HOT IMFs. For the $z>12$ sample where we use the HOT 60K IMF we include our full parent sample, however for the $z<12$ model we fit only galaxies with either $z \geq 8.5$ or fiducial \bagpipes{} stellar mass $\log_{10} M_{\star,\mathrm{fid}} > 9.5$. In terms of numbers, there are 221 galaxies in the HOT 45K IMF group, and 21 galaxies in the HOT IMF 60K group. Given that this IMF model is predicted to be applicable in the region of  $8 \leq z \leq 12$ \citep{Steinhardt2022}, and the lack of mass-dependence in our results, we do not expect the excluded lower mass galaxies at $8 \leq z \leq 8.5$ to significantly impact our findings. 

For the non-parametric ``delayed" SFH model, we find a median decrease in stellar mass of 0.33~dex for the $z \leq 12$ (HOT 45K IMF) galaxy sample. For the $z \geq 12$ sample, we find a median decrease of 0.46~dex. For the ``continuity bursty" SFH model, we find a median stellar mass decrease of 0.26~dex for the $z \leq 12$ sample (HOT 45K IMF) and a decrease of 0.36~dex for the $z \geq 12$ sample (HOT 60K IMF). As expected, we see a larger decrease in stellar mass for the HOT 60K model, which is more `top-heavy' than the HOT 45K model. 

For the galaxies with the largest fiducial \bagpipes{} stellar masses, which have the most tension with $\Lambda$CDM, we show the possible decrease in stellar mass using a top-heavy HOT IMF in  \autoref{fig:evstats}. 

\section{Galaxy Stellar Mass Functions}
\label{sec:gsmf}

The \bagpipes{} and \prospector{} fitting we perform in \autoref{sec:mass_comparisons} explores the consistency of stellar mass estimates on an individual galaxy basis. In order to explore the overall effect of a particular choice of SED fitting tool and model we look at the overall distribution of galaxy masses via the Galaxy Stellar Mass Function (GSMF). We focus primarily on constructing a GSMF from our fiducial \bagpipes{} results, and then demonstrate the effect of changing this to an alternative \bagpipes{} or \prospector{} model. 

We make a further cut to our galaxy sample here, removing the fields of El Gordo, SMACS-0723 and Clio. The depths and available filters of these datasets mean that they do not contribute significant volume to our GSMF estimates, but increase the redshift uncertainties \citep{adams2023epochs}. This reduces the number of galaxies used in the GSMF to 1092, and the total area used to 175 square arcminutes.

The GSMF measures the abundance of galaxies of different masses at a given redshift. A stellar mass function $\Phi (M, z) \Delta M $ is formally defined as the number density of galaxies in a mass bin $\delta M$ at a given redshift $z$. The evolution of the shape and normalisation of the stellar mass function traces the global abundance of baryons across cosmic time, and hence indirectly traces star formation. The integral of the stellar mass function over mass gives the galaxy stellar mass density, which is the cumulative formed stellar mass per unit volume at a given epoch.

We construct a galaxy stellar mass function (GSMF) from different mass estimates in order to investigate possible evolutions of the GSMF at high redshift. To derive the GSMF we use the 1/V$_{\textrm{max}}$ methodology \citep{rowan1968determination, schmidt1968space}:

\begin{equation}
    \phi(M) d\log(M) = \frac{1}{\delta M} \sum_i \frac{1}{C_i V_{max, i}}
    \label{eq:gsmf}
\end{equation}
with associated uncertainty
\begin{equation}
    d\phi(M) = \frac{1}{\delta M} \sqrt{\left( \sum_i \frac{1}{C_i V_{max, i}}\right)^2}
    \label{eq:gsmf_err}
\end{equation}
where $\delta M$ is the bin width in stellar mass, $C_i$ is the completeness of the galaxy in bin $i$, and $V_{max,i}$ is the total observable volume of the galaxy across all the fields. We use \autoref{eq:gsmf_err} to calculate the uncertainty in the bin, except in the case where there are the bin has very low occupancy (N $\leq$ 4) where we instead directly calculate the Poisson confidence interval with a more accurate estimator based on the $\chi^2$ distribution $I = [0.5\chi^2_{2N,a/2}, 0.5\chi^2_{2(N+1),1-a/2}]$ \citep{Ulm1990,adams2023epochs}, which avoids uncertainties such as $1\pm 1$, which appear infinite on a log-scale.

Other distribution function estimators can been used that are more statistically robust that 1/$V_{\textrm{max}}$, but 1/$V_{\textrm{max}}$ is very commonly used in estimates of the UV LF and GSMF \citep[e.g.][]{adams2023epochs, bhatawdekar2019evolution, navarro2023constraints, stefanon2021galaxy, 2024arXiv240308872W, weaver2022cosmos2020, McLeod2023}, so we choose to use it in order to compare our results directly with the literature. In \cite{adams2023epochs} we compare 1/$V_{\textrm{max}}$, to the Lynden-Bell C-method, which is more statistically robust, when estimating the UV LF and find results consistent within 0.5 dex \citep{lynden1971method, woodroofe1985estimating}.

We iteratively shift the galaxy SED from the fiducial \bagpipes{} SED fitting in small steps of $\Delta z = 0.01$ before recalculating the bandpass-averaged fluxes in the available NIRCam and HST filters for a given field. With these fluxes we test if the galaxy would still be selected at every redshift step given our selection criteria detailed in \autoref{sec:sample_select}. For the selection criteria which are dependent on SNR requirements we base the detection strength on the average depth for each field, as given in \autoref{tab:areas_depths}. This allows us to calculate a $z_{\mathrm{max}}$ and $z_{\mathrm{min}}$ redshift for each galaxy in each field, capped at the edges of each redshift bin. Accounting for a minimum redshift is essential for accurately measuring the detectable volume given our requirement for the shortest wavelength filter bandpass to fall blueward of the Lyman break. We convert these maximum and minimum redshifts within each bin to a volume by

\begin{equation}
    V_{\mathrm{max}, i} = \sum_{\mathrm{fields}}\frac{4\pi}{3} \frac{A_s}{4\pi \textrm{ sr}} (d_{z_{\mathrm{max}}}^3 - d_{z_{\mathrm{min}}}^3)
\end{equation} where $d_{z_{\mathrm{max/min}}}$ are the co-moving distances at the maximum/minimum detectable redshifts (capped at bin edges). $A_s$ are the survey areas, which are given in \autoref{tab:areas_depths}. 

The only field containing a significant lensing cluster that we include in the GSMF is the MACS-0416 field. We exclude the cluster itself, and assume there is no significant magnification for our galaxy candidates in the surrounding NIRCam pointings. None of our candidates show evidence of strong lensing, and we do not expect weak lensing by foreground neighbours to significantly affect our sample. 

We account for the posterior stellar mass uncertainties using a Monte Carlo bootstrap methodology to bootstrap our GSMF. We draw stellar mass estimates for each galaxy from the posterior stellar mass probability distribution functions and compute 1000 independent realisations of the galaxy stellar mass function from these posterior PDFs. We compute the 16th, 50th and 84th percentiles of the distribution in order to quantify the uncertainty introduced by our stellar mass estimates, and indirectly by the uncertainties on the photometry. Typically these uncertainties are smaller than the Poisson error, but can be large at the highest mass where individual objects near bin edges can have a large impact. We require a bin to be occupied in more than 20\% of realizations to avoid highly uncertain bins which have very low occupancy.

\subsection{Detection Completeness}
\label{sec:det_comp}
We carry out completeness simulations on JWST data by inserting simulated galaxies with an exponential light profile (Sérsic index of $n=1$) as galaxies at high-$z$ are typically not concentrated \citep[see][]{Leo2022,ferreira2023jwst,ormerod2024epochs} and with absolute UV magnitudes ranging from -16 to -24 in the detection image (inverse variance weighted stack of F277W, F356W and F444W). We then run {\tt SExtractor} on them using the same configuration as our normal catalogue creation pipeline in order to measure the fraction recovered as a function of apparent magnitude.

The magnitude range was swept in steps of 0.2, with 1000 sources inserted into the image at a time in order to prevent introducing artificial overdensities. We assume a size - luminosity relation of $r\propto L^{0.5}$ \citep{Grazian2012}, with reference size set to $r = 800$~pc at $M_{\mathrm{UV}} = -21$, with M$_{\rm UV}$ dependent intrinsic scatter modelled with a log-normal distribution calibrated from the results of \cite{yang2022early}. The inclination angle of these sources was assumed to be uniformly random. We sweep a redshift range of 6.5 to 13.5 in steps of 0.5 to account for the changes in angular diameter distance and obtain a 2D dependence of completeness on redshift and absolute magnitude.

The above procedure was carried out on all fields considered in this paper. Given that we perform our detection on a stacked image of the LW bandpasses, our detection completeness is $>90$\% across the redshift ranges and apparent magnitudes of our high-$z$ sample, and the overall completeness is dominated by the impact of our selection criteria. 

\subsection{Selection Completeness and Contamination}
\label{sec:sec_comp}
\begin{figure*}
\centering

\gridline{\fig{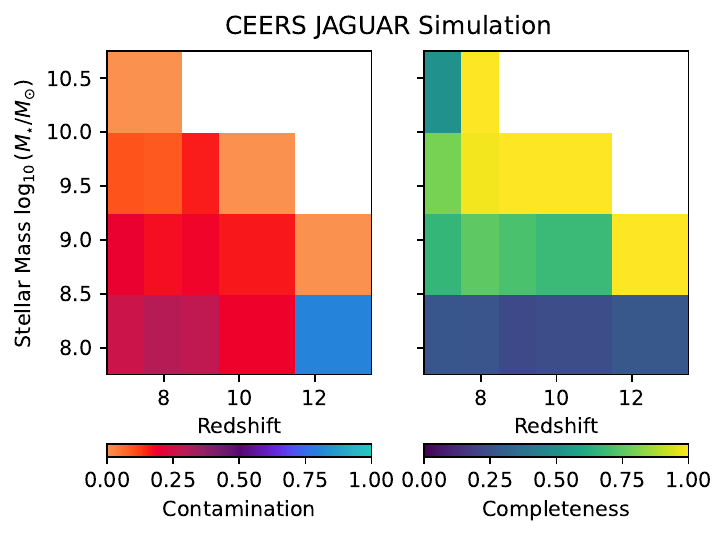}{0.49\textwidth}{}
          \fig{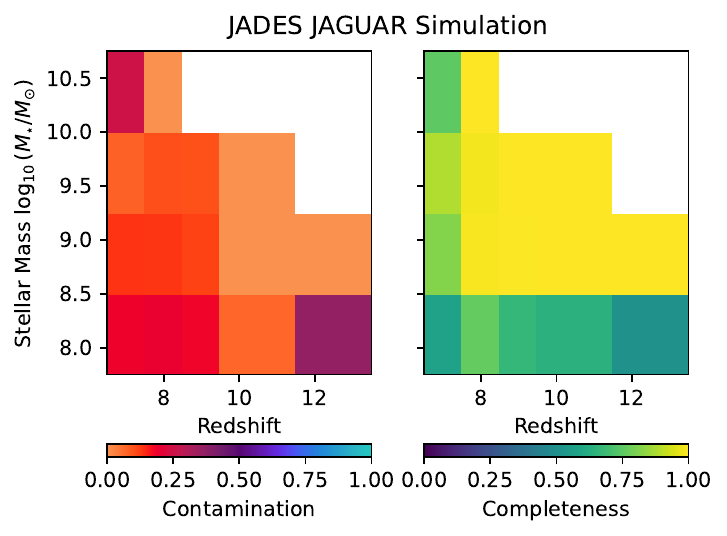}{0.49\textwidth}{}}    

\gridline{\fig{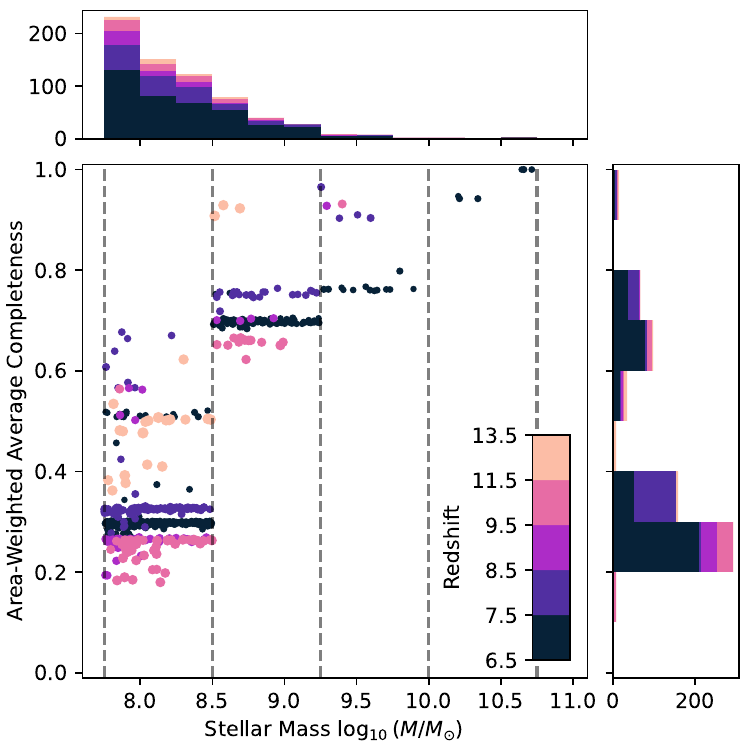}{0.49\textwidth}{}
          \fig{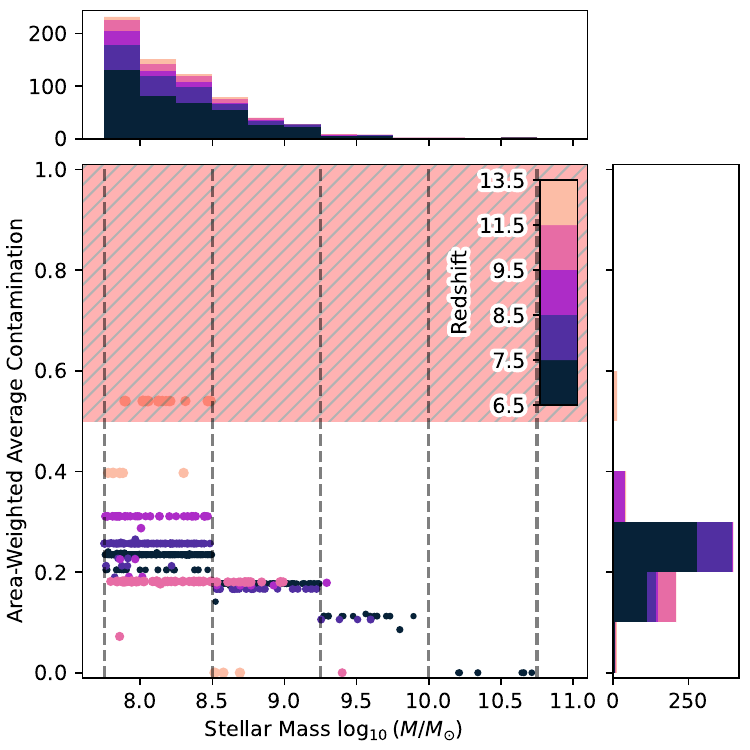}{0.49\textwidth}{}}     

\caption{\textbf{Top}) Selection completeness as a function of redshift and stellar mass for two example fields from our sample, based on our fiducial \bagpipes{} SED fitting and full selection procedure. Completeness and contamination are labelled as fractions, where $1 = 100\%$. \label{fig:select_comp} \textbf{Bottom}) Total area-weighted completeness and contamination for each of our galaxies as a function of stellar mass, colored by redshift bin. We see higher completeness and lower contamination at higher mass. The red shaded region in the contamination plot shows our 50\% contamination limit. }
\end{figure*}

To estimate the contamination and completeness of our sample selection as a function of stellar mass we use version 1.2 of the JADES Extragalactic Ultra-deep Artificial Realization (\jaguar) catalogue of synthetic galaxies released by the JADES Collaboration \citep{2018ApJS..236...33W}. We choose this model for our completeness simulations as the catalogues and spectra are readily available, and \cite{2023arXiv231108065W} have shown that the \jaguar{} model reproduces the observed color evolution of galaxies at high redshift. 

\jaguar{} is a empirical model based on observational constraints on the mass and luminosity functions at $z \leq 10$ from HST \citep{2014ApJ...783...85T, Bouwens2015, stefanon2017rest}. Mock spectra and photometry are generated using \beagle{} \citep{chevallard2016modelling}. We combine 5 different realisations of the simulation in order to improve the statistics for rare high-$z$ and high-mass sources. We filter the mock catalogues to the sources which are detectable given the average depths of our observations. This ranges between $8\times 10^5$ and $1.2\times 10^6$ realistic mock galaxies, which we run through our full galaxy selection procedure.

We generate mock observations from these catalogues by estimating 1$\sigma$ errors in the measurements based on the average depths of our JWST reductions in each field (see \autoref{tab:areas_depths}). We then perturb the catalogue photometry in flux space within a Gaussian centered on the catalogue measurement and width equal to the 1$\sigma$ error in that filter. These perturbed fluxes are then run through our full selection procedure to measure how well our pipeline recovers the true redshifts of the sources. This allows us to compare the robustness of our selection criteria given the differing depths and filters in each set of observations.

We test our ability to reconstruct the stellar masses of the \jaguar{} galaxies by fitting their perturbed photometry with our Bayesian SED-fitting using \bagpipes. We estimate completeness and contamination for each redshift and mass bin in our mass function by testing the recovery of simulated galaxies into the correct bin. Completeness is defined as the number of galaxies which our pipeline places in the correct redshift and mass bin, divided by the total number of simulated galaxies within that bin. Contamination is defined as the number of galaxies we place in the incorrect bin, divided by the total number of galaxies in the bin. As an example, if there are 100 simulated galaxies in a given redshift/mass bin, and our selection criteria selects 50, of which 10 are actually from a different bin, then the completeness would be 50\% (50/100) and the contamination would be 20\% (10/50). 

We see reasonable agreement between the true stellar mass and our inferred stellar mass. Initially 50\% of galaxies fall within the correct stellar mass bin used in construction of the galaxy stellar mass function. In order to account for the different choice of IMF and SFH parametrization between the \jaguar{} SEDs and our \bagpipes{} fitting, we derive a corrective scaling factor by which we scale the \jaguar{} masses in order to increase the agreement between the mass estimates to 65\%. As we bootstrap the GSMF using the stellar mass PDFs, we find that $\>$80\% of galaxies will contribute to the correct mass bin. The average offset between our recovered stellar mass and the simulated mass is 0.02~dex, with a standard deviation of 0.28~dex. Only 2\% of the time is the deviation between the scattered and recovered masses greater than 0.75~dex, which is the bin width used in the stellar mass functions. 

\autoref{fig:select_comp} shows the completeness and contamination for the CEERS and JADES fields, which make up a significant fraction of our total volume. 

To derive a single estimate of completeness and contamination for each galaxy, we calculate the area-averaged completeness and contamination for each galaxy in our sample, based on the fields in which the galaxy is found to be observable in when we calculate V$_{\textrm{max}}$. The bottom row of \autoref{fig:select_comp} shows the total completeness, including both detection and selection, for each galaxy as as function of stellar mass. Completeness is strongly dependent on the stellar mass of the galaxy, and at masses below $\approx 10^8 M_{\odot}$ we are $\leq 50 \%$ complete in the majority of our area, and dependent on the derived completeness corrections.  

In order to reduce the possible effects of contamination on our derived mass functions, we exclude objects where the estimated contamination fraction is $\geq$50\%. In the cases where no simulated galaxies fall within a given bin, which occurs only for the highest mass bins where we expect high completeness, we do not apply any correction factor. We also assume 100\% completeness when the total number of simulated galaxies in a mass-redshift bin is $\leq$5 due to the high uncertainty in deriving correction factors with low statistics.

Multiple classes of interlopers with photometry similar to high-redshift galaxies can contaminate estimates of the GSMF. In this section we detail our methodology to remove common types of interlopers, including low-$z$ galaxies, brown dwarfs, and AGN. 

\subsubsection{Low-z Galaxies}

The misclassification of low-$z$ galaxies as high-$z$ through catastrophic errors in photo-$z$ are commonly caused by misidentification of different spectral features, e.g. confusion of the Lyman and Balmer breaks, or strong rest-frame optical emission lines contributing to multiple wideband photometric observations, giving the appearance of continuum emission (e.g. the $z=16$ candidate in \cite{Donnan2022,arrabal2023confirmation}, which we do not select with as a robust high-$z$ galaxy with our \eazy{} SED fitting and selection criteria). 

Firstly, we compare our \eazy{} photo-$z$ estimates with spectroscopic redshifts (spec-$z$'s) in \cite{adams2023epochs}. In summary, we typically find $>90\%$ accuracy (within 15\%). With our \eazy{} photo-$z$'s, which we use as an informative prior for \bagpipes{} and \prospector, we do not observe the systematic photo-$z$ overestimation observed in other studies \citep{2023arXiv230109482F,Haro2023b}.

We are able to fully test the rate of catastrophic photo-$z$ failure using our \jaguar{} completeness and contamination simulations. The rate of contamination increases as we approach our 5$ \sigma$ detection limits in the rest-frame UV, and the vast majority of the contaminants have $<8 \sigma$ detections in the rest-frame UV. If we parametrize in terms of stellar mass, the highest rates of contamination are seen at lower stellar mass where individual galaxies have less impact on the GSMF. Candidates with SNR close to our selection limit are also those most commonly rejected in our manual inspection of all candidates, providing another layer of protection against possible interlopers. 

We restrict the redshift range and fields used when constructing the GSMF to exclude uncertain high-$z$ candidates. We do not include observations of El Gordo, Clio or SMACS-0723 when constructing the GSMF because they are shallow and show high rates of contamination in our \jaguar{} simulations. Clio and SMACS-0723 also lack observations in F115W, which \cite{trussler2023seeing} shows lead to inaccurate photometric redshifts at $8<z<10$. Despite our sample containing galaxies at $z \geq 14$, we do not attempt to measure the GSMF above $z\geq$13.5 because we lack the appropriate medium band observations (or spectroscopy) to remove interlopers such as the $z=16$ candidate in \cite{Donnan2022}. Below redshift 13.5 we have seen high accuracy in photo-$z$ estimates when compared to NIRSpec spectroscopy \citep[e.g.][]{2023NatAs.tmp...66C,Bunker2023,2023arXiv230315431A,wang2023uncover}.

\subsubsection{Brown Dwarfs}
\label{sec:brown_dwarfs}
Y and T-type brown dwarfs within the Milky Way can masquerade as high-redshift galaxies due to the appearance of a Lyman-break like dropout caused by strong molecular absorption in the optical, coupled with bright near-infrared emission. When mistaken for high-$z$ galaxies brown dwarfs can have a large impact on the galaxy stellar mass function due to their high inferred stellar masses.

\begin{figure*}
\centering

\gridline{\fig{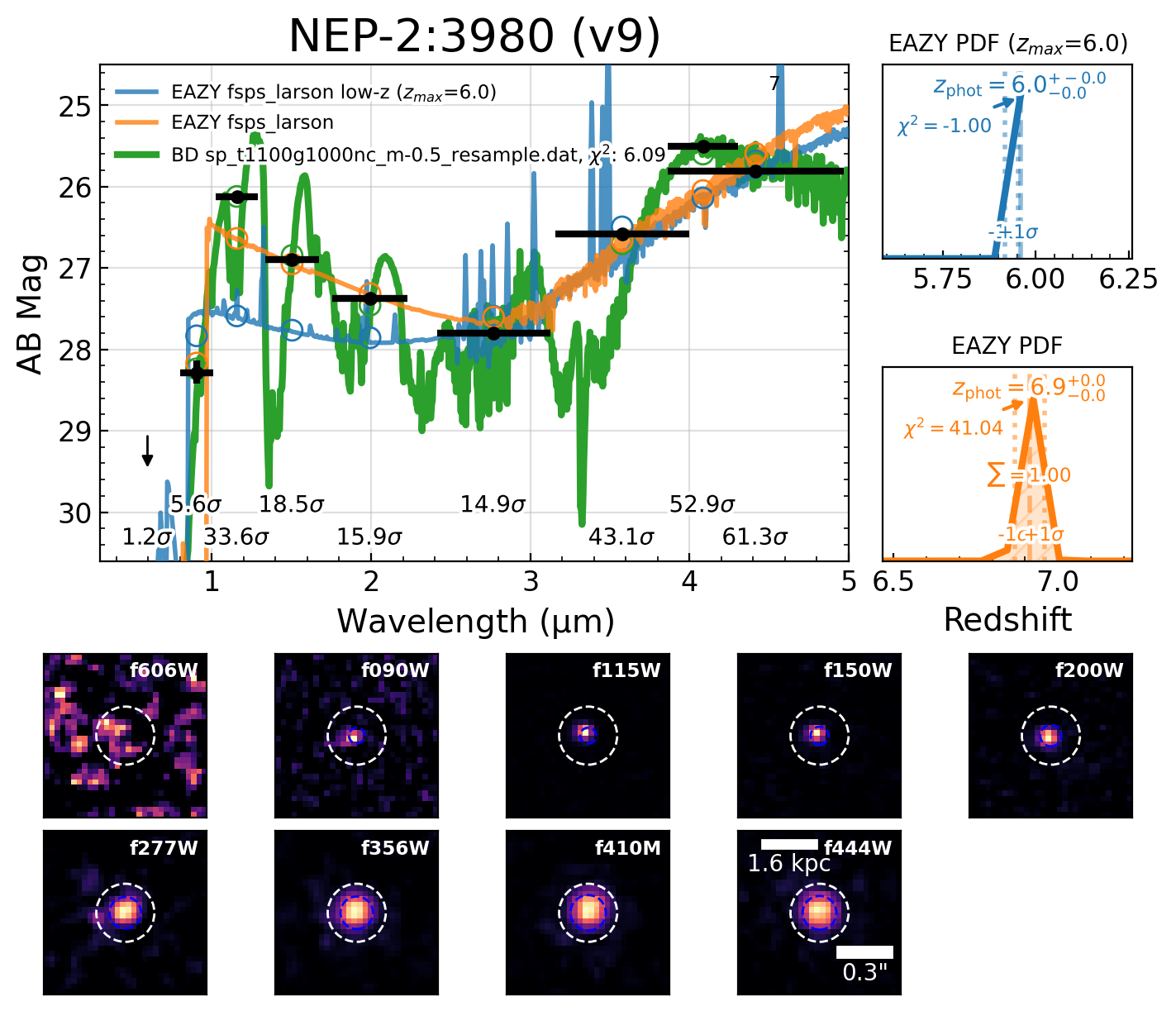}{0.50\textwidth}{}
          \fig{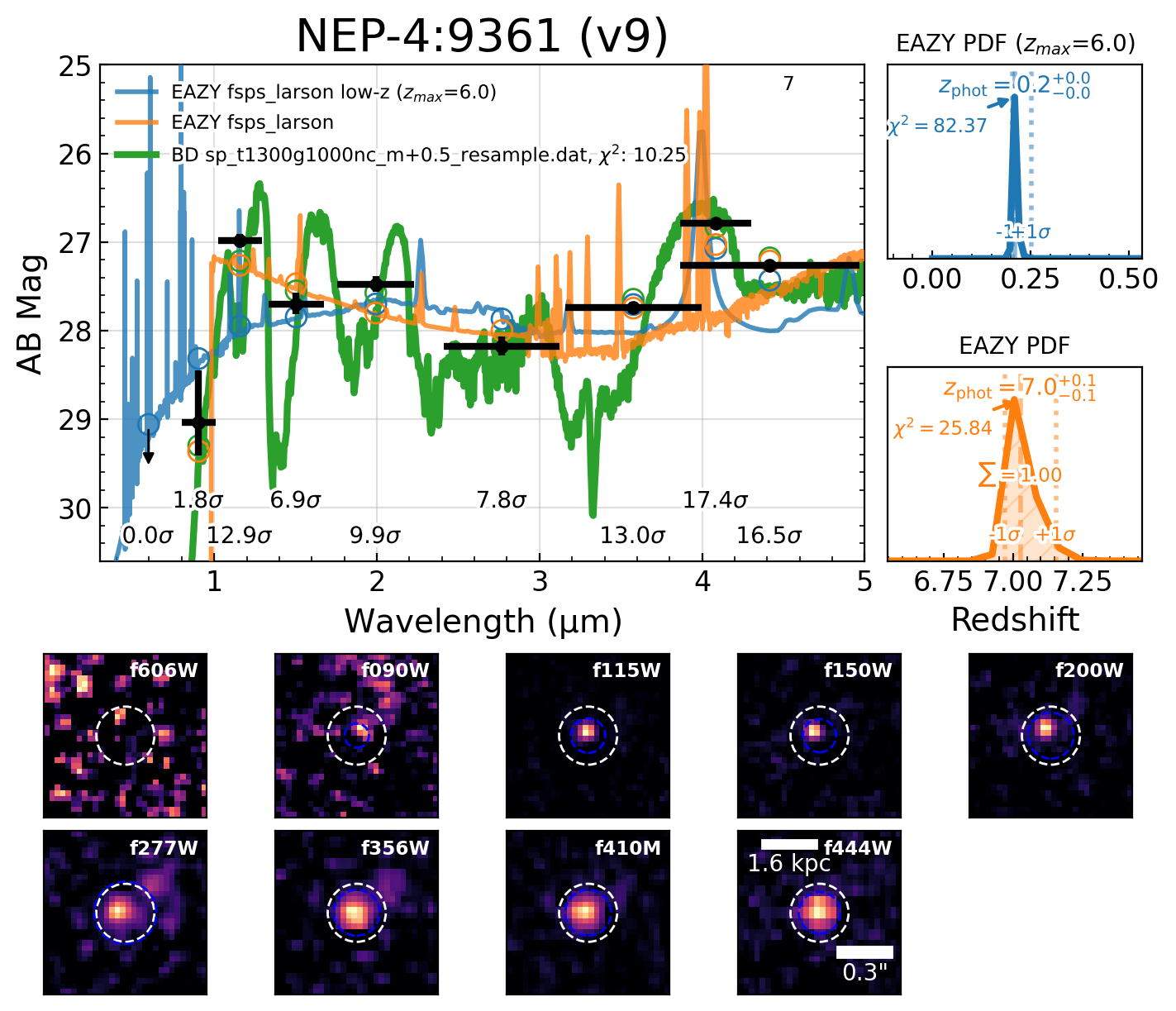}{0.50\textwidth}{}}     
\caption{Two example SEDs from our sample of 59 brown dwarf candidates (in green) selected with the Sonora Bobcat models, shown along with their best-fitting high-$z$ galaxy SEDs (in orange). The low redshift galaxy solution (with an upper redshift limit of $z \leq 6$), which is used in our robust galaxy selection criteria, is shown in blue. The redshift posteriors from \eazy{} are shown for both the high and low redshift galaxy models, and are overlaid with the statistics used within our selection criteria, indicating the $\chi^2$ and integral of the primary PDF peak. The cutouts of both brown dwarf candidates are shown below the best-fitting SEDs, demonstrating their compact PSF-like morphology. The cutouts are 0\farcs9 across, as shown with the scalebar and the white circle shows the extraction aperture used for SED-fitting (0\farcs16 radius) The scalebar showing physical size in kpc is calculated at the best-fitting galaxy photo-$z$ and is not applicable to the brown dwarf solution. The coordinates for these brown dwarf candidates are 17h23m12.36s +65d49m38.8s and 17h21m53.00s +65d49m20.82s respectively. \label{fig:bd_seds}}

\end{figure*}

We fit the photometry of all of our galaxy candidates with the synthetic {\tt Sonora Bobcat} brown dwarf templates \citep{marley_mark_2021_5063476} using a simple $\chi^2$ minimization to scale the flux of the templates to each candidate. We fit all the provided templates to each candidate, selecting the best-fitting template for each galaxy. We remove candidates from our sample which are both better fit by a brown dwarf template than a galaxy template, and show a PSF-like morphology (50\% encircled flux radius in F444W $<$ F444W PSF FWHM). This removes 59 candidates in total (4\% of the total sample). \autoref{fig:bd_seds} shows an example of several brown dwarf candidates which would otherwise have been selected as high-$z$ galaxy candidates. 

None of the brown dwarf candidates selected in \cite{hainline2023brown} are contained within our robust galaxy sample in the Goods-South or CEERS fields. As a check of our selection criteria we confirm that all of \cite{hainline2023brown}'s candidates located within the footprint of our datasets (GOODS-South, CEERS) are selected as brown dwarfs by our brown dwarf selection criteria. 

\subsubsection{Active Galactic Nuclei} \label{sec:agn}

Given the apparent prevalence of high-$z$ AGN detected by multi-wavelength and spectroscopic studies it is likely there is a population of galaxies containing AGN within this sample. Our modelling does not fit AGN components to the photometry, so if galaxies contain a significant AGN component it is likely we will overestimate the stellar mass. In particular, obscured AGN with red colors \cite[e.g.][]{kocevski2023hidden, 2023Natur.616..266L} can mimic strong Balmer breaks, leading to inference of an aged stellar population and high stellar mass. Known as LRDs, a number of studies have found numerous candidates up to $z\leq9$ \citep[see e.g.][]{greene2023uncover,furtak2023jwst,furtak2023jwstnirspec,2023arXiv230514418B,matthee2023little,labbe2023uncover,2024arXiv240109981K}.
Different selection techniques have been proposed for LRDs and it is not clear that all the objects grouped as LRDs with a given criteria indeed have the same origin. The term originated to categorise spectroscopically identified obscured broad-line AGN at high redshift, but has become more generally used to describe photometrically selected compact red galaxies, which may or may not be dominated by AGN emission. Photometric selections typically rely on a rest-optical color cut (F277W-F444W for $z\approx6-7$) as well as a compactness criteria, with spectroscopic analysis from \cite{greene2023uncover} suggesting that a F277W-F444W break exceeding 1.6 magnitudes identifies LRDs with an 80\% AGN purity, without requiring an additional compactness criteria. The criteria of \cite{2023Natur.616..266L} and \cite{2024arXiv240109981K} requires a weaker color selection, which may be selecting a mix of AGN and dusty star-forming systems. \cite{2024ApJ...968...34W} have used MIRI observations from 5 - 25$\mu$m of a sample of likely AGN-dominated LRDs with strong breaks, finding a flattening of the red colors (in F$_\nu$) at longer wavelengths, which they suggest as evidence for obscured AGN which lack toroidal hot dust. In contrast \cite{perez2024nature} uses MIRI observations of a wider population of LRDs, which may not be a single class of object, to suggest that the characteristic photometry is driven mostly by stellar emission, and that LRDs are mostly compact and highly obscured starburst galaxies with young stellar ages. Other studies have struggled to distinguish between AGN and stellar models \citep{2023arXiv230514418B} or favour an AGN model \citep{noboriguchi2023similarity}. Stacking deep X-Ray observations of LRDs have also revealed a surprising lack of X-Ray emission \cite{2024arXiv240413290Y, 2024arXiv240419010A, 2024arXiv240500504M} which is somewhat puzzling to explain via the AGN scenario, although high obscuration or intrinsic X-ray weakness in the accretion disc have been proposed \citep{2024arXiv240500504M}.
Further spectroscopic or MIRI observations can break the degeneracy between AGN and old stellar populations. 

Whilst it is essentially impossible to distinguish between the contribution of AGN and stellar emission to an SED with only rest-optical photometry we attempt to perform additional SED fitting using \prospector{} by including an AGN component, which models AGN emission lines for the most massive galaxies in our sample (M$_\star \geq 10^{9.5} \ \textrm{M}_\star$). In almost all cases the AGN fraction is less than 10\%, and we see no reduction in stellar mass due to the inclusion of AGN component. The addition of an AGN model does not statistically improve the fit for these galaxies, although we can not strongly exclude the presence of AGN in some of our candidates as the AGN model within \prospector{} models only obscured AGN \citep{2018ApJ...854...62L}. Further fitting with a wider range of AGN models, which is outside the scope of this paper, is required to constrain the potential contribution of AGN emission to the observed photometry. Given the uncertainty around the true nature of these sources, we choose not to remove them from our sample. We discuss the impact of LRDs on the GSMF in \autoref{sec:gsmf_no_agn}. 

We do however remove two galaxies from our sample which are spectroscopically confirmed as AGN, or with significant AGN components within the photometry. This includes the AGN within the CEERS field presented in \cite{larson2023ceers}.   

\subsection{Cosmic Variance}

Cosmic variance is the field-to-field variance of the distribution of galaxies at a given redshift due to galaxy clustering. Empirical measurements of cosmic variance do not exist at $z \geq 7$. Estimates of cosmic variance from simulations \citep[e.g. \emph{Bluetides},][]{2020MNRAS.496..754B} are available, but the accuracy of these estimates is difficult to measure given the limited cosmological volume and  dependence on the assumed cosmology and galaxy formation physics. \cite{2020MNRAS.496..754B} argues that cosmic variance will be a dominant source of uncertainty at these redshifts, as galaxies are predicted to be highly clustered \citep{bhowmick2018clustering}. 

We caution that the following methodology provides only an approximation of the true cosmic variance as the underlying quantities, such as the galaxy bias, are not well-quantified at $z \geq 7$ and are generally extrapolated from wider area clustering results at lower redshift or taken from N-body simulations \citep{2010ApJ...710..903M}. We estimate the uncertainty on the GSMF due to cosmic variance following the prescription of \cite{Driver2010}, combining the cosmic variance from different surveys using the volume weighted sum of squares from \cite{Moster2011}. We treat nearby pointings as one observation and sum their areas (e.g. the multiple pointings of CEERS or the NEP, which are not widely separated enough on the sky to be considered independent). We add the uncertainty in quadrature to \autoref{eq:gsmf_err}. 

We compute a cosmic variance estimate for each galaxy individually based on which of our fields they are found to be detectable in. Given that this work combines multiple widely separated fields for the majority of our sample cosmic variance is small compared to studies incorporating only a single field.  We see little difference in cosmic variance across our redshift bins. In the best case, where a galaxy is detectable across every field, the average cosmic variance across all redshift bins is found to be 18\%. The highest cosmic variance estimates are for faint galaxies which, given our selection criteria, are only detectable in the JADES GOODS-South field, where we find a cosmic variance of 42\%.

\subsection{Fitting the GSMF}

The overall scientific consensus is that the stellar mass function at high-$z$ is well-described by the \textit{Schechter} function \citep{Schechter1976}, given in logarithmic form in \autoref{eq:schec}.

\begin{equation}
\begin{aligned}
\Phi (\mathrm{d} \log M)= \ln (10) \Phi^\star \mathrm{e}^{-10^{\log M-\log M^\star}} \times \\ \left(10^{\log M-\log M^\star}\right)^{\alpha+1} \mathrm{~d} \log M
\label{eq:schec}
\end{aligned}
\end{equation}

The Schechter function, which consists of a power-law with an exponential fall-off, is parameterised by 3 parameters: $\alpha$, M$^\star$, and $\Phi^\star$. $\alpha$ controls the slope of the low-mass end of the SMF, while M$^\star$ gives the turnover point at which the function turns from a power-law to an exponential fall-off. $\Phi^\star$ controls the overall scaling of the SMF, and is the only parameter that has been confidently shown to evolve with cosmic time \citep{popesso2023main}. 

Before JWST the mass function was constrained up to $z \approx 9$ using HST WFC3/IR, Spitzer/IRAC and ground-based data (\textit{Visible and Infrared Survey Telescope for Astronomy} (VISTA), \textit{United Kingdom Infrared Telescope} (UKIRT)) observations of the HUDF, CANDELS, \textit{Cluster Lensing and Supernova survey with Hubble} (CLASH) and \textit{Hubble Frontier Fields} (HFF) surveys \citep{stark2013keck, bradavc2014spitzer, duncan2014mass, bouwens2016bright, Song2016, stefanon2017rest, bhatawdekar2019evolution, kikuchihara2020early}. These studies generally agree that the mass function steepens out to $z\approx10$, with the low-mass slope, $\alpha$, $\rightarrow -2$. \cite{duncan2014mass,bhatawdekar2019evolution,stefanon2021galaxy} find that the stellar mass to halo mass ratio in galaxies (M$_\star$-M$_{h}$) shows no evolution over $6 \leq z \leq 10$ despite a 3~dex increase in overall stellar mass, suggesting that stellar and halo mass grew together during reionization. Pre-JWST, the accuracy of stellar mass functions at these redshifts is limited by the uncertainty in stellar mass estimates from Spitzer data, where the broadband photometry can be affected by strong nebular emission causing scatter in stellar mass estimates, which is compounded by the limited number of spectroscopically confirmed galaxies at these redshifts \citep{roberts2016z, strait2020stellar}. Previously, measurements of the mass function above $z\geq 6$ have relied on UV-selected samples of galaxies, with mass indirectly inferred through a calibrated L$_{\textrm{UV}}$-M$_\star$ relationship \citep{Song2016, harikane2016evolution} rather than direct measurements of the rest-frame optical wavelengths. Photometric mass-selected samples above $z \geq 2$ often have significant unavoidable mass uncertainties at higher redshift \citep{retzlaff2010great, laigle2016cosmos2015, weaver2022cosmos2020}, due to the lack of available rest optical or NIR photometry. Many studies combine ground and space-based observations in order to probe the entire mass function, although combining these datasets robustly is difficult due to systematics between different surveys, including differing selection functions, SED fitting techniques, detection bands, and survey depths. 

We derive a mass function for 5 redshift bins: 6.5 $\leq z <$ 7.5,  7.5 $\leq z <$ 8.5,  8.5 $\leq z <$ 9.5, 9.5 $\leq z <$ 11.5 and  11.5 $\leq z <$ 13.5, which covers a time period of $\sim$500~Myr. We cover a mass range from $7.75 \leq \log_{10} M_\star/M_{\odot} \leq 11.5$ in steps of 0.75 $\log_{10} M_\star/M_{\odot}$. Our lower mass limit is driven by our completeness simulations, and the bin width is chosen to ensure adequate statistics in the majority of bins. Given the quantisation of our completeness correction to each stellar mass/redshift bin, we ensure our mass bins are wide enough to ensure our SED fitting places galaxies within the correct mass bin, so that the completeness corrections are calculated correctly. by ensuring that we can recover the masses of the simulated galaxies to within the correct stellar mass bin. We fit the GSMF using Bayesian methods; specifically Markov-Chain Monte Carlo via the {\tt emcee} package \citep{emcee}. 

For our highest redshift bin (11.5 $\leq z < $ 13.5), we do not attempt to fit the mass function for a number of reasons; firstly given the low number of available data point points a meaningful fit is not possible without fixing multiple parameters, and secondly due to the increasing uncertainties in stellar mass estimates and reliability of candidates at these redshifts. Furthermore the stellar mass estimates are less reliable when there is less rest-frame optical photometry available during SED fitting. Our \jaguar{} simulations also suggest that this bin has high contamination at lower mass, with many galaxies below 10$^{8.5}$ M$_\odot$ having $\geq$50\% average contamination across the fields they are detectable in. 
 
We use wide uniform priors on $\Phi^\star$, M$^\star$ and $\alpha$, with $-8 \leq \ \log \Phi^\star \leq -2, \ 8.0 \leq \ \log M^\star \leq 12.5$ and $-3.5 \leq \alpha \leq -1.0$ in our MCMC fitting procedure and run until it has converged before taking $50,000$ independent draws from the posterior. We calculate the median posterior, the $\pm 1 \sigma$ uncertainty from the 16th/84th percentiles as well as the maximum likelihood draw.

\subsection{Mass Functions at \texorpdfstring{6.5 $< z <$ 13.5}{6.5 \textless z \textless 13.5}}

\autoref{tab:params} gives our tabulated GSMF for each redshift bin, as well as the associated 68\% (1$\sigma$) uncertainties. We also list the median number of galaxies which contribute to each mass bin, as well as the average completeness and contamination estimates. As discussed in \autoref{sec:gsmf}, bootstrapping the GSMF means that the estimates for completeness, contamination, and occupation shown are only averages as they will vary in each realization of the GSMF.

\begin{figure*}
    \centering
    \includegraphics[width=0.93\textwidth]{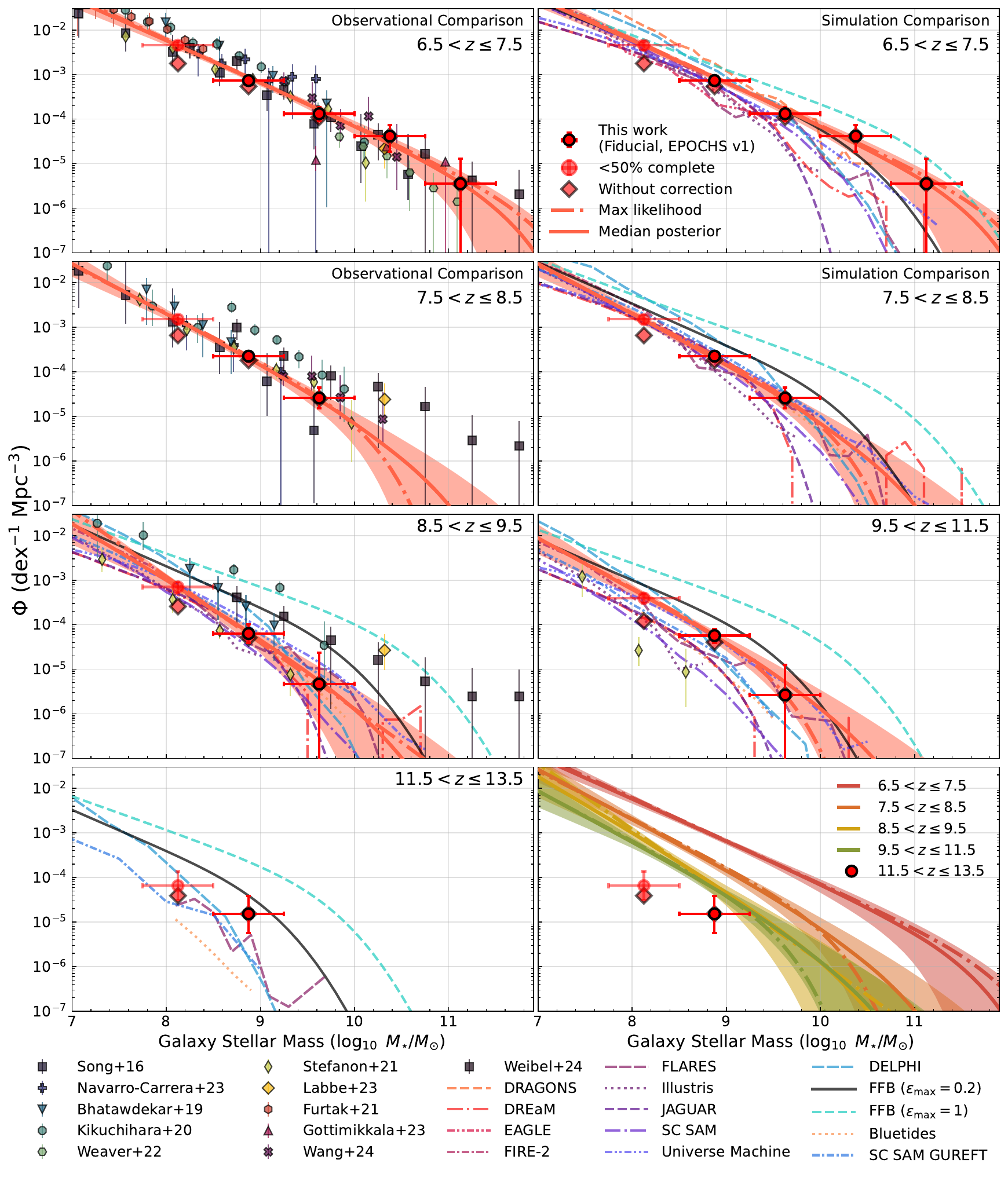}
    \caption{Galaxy Stellar Mass Functions (red markers) and the best fitting Schechter functions (red solid and dash-dot lines indicating the median and maximum likelihood draws from the fit posterior, and the 16$^{\mathrm{th}}$ and 84$^{\mathrm{th}}$ percentiles are shown by the red shaded region) for the EPOCHS v1 sample derived from the fiducial \bagpipes{} masses and photometric redshifts. We do not fit the $z\sim 12.5$ bin due to low statistics, uncertain stellar masses and the potential for high contamination. The bottom-right plot shows a comparison of the best-fitting Schechter functions for all redshift bins. Comparison to the following simulations are shown; \emph{BLUETIDES} (z$\leq$13, \cite{2016MNRAS.455.2778F, 2017MNRAS.469.2517W}), \emph{DRAGONS} ($z\leq$7, \cite{mutch2016dark}), \emph{DREaM} ($z\leq$10, \cite{2022ApJ...926..194D}), \emph{EAGLE} ($z\leq$7, \cite{2015MNRAS.450.4486F, 2015MNRAS.446..521S}), \emph{FIRE-2} ($z\leq$10, \cite{2018MNRAS.478.1694M}), \emph{FLARES} ($z\leq$15, \cite{2021MNRAS.500.2127L, 2023MNRAS.519.3118W}), \emph{Illustris} ($z\leq$10, \cite{2014MNRAS.445..175G}), \emph{Jaguar} ($z\leq$10,\cite{2018ApJS..236...33W}),  \emph{DELPHI} ($z\leq20$, \cite{mauerhofer2023dust}), \emph{Santa Cruz SAM (GUREFT}, ($z\leq$17, \cite{2019MNRAS.490.2855Y, 2023arXiv230404348Y}), \emph{Universe Machine} ($z\leq$10, \cite{2019MNRAS.488.3143B}) and the Feedback Free Burst  model of \cite{li2023feedback} (FFB, $\epsilon_{\rm max}=0.2$, $5 \leq z \leq 20$). We also show the SMF upper limit from \cite{li2023feedback}, assuming a maximum star formation efficiency ($\epsilon_{\rm max}$) of unity. Comparisons to observational results from \cite{Song2016}, \cite{bhatawdekar2019evolution} (disc-like galaxies), \cite{kikuchihara2020early}, \cite{stefanon2021galaxy}, \cite{2021MNRAS.501.1568F}, \cite{weaver2022cosmos2020}, \cite{navarro2023constraints}, \cite{gottumukkala2023unveiling}, \cite{wang2024true} and \cite{2024arXiv240308872W}. The observational results have been converted to a \cite{kroupa2001variation} IMF where necessary. In a minority of cases there is $\Delta z \leq 0.5$ between the redshift of our SMF and literature comparisons. }
    \label{fig:mf_zfit}
\end{figure*}

\autoref{fig:mf_zfit} shows our derived GSMF (red circles) and associated Schechter fit for each redshift bin, with a comparison to relevant literature results. For the lower redshift bins, which have the most comparisons in the literature, we split our comparison into observational and simulated results to aid readability. The shaded regions shown represent the 68\% (1$\sigma$) uncertainty combining the cosmic variance, Poisson error, and bootstrapping via the mass PDFs. Mass bins with $\leq50\%$ completeness are plotted without a black border. We also show GSMF estimates without completeness corrections applied with red diamonds. We convert the GSMF estimates from the literature to a \cite{kroupa2001variation} where necessary, using the assumptions of \cite{Madau2014}.  

The uncertainty regions are likely underestimated; \citep{wang2023quantifying} argue the statistical uncertainties encoded in the posterior distributions used for bootstrapping the GSMF do not represent the true uncertainty on the stellar mass estimates due to underlying modelling assumptions, such as the SPS model, IMF, nebular modelling and assumed SFH. We explore the impact of these assumptions on the GSMF further in \autoref{sec:alt_gsmf}, showing that alternative GSMF estimates fall outside the uncertainty region. 

\begin{figure*}
    \centering
    \includegraphics[width=\textwidth]{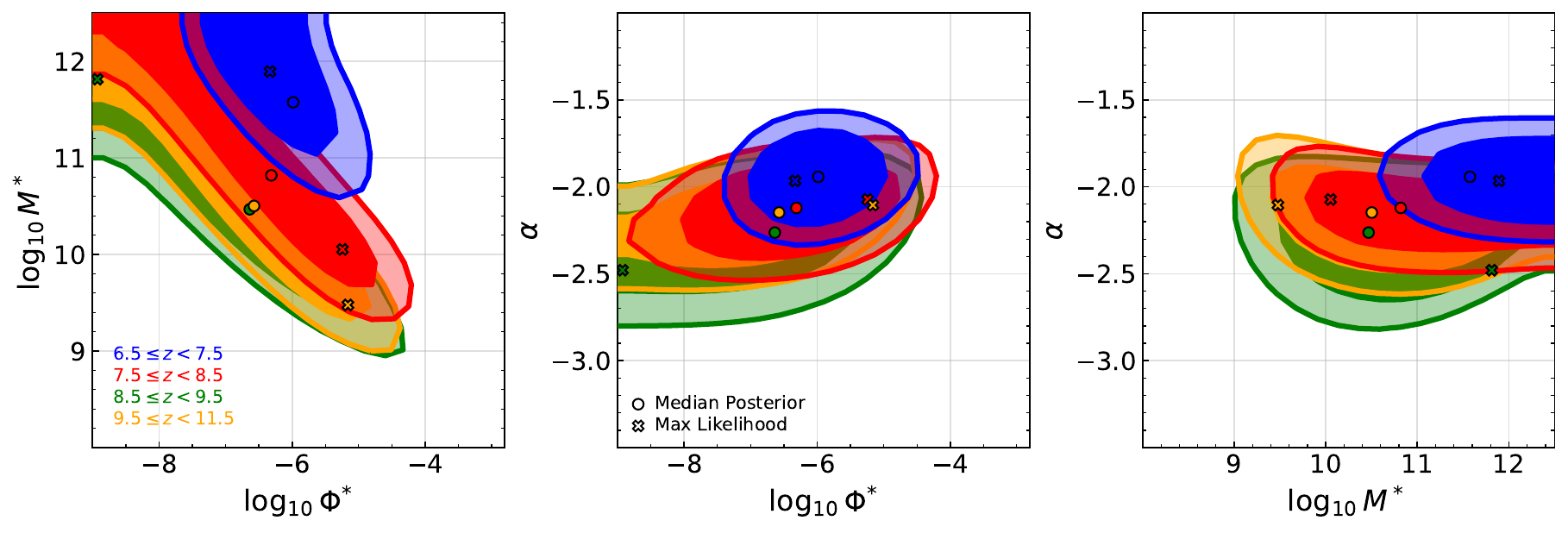}
    \label{fig:posterior}
    \caption{Confidence intervals for our best-fitting Schechter parameters for all fitted redshift bins. Overlaid are the locations of the median posterior value (filled circles) and maximum likelihood draw (filled crosses). Filled (shaded) regions show the 68\% (95\%) confidence levels.}
\end{figure*}

\autoref{tab:schechter} gives the maximum likelihood and median posterior estimates for our Schechter function parameters for each redshift bin. Uncertainties correspond to the 16\textbackslash84th percentile of each parameter distribution. \autoref{fig:posterior} shows the contours which correspond to the 68\% (1$\sigma$) and 95\% (2$\sigma$) confidence levels. 

\autoref{fig:posterior} shows the covariance of the Schechter parameters for each GSMF fit. A range of parameters are able to fit our GSMF estimates, demonstrating the highly covariant nature of the Schechter parameters. We note that the confidence intervals for the turnover mass, M$^\star$, and the normalization, $\phi^\star$, appear somewhat unconstrained at the high and low ends respectively. The lack of constraint on M$^\star$ is a consequence of the wide stellar mass bins used and consequential low number of overall mass bins. Wherever we find M$^\star >$11.5, the GSMF is fit by a pure power-law only, and there is no reason to continually extend the prior outside the range where we have available data to constrain M$^\star$.  

\begin{figure}
    \centering
    \includegraphics[width=\columnwidth]{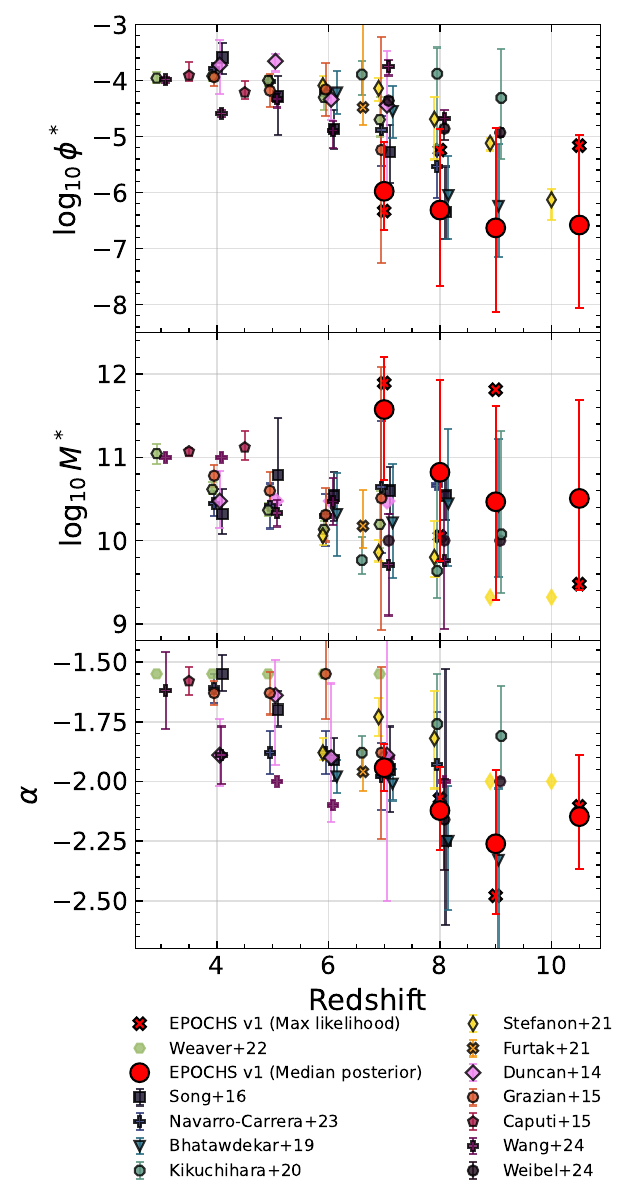}
    \caption{Evolution of best-fitting Schechter function parameters with redshift, with the results of this study shown by the red circles (median posterior) and red crosses (maximum likelihood). Comparisons to \cite{duncan2014mass}, 
    \cite{grazian2015galaxy}, \cite{caputi2015spitzer},
    \cite{Song2016}, \cite{bhatawdekar2019evolution}, \cite{kikuchihara2020early},
    \cite{stefanon2021galaxy}, \cite{2021MNRAS.501.1568F},
    \cite{weaver2022cosmos2020}, \cite{navarro2023constraints} and \cite{2024arXiv240308872W} are shown. Markers showing literature values without black outlines were fixed during fitting.  M$^\star$ values have been adjusted to reflect a \cite{kroupa2001variation} IMF.}
    \label{fig:schechter_param}
\end{figure}

\autoref{fig:schechter_param} shows the redshift evolution of these Schechter parameters, with a comparison to those derived in the literature. It is difficult to measure any significant redshift evolution, given the large uncertainty and covariance of the results. We observe some evidence of evolution of the low-mass slope $\alpha$, which steepens toward higher redshift in all but the highest redshift bin. 

In \autoref{fig:schechter_param} and \autoref{fig:posterior} we see that often our highest-likelihood Schechter function does not coincide with the median posterior, despite the model being fully-converged and a large number of posterior sample drawn. 

\begin{table*}
\centering
\caption{Tabulated GSMF from our fiducial \bagpipes{} SED-fitting. We give the average number of galaxies in each bin as well as the estimated average completeness and contamination estimates based on our \jaguar{} simulations, along with the median redshift (in brackets) and stellar mass for all the objects in a given bin. See \autoref{sec:sec_comp} for the definition of completeness and contamination used in this work. This table is available for download at \url{https://github.com/tHarvey303/EpochsIV}.}
\label{tab:params}

\begin{tabular}{c|cccccc}
\hline
Redshift Bin & $\log_{{10}}(\frac{{M_{{\star}}}}{{M_{{\odot}}}})$ & Med($\log_{10} M_{\star}$) & $\Phi \ (10^{-4}\textrm{dex}^{-1}$ Mpc$^{-3}$) & Comp (\%) & Cont (\%) & N$_{gal}$\\
\hline
6.5 $ \ <z \ \leq$ \ 7.5 & $8.125\pm0.375$ & 8.07 &$49.65^{+10.69}_{-10.29}$ & 31 & 23 & 273 \\
(6.94) & $8.875\pm0.375$ & 8.75 &$7.37^{+1.79}_{-1.57}$ & 70 & 18 & 103 \\
 & $9.625\pm0.375$ & 9.47 &$1.32^{+0.53}_{-0.42}$ & 79 & 10 & 18 \\
 & $10.375\pm0.375$ & 10.40 &$0.64^{+0.46}_{-0.35}$ & 60 & 0 & 5 \\
 & $11.125\pm0.375$ & 10.81 &$0.04^{+0.09}_{-0.04}$ & 100 & 0 & 1 \\
\hline
7.5 $ \ <z \ \leq$ \ 8.5 & $8.125\pm0.375$ & 8.09 &$16.06^{+3.68}_{-3.48}$ & 35 & 25 & 109 \\
(8.02) & $8.875\pm0.375$ & 8.73 &$1.64^{+0.60}_{-0.45}$ & 75 & 17 & 30 \\
 & $9.625\pm0.375$ & 9.45 &$0.17^{+0.17}_{-0.08}$ & 91 & 11 & 4 \\
\hline
8.5 $ \ <z \ \leq$ \ 9.5 & $8.125\pm0.375$ & 8.02 &$8.16^{+2.30}_{-2.13}$ & 28 & 30 & 40 \\
(8.86) & $8.875\pm0.375$ & 8.71 &$0.50^{+0.35}_{-0.20}$ & 70 & 17 & 8 \\
 & $9.625\pm0.375$ & 9.48 &$0.05^{+0.10}_{-0.05}$ & 93 & 18 & 1 \\
\hline
9.5 $ \ <z \ \leq$ \ 11.5 & $8.125\pm0.375$ & 8.10 &$4.53^{+1.12}_{-1.00}$ & 26 & 18 & 42 \\
(10.40) & $8.875\pm0.375$ & 8.73 &$0.64^{+0.25}_{-0.19}$ & 65 & 18 & 16 \\
 & $9.625\pm0.375$ & 9.42 &$0.03^{+0.10}_{-0.02}$ & 93 & 0 & 1 \\
\hline
11.5 $ \ <z \ \leq$ \ 13.5 & $8.125\pm0.375$ & 8.09 &$2.13^{+0.99}_{-0.90}$ & 34 & 40 & 5 \\
(11.94) & $8.875\pm0.375$ & 8.70 &$0.22^{+0.22}_{-0.12}$ & 90 & 0 & 5 \\
\hline
\end{tabular}

\end{table*}

\begin{table*}[]
\caption{Best-fitting Schechter function parameters and uncertainties derived from fitting the derived GSMF. We give both the median posterior parameter values (with 1$\sigma$ uncertainties derived from the $16^{\mathrm{th}}-84^{\mathrm{th}}$ percentiles), as well as the values corresponding to the maximum likelihood draw from the posterior (given in brackets). This table is available for download at \url{https://github.com/tHarvey303/EpochsIV}.}
\centering
\label{tab:schechter}
\begin{tabular}{c|ccc}
\hline
Redshift Bin & $\alpha$ & M$^{\star}$ & $\log_{10} \phi^\star$ \\
\hline
$6.5< z \leq 7.5$ & $-1.94^{+0.10}_{-0.10} (-1.97) $ & $11.57^{+0.63}_{-0.85} (11.89)$ & $-5.98^{+0.87}_{-0.69} (-6.33)$ \\
\hline
$7.5< z \leq 8.5$ & $-2.12^{+0.18}_{-0.17} (-2.07) $ & $10.82^{+1.11}_{-1.07} (10.05)$ & $-6.31^{+1.45}_{-1.36} (-5.24)$ \\
\hline
$8.5< z \leq 9.5$ & $-2.26^{+0.31}_{-0.30} (-2.48) $ & $10.47^{+1.15}_{-1.18} (11.81)$ & $-6.63^{+1.78}_{-1.50} (-8.93)$ \\
\hline
$9.5< z \leq 11.5$ & $-2.15^{+0.26}_{-0.22} (-2.10) $ & $10.51^{+1.18}_{-1.10} (9.48)$ & $-6.58^{+1.61}_{-1.48} (-5.16)$ \\
\hline
\end{tabular}
\end{table*}

\subsection{Stellar Mass Density}

The stellar mass density measures the cumulative build-up of stellar mass across cosmic time. This is given by the integration of the stellar mass function $\phi(M, z)$ at a given epoch (redshift) between two limiting masses, as shown in \autoref{eq:smd}. Care must be taken when comparing different estimates of the stellar mass density (SMD), as the choice of IMF means mass estimates must be converted; for example, to convert from a \cite{Chabrier2000} IMF to a \cite{salpeter1955luminosity} IMF masses must be multiplied by a factor of 1.64 \citep{Madau2014}. 

\begin{equation}
    \rho_\star (M, z) = \int_{M_{\textrm{l}}}^{M_{\textrm{u}}} M \phi(M, z) dM 
\label{eq:smd}
\end{equation}

The upper and lower integration limits $M_{\textrm{u}}$ and $M_{\textrm{l}}$ are normally taken to be 10$^{13} M_{\odot}$ and 10$^{8} M_{\odot}$ \citep[see e.g.][]{davidzon2017cosmos2015, bhatawdekar2019evolution, weaver2022cosmos2020}. 

We perform the integration given in \autoref{eq:smd}, where $\phi(M, z)$ is the best-fitting Schechter function fit for each redshift bin. We make use of the {\tt quad} function from SciPy \citep{2020SciPy-NMeth}, integrating between 10$^8$ M$_\odot$ and 10$^{13}$ M$_\odot$. We compare the integration of both the maximum likelihood fit as well as the median posterior fit. For comparison we also integrate the best-fitting Schechter functions from the literature in each redshift bin, converting to a \cite{kroupa2001variation} IMF if necessary. 

\autoref{fig:smd} shows our estimate for the stellar mass density at 6.5~$\leq z \leq 11.5$ from integration of the GSMF derived from our fiducial \bagpipes{} mass functions. We show comparisons to observational SMD results at $z \geq 5.5$. Estimates of the stellar mass density at $z \leq 5$ can be found in the literature \cite{Madau2014,2018MNRAS.475.2891D}. We also tabulate our SMD values in \autoref{tab:smd}. 

\begin{table}
\centering
\caption{Stellar mass density results from our fiducial \bagpipes{} SED fits calculated from the integral of the Schechter function for each redshift bin. Values with uncertainties come from the 16th, 50th and 84th percentiles of the posterior, and values in brackets are for the highest likelihood Schechter function. This table is available for download at \url{https://github.com/tHarvey303/EpochsIV}. \label{tab:smd}}

\begin{tabular}{cc}
\hline
Redshift Bin & $ \rho_{\star}$ ($\log_{10}$ M$_{\odot}$ Mpc$^{-3}$) \\
\hline
$6.5< z \leq 7.5$ & 6.36$^{+0.14}_{-0.17}$(6.42) \\
$7.5< z \leq 8.5$ & 5.52$^{+0.14}_{-0.13}$(5.51) \\
$8.5< z \leq 9.5$ & 5.03$^{+0.18}_{-0.18}$(5.02) \\
$9.5< z \leq 11.5$ & 4.93$^{+0.18}_{-0.15}$(4.85) \\
\hline
\end{tabular}

\end{table}

We integrate all independent posterior draws in order to propagate our mass function uncertainties into the stellar mass density. We show comparisons to predictions and measurements of the cosmic star formation rate density (SFRD, $\psi$) using \autoref{eq:sfrd_int} from \cite{Madau2014} in order to estimate the inferred stellar mass density:

\begin{figure}
    \centering
    \includegraphics[width=\columnwidth]{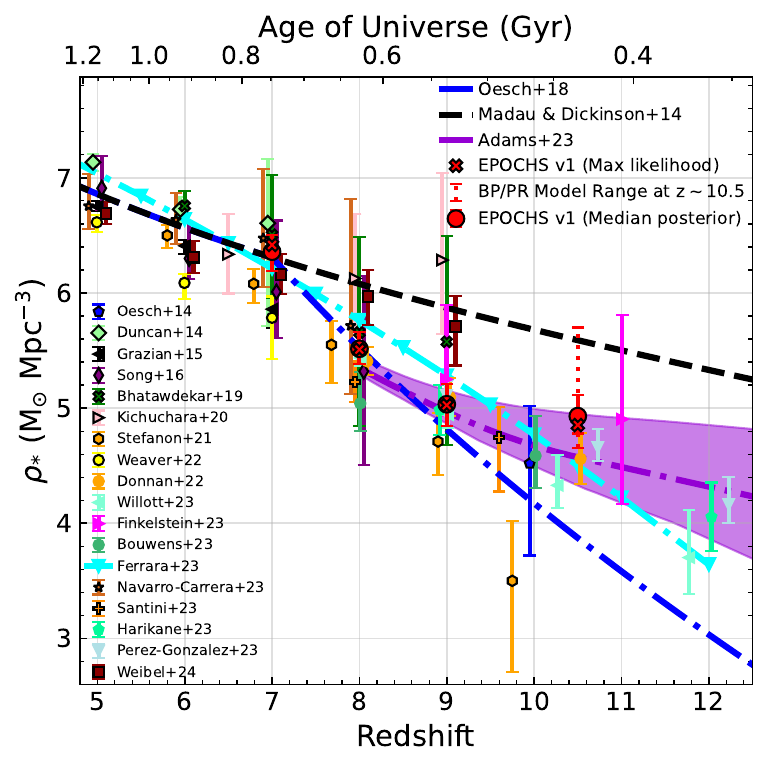}
    \caption{Stellar mass density as a function of redshift derived from the integration of the \bagpipes{} fiducial GSMF. Markers with black borders show comparisons to observational SMD results from \cite{oesch2014most, duncan2014mass, grazian2015galaxy, Song2016, bhatawdekar2019evolution, kikuchihara2020early, weaver2022cosmos2020,stefanon2021galaxy, 2024arXiv240308872W}. Markers with colored borders show the results of integrating the SFRD predictions of JWST-era studies, which includes \cite{Donnan2022,Santini2022,bouwens2023uv,Finkelstein2023,Willott2023,Harikane2023,perez2023life}. Where necessary we convert literature results to a \cite{kroupa2001variation} IMF. Markers may be shifted by up to $\Delta z = 0.1$ for clarity. We also show the theoretical predictions of the \cite{Madau2015} (constant star formation efficiency), \cite{Oesch2018} (DM halo evolution) models, as well as the model of \cite{2023MNRAS.522.3986F}. The purple shaded area shows the integral of the SFRD presented by \cite{adams2023epochs} for our sample, which is consistent with the SMD derived from our \bagpipes{} SED fitting. The red dotted errorbar at z$\sim$10.5 shows the SMD range we find using different SED fitting tools and models, calculated from the Schechter fits in \autoref{fig:gsmf_z10.5_compar}. We do not show the range of models at every redshift, but it typically exceeds the uncertainty derived from the fit itself. }
    \label{fig:smd}
\end{figure}

\begin{equation}
    \rho_\star(z) = (1-R) \int^{\infty}_{z} \psi \frac{dz}{H(z')(1+z')}
    \label{eq:sfrd_int}
\end{equation}

The return fraction $R$ is estimated as 0.423 under the assumptions of \cite{Madau2014} for a \cite{kroupa2001variation} IMF (a well-mixed closed-box model with constant IMF and metal yield and instantaneous recycling of metals). We integrate the predictions of \cite{Madau2014}, \cite{Oesch2018} and the UV luminosity density derived results of \cite{adams2023epochs}, as well as other JWST-era measurements.  For \cite{Adams2023} we propagate uncertainties in the inferred stellar mass density via a Monte Carlo integration of their SFRD measurements and uncertainties. Star formation is assumed to begin at $z = 20$, but the results are relatively insensitive to the exact formation redshift as long as it exceeds the redshift $z \geq 13.5$ limit used in this work. 

The inferred stellar mass density is heavily dependent on the best-fitting Schechter function, and therefore indirectly dependent on the GSMF and stellar mass estimates. For the alternative GSMF estimates in \autoref{fig:gsmf_z10.5_compar}, we find that the use of the ``continuity bursty" SFH, or stellar masses derived with our \prospector{} fitting, can increase the stellar mass density at redshift 10.5 by up to 0.75 dex, which is shown with a dashed errorbar. Similar variations are possible in the other redshift bins, but for simplicity we do not show every alternate GSMF and SMD derived for all redshift bins. 
    
\section{Discussion}
\label{sec:discuss}

We have presented our fiducial GSMF using \bagpipes{} at redshifts 6.5 $\leq z \leq 13.5$, finding a steep low mass slope ($\alpha \lesssim -1.95$ and a high stellar mass exponential cutoff (M$^{\star}\gtrsim 10.5$) in all redshift bins. We have also estimated the stellar mass density implied by our results, finding an apparent flattening of the stellar mass density at $z > 9$. In the following section we will discuss our results in context of theory and previous work. This includes examining the uncertainties in the GSMF derived from our other SED fitting results, as well as considering how a modified top-heavy IMF or contamination by hidden AGN would affect our results. 

\subsection{Massive Galaxies in the Early Universe}
\label{sec:evstats}

Numerous studies have reported an excess of galaxies with high inferred stellar mass at high-redshift \citep{2023Natur.616..266L, endsley2023alma,akins2023two,xiao2023massive}. If galaxies of the inferred masses do exist in the number densities implied by these studies, this could represent a challenge to $\Lambda$CDM cosmology or our understanding of high-redshift galaxy astrophysics, given the available timescale and available gas reservoirs \citep{2023Natur.616..266L,lovell2023extreme, boylan2023stress}. 

In order to test whether our derived stellar mass and redshift estimates are in tension with $\Lambda$CDM we use the Extreme-Value Statistics (EVS) approach, presented in \cite{lovell2023extreme} and available online\footnote{\url{github.com/christopherlovell/evstats}}. We follow the methodology of \cite{lovell2023extreme} throughout this section. EVS is a method for predicting the distribution of the most extreme values from a given distribution. In this case we use a parametrisation of the high-z halo mass function \citep{behroozi2013average}, and predict the PDF of the highest mass halo in some volume (given by a redshift interval, sky fraction and assumed cosmology).

\begin{figure}
    \centering
    \includegraphics[width=\columnwidth]{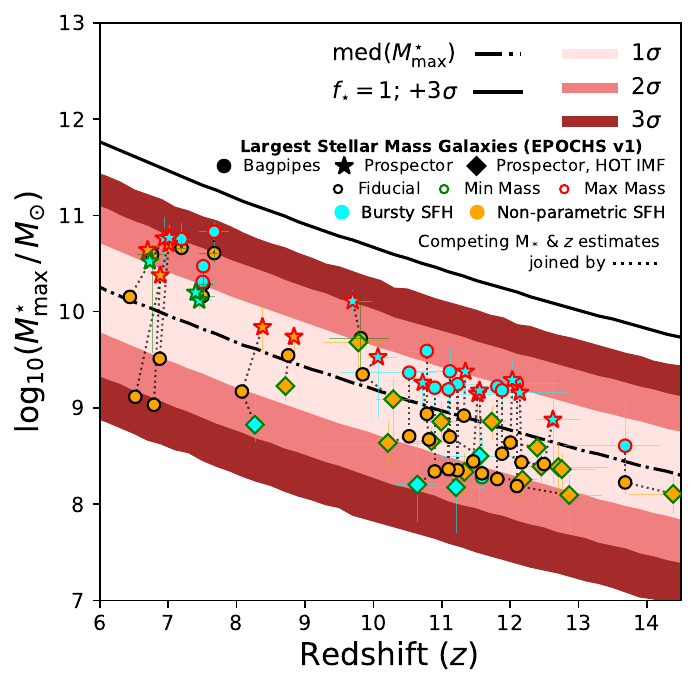}
    \caption{Photometric redshifts and stellar mass estimates for the most massive galaxies in the EPOCHS v1 galaxy sample. We use the Extreme-value Statistics (EVS) methodology of \cite{lovell2023extreme} to place constraints on the most massive galaxy at given redshift expected for our given survey area (187 arcmin$^{2}$) and a fiducial $\Lambda$CDM cosmology (with fixed baryon fraction of 0.16; \cite{2016A&A...594A..13P}). The contours show the upper and lower-limits for the most massive galaxy assuming a truncated log-normal distribution for the stellar fraction, with median shown by the dash-dot line. The solid line shows the extreme upper 3-$\sigma$ limit assuming a stellar fraction of unity. For galaxies in the EPOCHS v1 sample with a stellar mass estimate greater than 1.5$\times$ the limit for a realistic stellar fraction (dash-dot line) we show the maximum, minimum and fiducial stellar mass and redshift estimates for our \bagpipes{} and \prospector{} fits, joined with a dotted black line. The shape and color of the points references the SED-fitting tool, SFH and IMF used, as explained by the inset legend. No galaxies exceed the maximum stellar mass predicted by $\Lambda$CDM, although some galaxies do require a high stellar fraction, particularly the 'Little Red Dots' with high inferred masses at $z \sim 7$.} 
    \label{fig:evstats}
\end{figure}

\autoref{fig:evstats} shows these individual PDF contours projected into the mass-redshift plane. We show the 1, 2 and 3 sigma contours around the median prediction for the most massive galaxy at a given redshift with a dot-dash line and shaded contours. A unique aspect of the EVS approach is that it predicts both upper and lower bounds to the most massive object; the most massive galaxy can be too small for the assumed cosmology and astrophysics, as well as too massive. A universal baryon fraction of 0.16 is assumed based on cosmological results \citep{2016A&A...594A..13P}, and we compute the EVS limits for a range of stellar fractions (fraction of baryons which form stars). In \autoref{fig:evstats} we show the 3$\sigma$ upper limit, assuming a stellar fraction of unity, with a solid black line.



Overlaid on \autoref{fig:evstats} we show the SED-fitting derived redshifts and corrected stellar masses for our sample. We correct our derived stellar masses for Eddington bias \citep{eddington1913formula} following $\ln M_{Edd} = \ln M_{obs} + \frac{1}{2}\epsilon\sigma_{\ln M}^2$. $\ln M_{obs}$ is the stellar mass estimate, $\sigma_{\ln M}$ is the uncertainty in the stellar mass taken from the posterior PDF, and $\epsilon$ is the local slope of the halo mass function. \cite{jespersen2024significance} looks at the impact of cosmic variance on the predictions of EVS, as this is not accounted for in the method of \cite{lovell2023extreme} we use here. Given that our sample consists of multiple widely-separated fields, we do not believe cosmic variance will play a significant role on this EVS analysis. 
No galaxies in our sample fall above the 3$\sigma$ limit given a stellar fraction of unity, shown with a black line, which would require more stellar mass than the available baryons to form stars. 
For galaxies with redshift and stellar mass estimates which place them in possible tension with the EVS limits assuming a realistic stellar fraction (more than 1.5$\times$ the mean value of the lognormal distribution, shown by the dashed-dot line in \autoref{fig:evstats}) we show the individual mass measurements on the figure. Galaxies below this limit are not shown. 
For each of these galaxies we show the `fiducial' \bagpipes{} photo-$z$ and stellar mass, as well as the maximum and minimum stellar mass estimate for that galaxy in our \bagpipes{} and \prospector{} fitting, including the modified IMF results we discuss in \autoref{sec:imf_results}. We link individual mass estimates for the same object with dotted lines. In the majority of cases, whilst the maximum stellar mass estimate may suggest a possible tension, the minimum and often the fiducial mass estimates are not in tension. 

It is worth noting that all galaxies which show a possible tension at redshift $z \leq 8 $ in this figure are classed as `little red dots' (LRDs), which are discussed further in \autoref{sec:agn} and potentially contain a significant AGN component we do not account for. The mass estimates for these galaxies may be overestimated by up to 1~dex \citep{2024ApJ...968...34W}, as it is impossible to disentangle the relative contributions of stellar and AGN emission to the broadband photometry, and difficult even with spectroscopy \citep[e.g.][]{2024arXiv240302304W}.  It is clear from \autoref{fig:evstats} that none of our candidates are in tension with the predictions of $\Lambda$CDM, as seen by the lack of sources above the solid black line, however a number of objects do require very high stellar fractions at these redshifts. The galaxies which require the highest star formation efficiencies seem to be found at $z\sim 7 - 8$, rather than the higher redshift probed in this study. This has also been found for HST-dark galaxies observed as part of FRESCO, where the galaxies with highest implied stellar mass densities are between $5 \geq z \geq 6$ \citep{xiao2023massive}. 

Further evidence for the compatibility of our results with standard cosmological models can be seen in \autoref{fig:mf_zfit}, where we are below the SMF upper limit calculated by \cite{li2023feedback} in all redshift bins, which is shown with a black line. \cite{li2023feedback} present a bursty SFH model consisting of a series of feedback-free bursts (on timescales of $\sim$10 Myr per burst) for galaxies in halos above a given mass/redshift cutoff, resulting in higher star formation efficiency, cosmic SFR density and stellar mass density above $z \geq 8$. The upper limit shown is computed assuming a maximum star formation efficiency ($\epsilon_{\rm max}$) of unity. Our fiducial GSMF results at $z \approx 7$ are close to the FFB predictions with $\epsilon_{\rm max}$ = 0.2, but fall below this value at higher redshifts. 

\cite{wang2024true} have shown that stellar masses of high mass galaxies are typically overestimated by $\sim$0.4~dex when $> 1 \mu$m rest-frame emission is not used in SED fitting, which requires MIRI observations in this redshift regime, and as much as 0.6 - 1 dex for the reddest sources \citep{2024ApJ...968...34W}.As MIRI observations are not used in this study, the degeneracies in age-attenuation and the relative contributions of strong emission lines and the stellar continuum observed by \cite{wang2024true} are not constrained, possibly leading to an overestimation in stellar mass. Outshining and stochastic star formation are however still likely to have an effect on stellar masses derived from SED fitting, even when MIRI data is used \citep{narayanan2023outshining,shen2023impact}, and the derived discrepancy in stellar mass without MIRI will likely depend on the assumed SFH, dust law and parameter priors, as we have shown they can also systematically change stellar mass estimates. \cite{wang2024true} also find that high mass galaxies at $6 \leq z \leq 8$ may require a higher star formation efficiency ($\epsilon \sim 0.3$) than the local Universe, but they do not find any incompatibility with standard cosmological models. Alternative explanations which do not require high SFE, such as a blue-tilted primordial power spectrum, have also been proposed in the literature \citep[e.g.][]{parashari2023primordial}.

In \autoref{fig:high_mass_SEDs} we show examples of the photometry, best-fitting \bagpipes{} \& \prospector{} SEDs, and posterior redshift and stellar mass estimates for a few of the most massive galaxies in the EPOCHS v1 sample at a range of redshifts. For the galaxy labelled \textit{CEERSP1:7463}, which is representative of the LRDs we observe, the inferred star formation history for the non-parametric SED fits suggest the bulk of stellar mass was formed $\sim$100-200 Myr ago, which may put it and some of the other high-mass LRD galaxies in greater tension at $\Lambda$CDM at earlier points in their star formation histories, if the integrated SFH is considered. However we discuss in \autoref{sec:agn_discuss}, the masses and star formation histories of the LRDs are highly uncertain, particularly when they contribution of AGN is not considered.

\subsection{Contamination of the GSMF by hidden AGN} \label{sec:labbe} \label{sec:agn_discuss}
The study by \cite{2023Natur.616..266L} discovered unexpectedly massive, high-redshift galaxies in the CEERS field known as 'Little Red Dots', as discussed in \autoref{sec:agn}. \cite{2024arXiv240109981K} used similar selection criteria to extend  this to the GOODS-S field. In this section we compare our photometric redshifts and mass estimates and investigate the impact on the GSMF. We cross-matched their 13 candidates with our catalogue to compare masses and redshifts. Of these, 11 are in our catalogue, 7 of which meet our selection criteria and are among the most massive galaxies we identified. Two candidates were not recovered due to blending with neighboring sources in our {\tt SExtractor} segmentation maps. Four others we detected but did not select: \textit{5346\_CEERSP6 (11184)} is near a detector gap, and the other three did not meet our UV signal-to-noise requirement. Notably, \textit{2683\_CEERSP3 (13050)} is a spectroscopically confirmed broadline AGN at $z = 5.6$ \citep{kocevski2023hidden}. 

We did not detect \textit{37888} or \textit{39575}, the lowest mass candidates in \cite{2023Natur.616..266L} as they were blended with neighbours. Two galaxies were excluded from our study as their redshifts were below our limit. For these, we found lower redshifts than \cite{2023Natur.616..266L}: \textit{8750\_CEERSP3 (7274)} at $6.03_{-0.14}^{+0.36}$ and \textit{2499\_CEERSP1 (25666)} at $6.45_{-0.18}^{+0.10}$. For five galaxies, our photo-$z$ estimates matched well, except for \textit{1516\_CEERSP2 (21834)}, where our estimate was $z = 10.4^{+1.3}_{-0.6}$ with \eazy, differing to \bagpipes{} which found $ z = 8.63^{+0.24}_{-0.32}$, aligning with their result of $z = 8.54^{+0.32}_{-0.51}$. Our stellar mass estimates averaged 2.1$\times$ smaller than \cite{2023Natur.616..266L}, largely due to differing IMFs. The most massive candidate, \textit{7463\_CEERSP1 (38094)}, with a stellar mass of $\log_{10} (M_\star/M_{\odot}) = 10.89^{+0.09}_{-0.08}$, is the second most massive in our sample, with a \bagpipes{} mass of $10.65^{+0.09}_{-0.10}$. Given our larger survey area, this suggests the CEERS field might be overdense, as suggested by \cite{desprez2024lambdacdm}. We computed the GSMF implied by \cite{2023Natur.616..266L}, showing their datapoints at  $z = 8$ and $z=9$ are above our best-fitting Schechter functions, due to smaller cosmic volume and higher stellar masses. \cite{wang2024true} shows that including MIRI data in SED fitting reduces high-z galaxy masses by $\sim 0.4$ dex, indicating potential overestimation without MIRI data. When comparing to \cite{2023Natur.616..266L},  we find 8 of their CEERS \& GOODS-South candidates in our robust sample, with 28 more being detected but below our $z< 6.5$ cutoff or below our rest UV SNR requirements. We see good agreement in redshift, with a maximum offset of $\Delta z\approx0.4$, and a mean offset of $\Delta z\approx0.14$. These galaxies are found to have high stellar masses in our SED-fitting, with a median stellar mass of $\log_{10} M_\star/M_\odot$ = 10.30 in our fiducial \bagpipes{} results, with five of them forming the most massive galaxies in our fiducial sample. They are all at $z\leq 8$ in our sample.  
Further observations with NIRSpec or MIRI are essential to ascertain the true nature of these sources. In \autoref{sec:gsmf_no_agn} we recompute the $z=7$ GSMF without any LRDs and show that these sources dominate the high mass end of the GSMF. \cite{d2023star} has shown that accounting for the contribution of AGN lowers the cosmic star formation rate density by 0.4~dex at $z \geq 9.5$, which will also lower the inferred stellar mass density, and finds that a significant fraction of the LRDs are hidden AGN. 

Our sample is primarily selected in the rest-frame UV, potentially missing galaxies with weak UV emission but significant stellar mass, such as Submillimeter Galaxies (SMGs). These high-mass but low spatial density galaxies are excluded from our selection. Additionally, our high-resolution detection favors compact, high-surface brightness sources over extended, low-surface brightness ones, possibly excluding a population of extended, diffuse galaxies at these redshifts.

\subsection{Impact of Modified IMF}
Our modified top-heavy IMF implementation in \prospector{} reduces masses by up to $\sim$0.5 dex for galaxies at $z\geq$12 (HOT 60K) compared to the standard \cite{kroupa2001variation} IMF, as shown in \autoref{fig:prospect_imf}
. The mass decrease depends on the star formation model, with a larger reduction seen in the parametric "delayed" SFH model (0.46 dex) compared to the non-parametric "continuity bursty" SFH model (0.35 dex). The best-fitting models show the $\chi^2$ is almost unaffected by the IMF change, indicating the modified IMF models match the observed photometry as well as the original data. For the modified IMF model used at 8$\leq z \leq 12$ (HOT 45K), we see smaller mass decreases ($\approx$0.3 dex).

Comparable studies also examine the impact of a top-heavy IMF on high-redshift galaxy masses. \cite{Steinhardt2022}, using the same top-heavy IMF model with a different SED fitting tool, observed decreases of $0.5 - 1$ dex in stellar mass, while our implementation shows smaller decreases of $0.3 - 0.5$ dex. This discrepancy may be due to differences between the standard \eazy{} templates and their models, or variations in star formation histories resulting from the IMF change.
\cite{woodrum2023jades} also studied top-heavy IMF modifications using \prospector, focusing on a modification to the \cite{Chabrier2000} IMF rather than \cite{kroupa2001variation}. They found similar reductions in stellar mass (0.38 to 0.5 dex) with no change in the goodness of fit.

Our analysis shows that a modified top-heavy IMF can decrease the stellar masses of high-$z$ galaxies without altering the simulated photometry. However, our examination of the $\Lambda$CDM limits on stellar mass growth (\autoref{sec:evstats}
) does not require a non-standard IMF for compatibility with $\Lambda$CDM. As shown in \autoref{fig:bagpipes_compar}, stellar masses can vary significantly with other SED fitting assumptions (dust law, assumed SFH) before considering any IMF changes. For galaxies with high stellar masses across all models, it is challenging to distinguish between a modified IMF and high star formation efficiency based on photometry alone.

\subsection{Comparing the measured GSMF with other observations and theory/simulations}

In \autoref{fig:mf_zfit}, we compare our GSMF estimates to a wide-range of observational and theoretical/simulation based predictions of the GSMF. Here we briefly discuss our GSMF estimates for each of the redshift bins. In order to make direct comparisons where necessary all results have been converted to use a \cite{kroupa2001variation} IMF. The overall evolution of the derived Schechter parameters and a comparison to the results derived by other studies can be seen in \autoref{fig:schechter_param}. Whilst the Schechter parameters are highly covariant, and our results typically have large uncertainties, we observe an evolution in $\alpha$ and $\phi^\star$ compared, with both parameters decreasing compared to the results at $z \sim 4$ of \cite{caputi2015spitzer}, \cite{duncan2014mass} and \cite{grazian2015galaxy}. We see little evolution of M$^{\star}$ within the large range of uncertainties. 

\subsubsection{Redshift z=7 GSMF}
We derive a mass function at $z \sim 7$  primarily as a proof of concept of our method. We do not take advantage of galaxy lensing in this work, so any reasonable mass completeness limit is higher than previous studies, nor do we have the area of wide-field studies like \cite{weaver2022cosmos2020} in order to detect rare, bright and high mass galaxies. However with JWST we have seen a surprising excess of UV-faint LRD like objects with high inferred stellar masses, as discussed in \autoref{sec:agn_discuss}. The majority of these sources were previously undetected with HST due to relatively weak Lyman-breaks, and so do not appear in pre-JWST stellar mass estimates. Their inclusion in our GSMF has resulted in an excess at the high mass end of the GSMF when compared to other observational studies, and consequently a higher and poorly constrained estimate of M$^\star$, as we see little evidence for any exponential turnover. The highest mass GSMF data-points of \cite{weaver2022cosmos2020} fall within our 1$-\sigma$ uncertainty region, but our results are significantly above the measurements of \cite{stefanon2021galaxy}. At the low-mass end we fall below the results of \cite{kikuchihara2020early} and \cite{navarro2023constraints}, but agree within the uncertainties of \cite{2021MNRAS.501.1568F} and \cite{bhatawdekar2019evolution}. \cite{stefanon2021galaxy, kikuchihara2020early,bhatawdekar2019evolution} and \cite{2021MNRAS.501.1568F} are all based on HST+Spitzer observations of the Hubble Frontier Fields, and incorporate lensing, which means they probe the low mass end of the GSMF more accurately then this study. Our low mass slope $\alpha=-1.94^{+0.1}_{-0.1}$ is in good agreement with \cite{2021MNRAS.501.1568F}, but steeper than the results \cite{stefanon2021galaxy} and \cite{kikuchihara2020early}. 
At the time of writing, \cite{navarro2023constraints}, \cite{gottumukkala2023unveiling}, \cite{wang2024true} and \cite{2024arXiv240308872W} are the only other studies to incorporate JWST observations into their GSMF estimates. We see reasonable agreement with the results of  \cite{navarro2023constraints} as our data is within their GSMF uncertainties. This work relies almost entirely on JWST observations, whereas they combine deep JWST observations of small volumes ($\leq$20 arcmin$^2$) with HST and ground-based catalogues. This ground-based data allows them to find more rare, high-mass galaxies than our study, but at lower masses the small volumes probed with their JWST data are potentially vulnerable to cosmic variance. Reliance on ground-based and HST data also limits the maximum redshift they can probe to $z\leq 8$. \cite{wang2024true} uses PRIMER observations with NIRCam and MIRI to measure the GSMF, notably finding that the use of MIRI observations systematically reduces stellar masses measured with SED fitting. We see good agreement in the measured GSMF within the uncertainties of both studies, despite this work not incorporating MIRI data or correcting for any systematic offset in mass arising from the lack of restframe $> 1 \mu$m observations. Our GSMF does extend to higher stellar mass than the result of \cite{wang2024true}, resulting in a higher value for $M^{\star}$, as can be seen in \autoref{fig:schechter_param}. 

\cite{gottumukkala2023unveiling} examine the contribution of high-mass, dusty galaxies at $3 < z < 8$ to the GSMF using data from the CEERS survey. Given that our GSMF probes a wider galaxy population, we do not expect to see overlap at all stellar masses. We see good overlap at the highest stellar masses M$_{\odot} \sim 10^{10.5}$, where our SMF estimate is dominated by dusty LRD galaxies (as discussed in \autoref{sec:agn_discuss}). We see good agreement with \cite{2024arXiv240308872W} at $z=7$, who construct the GSMF at $4 \leq z \leq 9$ using data from JADES, CEERS and PRIMER. 

When we compare to predictions from models and simulations, we see agreement with the majority of models at the low-mass end but a significant excess at higher masses that is not reproduced by any of the models. We find in particular that the \jaguar{} model we use for our completeness simulations shows a more rapid decline at high stellar mass than the other models, but given that we are not reliant on our completeness correction in this mass regime this does not impact our estimate of the GSMF. We are closest to the prediction of \emph{Universe Machine} \citep{2019MNRAS.488.3143B} at the highest stellar mass bin. 

\subsubsection{Redshift z=8 GSMF}

Our fiducial GSMF estimate at $z \sim 8$ shows reasonable agreement with most predictions. As we do not bootstrap in redshift when constructing the GSMF, we do not account for galaxies scattering between redshift bins, and for example a galaxy found to be at $z=7.49$ with \bagpipes{} would contribute only to the $z = 7$ GSMF, even in a significant fraction of the redshift PDF is above $z \geq 7.50$. This does not affect the majority of galaxies within our sample, but it does explain some of the discrepancy between our results and the implied results of \cite{2023Natur.616..266L}, shown in purple in \autoref{fig:mf_zfit} assuming 100\% completeness. The LRD galaxies of \cite{2023Natur.616..266L} which we also select are all found to be at $z \leq 7.5$, meaning they do not contribute at all to our estimate of the $z=8$ GSMF. The best-fitting redshift for these objects in some cases is quite close to this boundary however, meaning that these objects could theoretically contribute to the $z=8$ GSMF instead, which would boost the high-mass end significantly. Our GSMF is also lower than the results of \cite{kikuchihara2020early}, which incorporates strong gravitational lensing in order to probe to lower stellar mass. Our GSMF agrees with the results of \cite{Song2016, bhatawdekar2019evolution,stefanon2021galaxy}, and appears to validate the majority of pre-JWST GSMF estimates. We can also draw a comparison to the results of \cite{wang2024true}, whose GSMF results at $z=8$ are systematically above our results, but within the derived uncertainties of both studies. We are also below the results of \cite{2024arXiv240308872W}.

We see good agreement with most theoretical predictions of the GSMF at this redshift, with \emph{Universe Machine} \citep{2019MNRAS.488.3143B}, \emph{SC SAM GUREFT} \citep{2023arXiv230404348Y} and \emph{FLARES} \citep{2021MNRAS.500.2127L, 2023MNRAS.519.3118W} having the most similar results.

\subsubsection{Redshift z=9 GSMF}

Our GSMF estimate at $z \sim 9 $ is below the results of \cite{bhatawdekar2019evolution} and \cite{kikuchihara2020early}, but within the uncertainties of \cite{stefanon2021galaxy}. We are below the implied result of \cite{2023Natur.616..266L}, which is derived from two galaxies in their sample in this redshift bin, but assuming 100\% completeness. We include one of these galaxies in our GSMF at this redshift, as the other does not meet our selection criteria, as we do not detect the Lyman break at 5$\sigma$. We additionally include one other candidate from their sample in this GSMF, as our fiducial \bagpipes{} photo-$z$ places it within this redshift bin, rather than the $z \sim 8$ redshift bin based on their photo-$z$. For both of their galaxies we do include in this redshift bin we find $\sim 0.4$ dex lower stellar masses, meaning they contribute to a lower stellar mass bin. Our reliance on the rest-frame UV to robustly detect sources is one limitation of this work, although the increased depth of JWST observations when compared to HST has reduced this in some fields. We investigated less-stringent constraints on the Lyman-break, but found that this dramatically increased rates of contamination within our sample.
We are also below the results of \cite{2024arXiv240308872W}, but within the uncertainties where they consider their $z=9$ GSMF reliable, at $8.5 \leq \log_{10}(M_\star/M_\odot) \leq 9.5$, as they suspect low-z contamination at higher inferred masses.
When compared to simulations, \emph{FLARES} and \emph{Universe Machine} are close to our GSMF estimate, but almost all the predictions are within our posterior region. Interestingly, in this redshift bin we are below the predictions of two recent JWST-era studies; \cite{mauerhofer2023dust} and \cite{li2023feedback}, both of which incorporate higher star formation efficiencies than typical models.

\subsubsection{Redshift z=10.5 GSMF}

At $z \sim 10.5$ observational comparisons can be made only to the pre-JWST results of \cite{stefanon2021galaxy}. In comparison to their results we find a significant excess of high mass galaxies in our observations. Our results show that above $z \geq 10$ JWST observations are essential to accurately sample the high-$z$ galaxy population. Our results are above the majority of theoretical and simulation-derived predictions, but do show good agreement with \emph{Universe Machine} \citep{2019MNRAS.488.3143B} and \emph{FLARES} \citep{2021MNRAS.500.2127L, 2023MNRAS.519.3118W}. 

\subsubsection{Redshift z=12.5 GSMF}

In our highest redshift bin, 11.5 $< z \leq 13.5$  which covers only $\approx$ 80 Myr, there are no published observationally-derived results and few theoretical or simulation-based GSMF comparisons at this redshift. Pre-JWST estimates of the GSMF were not possible at this redshift, and even with JWST our GSMF estimate is also uncertain due to the difficulty in accurate stellar mass estimates as well as the possible contribution of contaminants. At $z = 12.5$ the longest wavelength NIRCam filter falls within the rest-frame UV, which is dominated by young stars, leading to highly uncertain star formation histories and stellar masses. As we explored briefly in \autoref{sec:imf_results}, the more likely possibility of a top-heavy IMF or exotic stellar populations in these early galaxies further increases the systematic uncertainties in the stellar mass estimates. An example of three galaxies within this redshift bin is shown in the lower plot of \autoref{fig:high_mass_SEDs} and in \autoref{fig:discrepant_SEDs}, and the range of stellar mass estimates ($\sim0.8 - 1 dex$) for different \bagpipes{} and \prospector{} with very little difference in the fitted rest-UV spectra shows the difficulty in estimating stellar mass at these redshift. Previous studies at lower redshift with HST and Spitzer have found that stellar masses estimated by HST alone, with no measurement of the rest optical emission, typically underestimate stellar masses by 0.62~dex, compared to measurements including HST and Spitzer NIR observations \citep{2021MNRAS.501.1568F}. At this redshift range, our JWST NIRCam observations are probing comparable rest-frame UV wavelengths to HST observations at $z \sim 6-7$, and it possible that our stellar masses are also underestimated unless there is a significant change in stellar populations or IMF. The possibility of more stochastic star formation histories at this redshift compared to $z \sim 6-7$ may also lead to outshining, which further increases the stellar mass discrepancy \citep{narayanan2023outshining}. In order to test whether the reliance of the rest UV is impacting the stellar masses compared to lower redshifts, we refit all galaxies using \bagpipes{} in the $z=9$ bin without F410M or F444W, which gives equivalent rest-frame coverage to a $z=12$ galaxy. We see no systematic shift in the stellar masses, with the median offset being 0.07 dex and a standard deviation of 0.3 dex, when comparing to the stellar mass estimate including the longer wavelengths filters.

We note that several galaxies in this bin have been excluded from the GSMF in this case due to our requirement that the contamination is less than 50\%. The inclusion of these possible contaminants would result in an $\approx$ 0.3~dex increase in the lowest mass bin. The results of this are shown in \autoref{sec:z12.5_contam}.
Whilst we do not attempt to fit the GSMF, we can make approximate comparisons to the few available predictions. We see the closest agreement with \emph{DELPHI} \citep{mauerhofer2023dust} and are within 1.5$\sigma$ of \emph{FLARES} \citep{2021MNRAS.500.2127L, 2023MNRAS.519.3118W} in the higher mass bin, but have an excess of $\approx 10^{8}$ M$_{\odot}$ galaxies when comparing to \emph{FLARES} and \emph{SC SAM GUREFT} \citep{2023arXiv230404348Y}. Our results are significantly above the predictions of \emph{BLUETIDES} \citep{2016MNRAS.455.2778F, 2017MNRAS.469.2517W} at all stellar masses. Our fiducial GSMF prediction is slightly below the prediction of \cite{li2023feedback} with maximum star formation efficiency $\epsilon_{\rm max}$ = 0.2, and significantly below the upper limit of $\epsilon_{\rm max}$ = 1.

\subsubsection{Alternative GSMF Estimates}\label{sec:alt_gsmf}

\begin{figure}
    \label{fig:gsmf_z10.5_compar}
    \includegraphics[width=\columnwidth]{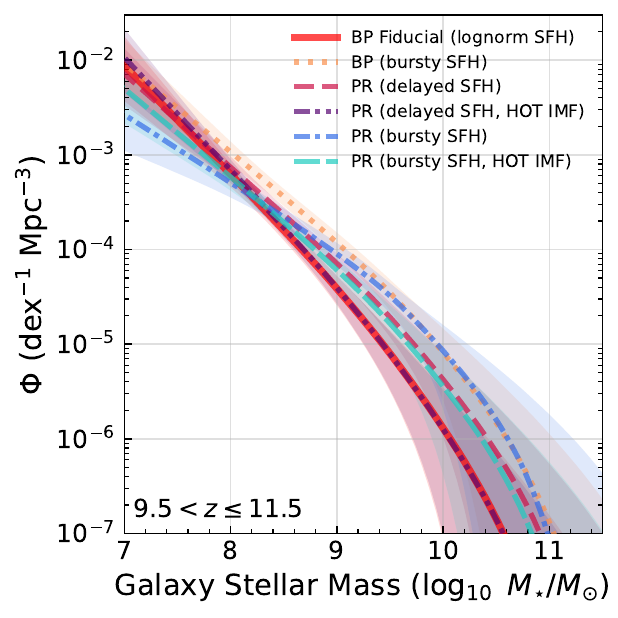}
    \caption{GSMF Schechter parametrization for the $z\sim$10.5 redshift bin derived for our fiducial \bagpipes{} SED-fitting compared to alternative GSMF estimates. We show the GSMF derived using \bagpipes{} with the non-parametric ``continuity bursty" SFH (labelled `bursty') as well as for two GSMFs derived using \prospector{} SED-fitting for a parametric and non-parametric SFH. We also show for comparison the GSMF inferred with the alternative top-heavy IMF model. The derived GSMF is clearly dependent on the choice of model and SED fitting tool, with the ``continuity bursty" SFH model typically shifting the GSMF towards higher stellar mass. These Schechter functions are tabulated in \autoref{sec:alternative_tabulated}.}
\end{figure}

Following on from the comparison of galaxy stellar mass estimates with different choices of SFH and priors in \autoref{sec:mass_comparisons}, it is possible to estimate the GSMF for any of the different SED-fitting models. A full comparison of the derived GSMF for every model, given the many possible combinations of possible IMF and SFH models, is beyond the scope of this work, but we give a representative example of the GSMF derived at z$\sim$10.5 for the models which show the most variation in stellar mass when compared to our fiducial \bagpipes{} fitting. Here we choose to investigate the GSMF dependence on the chosen SFH model and SED fitting tool, rather than the choice of parameter prior or dust law. This is because in \autoref{sec:mass_comparisons}, there was larger variation in stellar mass with little variation in $\chi^2$ for these alternative models.  Additionally, as discussed, the use of non-parametric SFHs is more common in the literature \citep[e.g.][]{tacchella2022stellar,gimenez-artega2022,jain2024motivation,giménezarteaga2024outshining} for high-$z$ galaxies due to problems such as outshining.

We derive these GSMF estimates using the same method as described in \autoref{sec:gsmf} for the fiducial \bagpipes{} GSMF, replacing the stellar mass PDFs, best-fitting SEDs and redshift estimates with those of the chosen model. \autoref{fig:gsmf_z10.5_compar} shows a comparison of the fiducial \bagpipes{} GSMF to a GSMF derived from the ``continuity bursty" non-parametric SFH model, which increases the stellar mass estimates by 0.2~dex on average, but $\geq 1$~dex in some cases. This results in the largest change in the overall GSMF when compared to the fiducial \bagpipes{} result.

We also show mass functions derived from our \prospector{} SED-fitting, which are offset above our fiducial \bagpipes{} GSMF for both the parametric and non-parametric results. These results are somewhat comparable to the spread seen in the \bagpipes{} results, although the low-mass end of the GSMF in the ``continuity bursty" SFH model produces a shallower slope than the other models. Crucially we can see that the derived GSMFs are in tension with each other, and do not typically fall within the confidence intervals across the majority of the stellar mass range. This is consistent with \cite{wang2023quantifying}, who argue that the stellar mass uncertainties are typically underestimated by SED fitting procedures. 

The change in inferred stellar mass we observe with a modified IMF does not appear to vary strongly with stellar mass, so the the impact on the GSMF can generally be seen as a shift towards lower stellar mass of $0.3 - 0.4$~dex. This is comparable in magnitude and opposite in direction to the shift seen when moving from \bagpipes{} to \prospector{} when using a parametric SFH, which results in little overall change in the resulting GSMF. 


These results demonstrate the overall systematic uncertainty different assumptions cause in the GSMF which are not represented by the uncertainty contours. Most GSMF estimates do not consider the overall uncertainty introduced by the assumptions of their modelling, which often dwarfs the statistical uncertainty in the fit itself. The variation in the derived GSMF can also significantly impact the implied SMD, as discussed in the next section.

\subsection{Stellar Mass Density Evolution in the Early Universe}

The growth of stellar mass density in the early Universe is highly uncertain. Some observational studies \cite[e.g.][]{oesch2014most,stefanon2021galaxy,Willott2023} have found a sharp decline in stellar mass density at $z \geq$ 8, whereas others see a flatter evolution \citep{kikuchihara2020early, bhatawdekar2019evolution}. On the theoretical side, \cite{Oesch2018} uses dark matter halo evolutionary models to predict a deviation from the constant star formation efficiency (CSFE) model of \cite{Madau2015}, which follows a significantly steeper slope at $z\geq$7. \cite{2023MNRAS.522.3986F} has also recently presented a new model for the UV LF and SMF incorporating obscured star formation as well as a lack of dust within high-$z$ galaxies, which most closely matches our SMD results, despite predicting a steeper slope than we observe at $z>9$. 

Our results from our fiducial \bagpipes{} model fall between the predictions of \cite{Madau2015} and \cite{Oesch2018}. We see a flatter evolution with redshift than predicted with \citep{Oesch2018}, but an overall lower stellar mass density than the CFSE model of \cite{Madau2015}. For our other GSMF estimates at $z=10.5$, shown in \autoref{fig:gsmf_z10.5_compar}, we find that $\rho_\star$ increases by up to 0.75~dex, which would bring it closer to the constant star formation efficiency prediction of \cite{Madau2015}. Whilst we do not show the SMD scatter measured in other redshift bins, we typically see the same behaviour at $z > 7$, with our fiducial \bagpipes{} SMD result producing lower $\rho_\star$ estimates than our alternative models. With our fiducial \bagpipes{} results we see significant evolution of the GSMF between $z=7$ and $z=8$, with $\rho_\star$ decreasing by $\sim 0.85$~dex. However we see a significantly flatter evolution in the SMD derived from the ``continuity bursty" model GSMF, with a decrease of only $\sim 0.4$~dex across the same redshift range. This is due partly to the overall increase in stellar mass estimates observed with this SFH model when compared to our fiducial model, as detailed in \autoref{sec:mass_comparisons}, but is also due to the scattering of the high-mass LRD galaxies scattering between the $z=7$ and $z=8$ redshift bins due to uncertain photo-$z$ estimates, which significantly impacts the GSMF at higher stellar masses. 

We see a good agreement between the integration of the star formation rate density of \cite{adams2023epochs}, which uses the same sample, and our fiducial SMD results. There are very few JWST-era GSMF estimates to directly compare against, and so we have computed the inferred stellar mass density based on the integral of the cosmic star formation rate density of other studies. We note however the numerous works showing the increased scatter in mass-to-light ratios observed  due to bursty star formation \citep{Santini2022,asada2023bursty}, which will impact the assumptions made to convert these UV luminosity densities into stellar mass densities. 

In \autoref{fig:smd} we show the overall stellar mass density range we find when we use a different SED fitting tool or star formation history model (dotted red uncertainty). This is significantly larger than the statistical uncertainty in the stellar mass density from our fiducial \bagpipes{} results. A change of up to 0.75~dex at $z\approx 10.5$ is possible, when only the SED fitting tool or SFH model is varied and the overall sample is unchanged. More significant variations are possible between the results of independent studies, which also have to consider differences in reduction, source detection, photo-$z$ estimation, selection procedure, cosmic variance, and completeness corrections. Not accounting for the contribution of AGN to the observed photometry may cause overestimation of the stellar mass density at high-redshift \citep{d2023star}. 

The range of stellar mass densities possible with our alternative GSMF estimates at $z \sim 10.5$ is mostly above the 1$\sigma$ range of \cite{adams2023epochs}. A discrepancy between the integrated star formation rate density and stellar mass density measured for the same sample could hint at a different IMF, since the assumed return fraction is strongly dependent on the chosen IMF, and the SMF and UVLF probe different stellar populations with different characteristic stellar mass. However there are a number of other possible issues with the conversion of the UVLF into a SMD estimate; the conversion of UV flux to SFR ($\kappa_{\rm UV}$), is often assumed to be a constant factor but is actually dependent on the age and metallicity \cite{Madau2014}, as well as the other assumptions used to calculate the return fraction (closed-box model, constant IMF \& metal yield and instantaneous recycling of metals) which may not be valid at high-redshift. 

As we show in \autoref{sec:mass_comparisons}, in some cases discrimination between models or priors based on the goodness of fit may be possible, but in others (e.g. assumed SFH model), significant scatter in stellar mass estimates are possible with no difference in $\chi^2$. Other studies which use only one method for measuring stellar mass estimates will underestimate the overall uncertainty in the derived GSMF and stellar mass density estimates.

\section{Conclusions}
\label{sec:conclusions}
In this paper we present an investigation into the properties of the EPOCHS v1 sample of 1120 high-redshift galaxies at 6.5 $\leq z \leq 13.5$ taken from a uniform reduction of 187 arcmin$^2$ of JWST data, including the GTO program PEARLS as well as other public ERS/GO JWST programs. 

We examine the consistency of galaxy properties, including stellar mass, under different assumptions and using different SED fitting tools, including \bagpipes{} and \prospector. In particular we examine the impact of different SFH parametrizations as well as switching between a parametric and non-parametric SFH models. We also also investigate the possible reduction in stellar mass when assuming a top-heavy IMF. We then use this sample and our range of stellar mass estimates to construct possible realisations of the galaxy stellar mass function. Lastly we integrate our mass function estimates to probe the buildup of stellar mass in the early Universe via the stellar mass density. 

The major conclusions from this study are as follows:

\noindent 1. We find that the stellar mass of high-redshift galaxies can depend strongly on assumed models, their priors and the SED fitting package used. In particular the estimated stellar mass can increase by $>$1~dex when a parametric SFH is exchanged for a non-parametric SFH, with no change in the goodness of fit. Higher stellar mass discrepancies are seen at $z > 10$ due to a lack of rest-optical emission. 

\noindent 2. We find that the assumption of a modified top-heavy \cite{kroupa2001variation} IMF, which may more accurately model the hot star-forming regions within high-$z$ galaxies, can reduce stellar mass estimates by up to 0.5~dex with no impact on the goodness of fit. 

\noindent 3. Whilst some of the stellar mass estimates imply a high star formation efficiency, in our analysis of the most massive galaxies in our sample using the Extreme-Value Statistics methodology of \cite{lovell2023extreme} we do not find any galaxies which are incompatible with the existing $\Lambda$CDM cosmology. The largest stellar mass estimates are typically found when fitting the non-parametric SFH models, and often can be significantly reduced with an alternative model. We not require a top-heavy IMF 

\noindent 4. Across all of the fitted models, the highest mass galaxies in our sample are `Little Red Dots', with inferred masses of $>$10$^{10}$ M$_{\odot}$ at $z\approx 7$. These galaxies dominate the highest mass bins of our galaxy stellar mass function (GSMF) estimates, so understanding their true stellar populations and accounting for the likely contribution of AGN \citep{greene2023uncover, furtak2023jwst,furtak2023jwstnirspec,2024ApJ...968...34W} will be essential to more accurately constrain further GSMF estimates. 

\noindent 5. With the GSMF derived from our fiducial \bagpipes{} results, we typically see good agreement with existing constraints on the GSMF at z$\leq$9.5. At the limits of HST+Spitzer (z$\geq$10) we see an excess of galaxies when compared to pre-JWST observations, but our GSMF results fall within predictions of simulations and theory. 

\noindent 5. The systematic variation in stellar mass estimates we find can dramatically impact the inferred galaxy stellar mass function and therefore the stellar mass density. We show that the choice of star formation history model or SED fitting tool can cause up to a 0.75~dex shift in the overall stellar mass density at z$\approx$10.5 with the same sample of galaxies. We predict larger offsets between independent samples, where different reductions, selection techniques and photo-$z$ estimates will increase the uncertainties. 

\noindent 6.  We see a flatter evolution of the cumulative stellar mass density than predicted by dark matter halo evolution models, whilst the slope of our results is more consistent with constant star formation efficiency models. Our results suggest that significant stellar mass had already formed at $z\geq$11.5. 

This is only the beginning of GSMF estimates at $z > 10$, and the use of ultra-deep observations (the second NGDEEP epoch \citep{Bagley2023ngdeep}, the JADES Origins Field \citep{2023arXiv231012340E,2023arXiv231210033R} and others) and magnification by lensing clusters \citep[PEARLS, UNCOVER, CANUCS; ][]{Windhorst2023,2022arXiv221204026B,2023jwst.prop.4527W} will help constrain the GSMF at stellar masses below our completeness limit of $\sim 10^{8}$ M$_{\odot}$, whilst widefield surveys \citep[e.g. PRIMER, UNCOVER, Cosmos-Webb; ][]{2021jwst.prop.1837D,2022arXiv221204026B,2021jwst.prop.1727K} will add area and rare higher-mass sources. Deep MIRI F560W or F770W observations \citep[e.g. the MIRI HUDF survey,][]{2017jwst.prop.1283N} will be crucial to provide better constraints on stellar mass estimates at these redshifts by extending the wavelength range further into the rest-frame optical, although the sensitivity of MIRI decreases rapidly with increasing wavelength \citep[e.g.][]{wang2024true}. More complete NIRSpec coverage is also important to identify interlopers, confirm photometric redshifts and distinguish between AGN emission and star forming galaxies. 

All of the raw JWST data used in this work are the same as used in \cite{adams2023epochs} and can be accessed via this MAST DOI: \href{https://archive.stsci.edu/doi/resolve/resolve.html?doi=10.17909/5h64-g193}{DOI 10.17909/5h64-g193}. All proprietary data from the PEARLS program will all become accessible over 2024. Catalogues for all high-$z$ galaxies will be published with the EPOCHS I paper \citep{2024arXiv240714973C}. The fiducial GSMF and SMD results from this work are availale on \href{https://github.com/tHarvey303/EpochsIV}{GitHub}, and results for our alternative models will be made available upon request.


\clearpage

\begin{acknowledgments}
TH, CC, NA, DA, QL, JT, LW acknowledge support from the ERC Advanced Investigator Grant EPOCHS (788113), as well as two studentships from the STFC. 

AZ acknowledges support by Grant No. 2020750 from the United States-Israel Binational Science Foundation (BSF) and Grant No. 2109066 from the United States National Science Foundation (NSF); by the Ministry of Science \& Technology, Israel; and by the Israel Science Foundation Grant No. 864/23. RAW, SHC, and RAJ acknowledge support from NASA JWST Interdisciplinary Scientist grants NAG5-12460, NNX14AN10G and 80NSSC18K0200 from GSFC. CCL acknowledges support from the Royal Society under grant RGF/EA/181016. The Cosmic Dawn Center (DAWN) is funded by the Danish National Research Foundation under grant No. 140. CNAW acknowledges funding from the JWST/NIRCam contract NASS-0215 to the University of Arizona. M.N. acknowledges INAF-Mainstreams 1.05.01.86.20. CNAW acknowledges support from the NIRCam Development Contract NAS5-02105 from NASA Goddard Space Flight Center to the University of Arizona.

This work is based on observations made with the NASA/ESA \textit{Hubble Space Telescope} (HST) and NASA/ESA/CSA \textit{James Webb Space Telescope} (JWST) obtained from the \texttt{Mikulski Archive for Space Telescopes} (\texttt{MAST}) at the \textit{Space Telescope Science Institute} (STScI), which is operated by the Association of Universities for Research in Astronomy, Inc., under NASA contract NAS 5-03127 for JWST, and NAS 5–26555 for HST. The authors thank all involved with the construction and operation of JWST, without whom this work would not be possible. We also thank the PI's and teams who designed and executed the ERS, GTO and GO programs used within this work, including PEARLS (1176, 2738), SMACS-0723 (2737),  GLASS (1324), CEERS (1345), JADES (1180, 1210, 1895, 1963)
and NGDEEP (2079).  

This work makes use of {\tt astropy} \citep{Astropy2013,Astropy2018,Astropy2022}, {\tt matplotlib} \citep{Hunter2007}, {\tt reproject}, {\tt DrizzlePac} \citep{Hoffmann2021}, {\tt SciPy} \citep{2020SciPy-NMeth} and {\tt photutils} \citep{larry_bradley_2022_6825092}.

\end{acknowledgments}

%

\vspace{5mm}
\facilities{JWST (STScI), HST (STScI)}



\appendix

\section{Effect of SED Fitting Assumptions on Derived  SPS Quantities}
\label{sec:appendix_sps}
\begin{figure}
    \centering
    \includegraphics[width=0.9\textwidth]{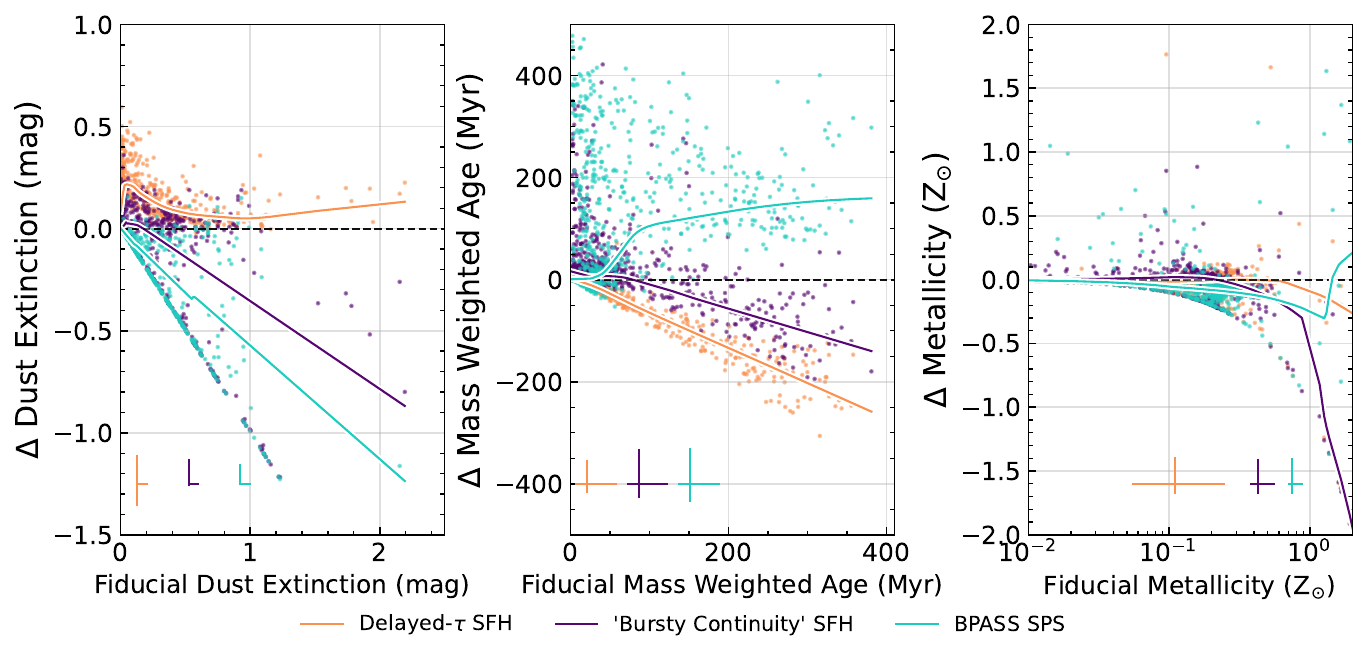}
    \includegraphics[width=0.9\textwidth]{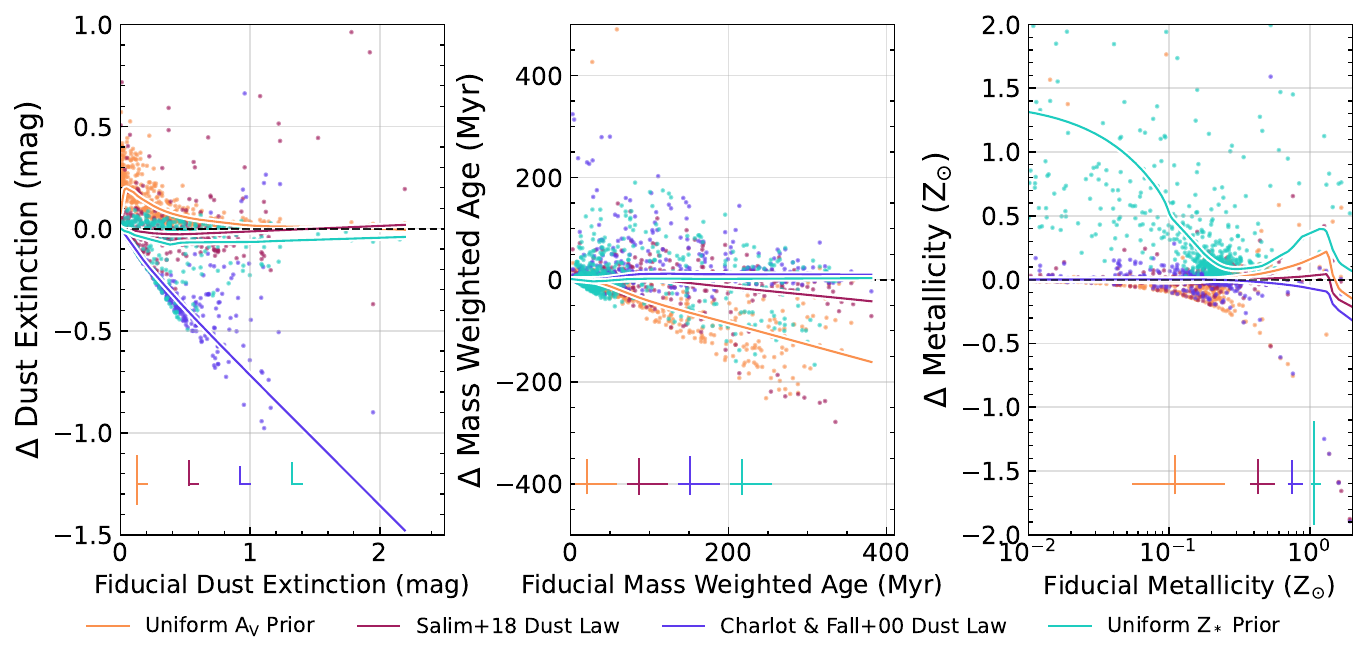}
    \caption{Comparison of the derived SPS parameters for our fiducial \bagpipes{} model when compared to the alternative models listed in \autoref{tab:bagpipes_table}. On the y-axis we plot $\Delta = $ fiducial model - alternative model for the given parameter. From left to right we show the dust extinction A$_{\rm V}$, the mass-weighted age, and the stellar metallicity. The top row compares SFH and SPS models whilst the bottom row considers the impact of priors and the chosen dust law. Average uncertanties are given in the bottom left of each point, and the LOWESS trend is shown with a line, where the color corresponds to the \bagpipes{} model.}
    \label{fig:bagpipes_sps_compare}
\end{figure}

Here we give a more detailed comparison between our fiducial Bagpipes results and the alternative models discussed in \autoref{sec:bagpipes}. In  \autoref{fig:bagpipes_sps_compare} we show the equivalent of \autoref{fig:bagpipes_compar} for other derived parameters, in order to understand what drives the observed discrepancies in stellar mass between different models. 

\subsection{Impact of Priors}
The first comparison we make is a substitution of the default logarithmic prior on the V-band dust extinction, A$_{\textrm{V}}$, to a uniform prior. A uniform prior favors higher dust extinctions, and the largest difference to the fiducial model is seen in galaxies with A$_{\textrm{V, fid}}<0.2$ mag, where the dust content is poorly constrained. We see good agreement in photo-$z$, along with a few extreme outliers, which is unsurprising given the informative redshift prior from \eazy{} we use. The stellar mass offset is shown in light blue in the top plot of \autoref{fig:bagpipes_compar} and shows good agreement within the posterior uncertainties. Galaxies with higher levels of dust are found to be significantly younger and more star-forming with a uniform dust prior, as can been see in \autoref{fig:bagpipes_sps_compare} suggesting the inferred star formation histories are dependent on the dust prior. \autoref{fig:bagpipes_sps_compare} also shows that the measured dust extinction which is influenced by the prior is primarily at $A_V < 0.3$, with both models agreeing on the dust extinction for the galaxies with higher dust extinction. Comparison of the best-fitting $\chi^2$ values shows that both models are equally well-fitted to the photometry. 

We also test the impact of our metallicity prior, which is logarithmic in our fiducial \bagpipes{} model. This favours low metallicity, which we expect in the early Universe. Here we exchange this prior for a uniform distribution which favours higher metallicity, shown in orange in the top plot of \autoref{fig:bagpipes_compar}. We see little overall impact on the stellar mass from the metallicity prior, with individual galaxies scattering up to 0.5~dex and the majority consistent with results from our fiducial \bagpipes{} run. The metallicity itself, which is extremely difficult to constrain from photometry, is highly influenced by the prior chosen, as can be seen in \autoref{fig:bagpipes_sps_compare}, although this doesn't appear to systematically impact the star formation history or dust extinction.

Overall the impact of the dust and metallicity priors alone appears to have only a small systematic effect on the derived galaxy masses. However, in a small number of individual cases masses can scatter by $\sim$0.5~dex with little difference in the goodness of fit. 

\subsection{The Assumed Dust Law}

Our fiducial \bagpipes{} model assumes a simple one-component \cite{calzetti2000dust} dust law. The slope of the dust law is known to vary in some galaxies, and numerous alternative models \cite[e.g.][]{charlot2000simple,2018ApJ...859...11S} including additional parameters have been suggested. \cite{charlot2000simple} fit a two-component dust law, with different amounts of extinction for young ($\leq 10$~Myr) and old stellar populations, to account for dust in stellar birth clouds. The resultant deviations in stellar mass are shown in magenta in the top plot of \autoref{fig:bagpipes_compar}, with a significant deviation at the highest stellar masses, which can be seen in the LOWESS trend. For galaxies with a fiducial stellar mass of $\geq 10^{10}$ M$_{\odot}$, the majority are found to have $\geq$1 dex larger stellar masses with the \cite{charlot2000simple} dust law. However for these galaxies the goodness of fit is considerably poorer, with $\Delta \chi^2 \geq 10$ with the \cite{charlot2000simple} model. 

Despite the same prior, the dust attenuation (A$_{\textrm{V}}$) posteriors are quite different between the two models, due to the degeneracy between A$_{\textrm{V}}$ and slope, as can be seen in \autoref{fig:bagpipes_sps_compare}. Galaxies with moderate dust extinction in the fiducial results ($\leq 1$ mag) are typically found to have very little dust extinction on the old stellar population, with higher dust extinction on the young stellar populations. The majority of galaxies favor a steeper slope ($n$) for the attenuation power-law than given by \cite{calzetti2000dust}, in this mode with a distribution centered on n$\approx 1.3$, which is 2$\sigma$ from the prior value of 0.70. 

\cite{2018ApJ...859...11S} allow a deviation in slope compared to the \cite{calzetti2000dust} and an additional UV bump at $2175~\mathrm{\AA}$. The UV bump in the \cite{2018ApJ...859...11S} model is driven primarily by Polycyclic Aromatic Hydrocarbon (PAH) emission, which is expected when emission is strongly reprocessed. We replace the \cite{calzetti2000dust} dust law with the \cite{2018ApJ...859...11S} model, and fit for these additional components, as shown in \autoref{tab:bagpipes_table}. We find that the inferred stellar mass can increase by $>$1~dex in some cases, particularly at the highest fiducial stellar mass, but the average offset (see the LOWESS fit in \autoref{fig:bagpipes_compar} is smaller than for the \cite{charlot2000simple} case. In a number of cases the goodness of fit for the \cite{2018ApJ...859...11S} model is considerably better than our fiducial model, with $\delta \chi^2 \leq -10$ in a number of galaxies. 


\subsection{Comparison to `delayed' SFH}

To test the consistency of galaxy stellar mass estimates with different parametric star formation histories, we replace our `lognormal' SFH with another commonly used SFH; a delayed-$\tau$ SFH. We find systematically slightly lower stellar masses above 10$^{8}$ M$_{\odot}$, but reasonable agreement at the lowest stellar masses. This is likely due to the log-uniform prior on age used in the delayed-$\tau$ model, which results in a systematically younger stellar population, as can be seen in \autoref{fig:bagpipes_sps_compare}. We also see some impact on the dust attenuation, likely driven by the change in inferred SFH, with systematically higher dust attenuation inferred with this SFH. Comparison of the goodness of fit via the $\chi^2$ parameter suggests the models typically have slightly poorer fits than the fiducial model, but in some cases the stellar masses are reduced by $\sim$ 0.5~dex with very little impact on the goodness of fit.

\subsection{Comparison to ``continuity bursty" SFH}

Here we replace our fiducial \bagpipes{} `lognormal' SFH with a ``continuity bursty" SFH model described in detail in \autoref{sec:bagpipes}. We reproduce the result of \cite{tacchella2022stellar} that this SFH increases the galaxy stellar mass, finding an average increase of 0.2~dex. The stellar mass discrepancy between the two models is shown in purple in the lower figure of \ref{fig:bagpipes_compar}. In individual cases the increase in stellar mass can reach $\approx$1~dex, with only small changes in photo-$z$. The largest offsets in stellar mass are typically seen for galaxies with a fiducial stellar mass of $10^{8}$ M$_{\odot}$, with the highest fiducial stellar mass galaxies ($\geq 10^{10}$ M$_{\odot}$) seeing considerably smaller increases. 

We see the largest discrepancies in stellar mass  between the two models when the $\chi^2$ significantly favours the fiducial model, suggesting the higher stellar mass estimate of the ``continuity bursty" model may not be accurate in these cases. However there are a small number of galaxies in which the offset exceeds 1~dex with minimal $\chi^2$ difference, and even a few galaxies where the offset exceeds 0.5~dex and the ``continuity bursty" model is significantly preferred. An example of the discrepancy between the stellar mass PDFs of our fiducial and the ``continuity bursty" model can be seen for the individual galaxy SED of \textit{JADES-Deep-GS:9075} shown in \autoref{fig:discrepant_SEDs}, where there is a 1.3~dex difference between the stellar mass estimates, with only a difference of $\Delta \chi^2 = 0.2$ between the two solutions.

Star-formation rates (SFRs) are typically higher, however the model does not reproduce the highest SFR estimates of the fiducial model. Individual inspection of these highly-star forming galaxies show that the ``continuity bursty" model struggles to reproduce SFRs high enough to match the measured \hbeta+\oiii \ equivalent widths inferred from the photometry. Despite the greater flexibility (and number of fitted parameters), the ``continuity bursty" model has a higher $\chi^2$ than our fiducial model for the majority of galaxies. This may suggest that the fitting procedure is struggling to accurately constrain the SFH in the non-parametric case. 

Looking at the other derived properties in \autoref{fig:bagpipes_sps_compare}, we also see significant scatter in dust extinction, where the galaxies with highest extinctions in our fiducial model found to contain significantly less dust with this SFH model. For galaxies older than 50 Myr in our fiducial model, they are typically found to be systematically younger with this non-parametric SFH, but some of the youngest galaxies in the fiducial model are found to be much older.

\subsection{Comparison to BPASS SPS Models}

By default \bagpipes{} uses the 2016 version of \cite{bruzual_charlot_2003}'s (BC03) stellar population synthesis (SPS) models. However \bagpipes{} can also employ \textit{Binary Population and Spectral Synthesis} \citep[BPASS,][]{stanway2018re} models. Specifically we use models generated with v2.2.1 of BPASS, with the default IMF (slope of 1.35, 300 M$_{\star}$). The IMF parametrization of this SPS model differs slightly from the \cite{kroupa2001variation} IMF used in the BC03 models, causing an intrinsic offset in stellar mass. However as shown in orange in the lower plot of \autoref{fig:bagpipes_compar}, the comparison of the mass estimates is more complex. At the lowest and highest stellar masses ($\log_{10} M_\star/M_\odot \leq 7.5$ and $\log_{10} M_\star/M_\odot \geq 10$), we see systematically lower stellar masses than the fiducial \bagpipes{} model, which is what we would expect based the IMF difference alone. However for the majority of the sample, which falls between these two mass extremes, we see a significantly larger offset in stellar masses, which increases with redshift, as shown by the LOWESS trend.

The SFRs and redshifts are broadly correlated, with some outliers. However the ages show a large scatter, and higher mass-weighted ages (MWA) are preferred in the majority of cases when using BPASS, which may be more reasonable that the young ages ($\langle$MWA$\rangle \sim 20$Myr) inferred with our fiducial \bagpipes{} results. In particular, there are a significant subset of galaxies found to have a $\geq$2 dex shift in mass-weighted age.

The best-fit $\chi^2$ shows some scatter, where fits with low $\chi^2$ in the fiducial \bagpipes{} results typically having a similar or worse fit, but some galaxies with higher $\chi^2$ having significantly improved. For these galaxies with improved $\chi^2$ ($\Delta \chi^{2} > 5$), the main difference is that the best-fitting SED reproduces the observed rest-UV fluxes more closely. Interestingly, this subset of galaxies with significantly improved $\chi^2$ also typically have considerably higher MWA using BPASS than BC03, with their recreated SFHs suggesting a constant SFH, rather than the recent burst preferred when fitting with the BC03 SPS models. 

In \autoref{fig:bagpipes_sps_compare} we also see the significant impact the BPASS SPS model has on the derived dust extinction, age, and metallicity. Despite the same prior and dust model, when the BPASS SPS model is used the dust extinction A$_{\rm V}$ is systematically much lower, often inferring essentially no dust extinction. Galaxy ages are also found to be systematically much larger, often by $\geq$ 100 Myr.

\subsection{Other Comparisons}

Numerous other \bagpipes{} models were tested, and we have presented in detail the results of a subset of them above. Here we summarise the effects of a few other variations which we do not include in \autoref{fig:bagpipes_compar}.

\noindent 1. Fixed redshift to \eazy{} max $P(z)$: Little overall effect on stellar masses or star formation rates, with some individual scatter. This is the expected behaviour as the majority of redshifts are consistent within the \eazy{} posterior uncertainty, given that we use this as a prior in our fiducial model.

\noindent 2. ``continuity" non-parametric SFH model': This variation tests the other non-parametric SFH model introduced in \cite{leja2019measure} and \cite{tacchella2022stellar}, which more tightly constrains the SFR in neighbouring time bins to force more smoothly varying star formation histories than the ``continuity bursty" model. We find overall similar behaviour to the ``continuity bursty" model, with systematically higher masses and higher mass-weighted age. The models typically have worse $\chi^2$ statistics than the ``continuity bursty" model fits potentially indicating that more stochastic star formation histories are preferred for the majority of galaxies, as suggested in the literature \citep{faucher2018model,Looser2023,asada2023bursty,Endsley2023}.

\subsection{Comparison to \prospector}

\begin{figure}
    \centering
    \includegraphics[width=0.6\textwidth]{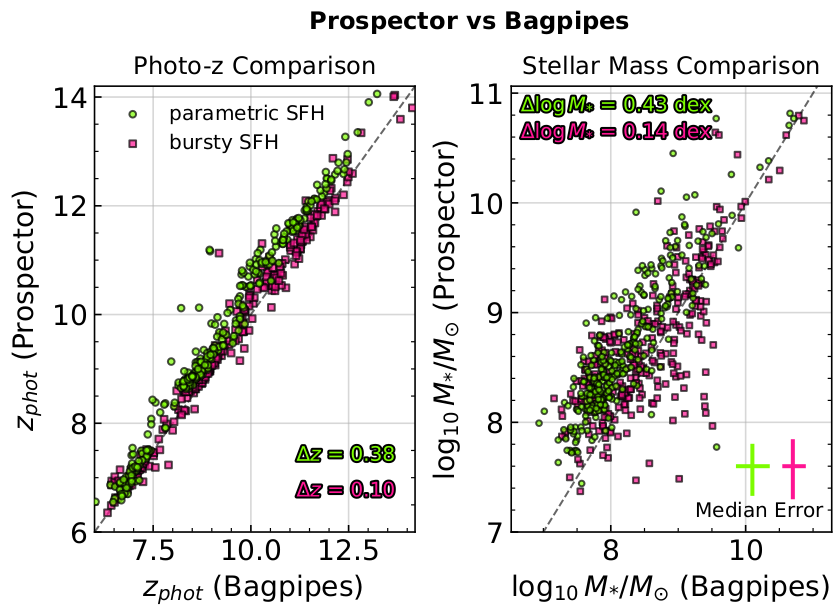}
    
    \caption{(\textbf{Top}) Comparison of redshift estimates between \bagpipes{} and \prospector{}, both with the same Gaussian photo-$z$ prior from \eazy. Photo-$z$ estimates are systematically larger in \prospector{} when comparing the parametric SFH models. Median photo-$z$ offset for each model is shown on the plot. (\textbf{Bottom}) Comparison of derived stellar mass estimates between \prospector{} and \bagpipes{} for both parametric and non-parametric SFH models. Median stellar mass offset is shown on the plot, as is the average uncertainty for both stellar mass estimates. \label{fig:prospect_compare}}
\end{figure}

In this section we compare our fiducial \bagpipes{} model to our results from \prospector. \autoref{fig:prospect_compare} shows a comparison of the stellar masses derived from \bagpipes{} and \prospector. We compare our `fiducial' \bagpipes{} model, as described in \autoref{sec:bagpipes}, to our parametric SFH \prospector{} model. These models are generally similar with a few key differences, but both provide a baseline for comparison to our other models. We also compare our non-parametric SFH models to each other, which both employ the same ``continuity bursty" SFH, with the same time bins and priors. We use the same \cite{kroupa2001variation} IMF for both \bagpipes{} and \prospector, so we do not expect any different in stellar mass estimates due to the IMF parametrization. 

For the comparison of the parametric SFH models (delayed-$\tau$ for \prospector, lognormal for \bagpipes{}), we see systematically larger photo-$z$ and stellar mass estimates with \prospector. Whilst in the \prospector{} model we allow the IGM attenuation to vary, which could impact photo-$z$ estimates, we do not see the same offset with the non-parametric SFH model, where the IGM attenuation is also allowed to vary. Both SED-fitting tools are given the same redshift prior from \eazy. For both example galaxy SEDs in \autoref{fig:discrepant_SEDs} that \prospector{} instead prefers higher-$z$ solutions than \bagpipes{} because it is inferring the presence of Lyman-$\alpha$ emission. Whilst Lyman-$\alpha$ emitters have been found at $z \geq 9$ \citep[e.g.][]{bunker2023jades}, we do not expect to observe Lyman-$\alpha$ from the majority of galaxies at these redshifts due to the attenuation from neutral hydrogen during the Epoch of Reionization, and a systematic photo-$z$ offset from \prospector{} is likely for galaxies without Lyman-$\alpha$ emission. However given that we see a photo$-z$ offset primarily for the 'parametric SFH' only, Lyman-$\alpha$ emission is unlikely to be the only cause of the offset as the \prospector{} SEDs in \autoref{fig:discrepant_SEDs} show Lyman-$\alpha$ emission for both SFH models. This offset will generally cause a slight increase in stellar mass estimates, as a more distant galaxy must be intrinsically brighter. We see a median increase of 0.43~dex in stellar mass.

For the ``continuity bursty" SFH models we see better agreement in redshift, with a median offset of only $\delta z=$ 0.1. Stellar mass estimates are also more consistent on average, although we see large scatter, with individual mass differences reaching $\approx$1.5~dex. 

We typically see comparable $\chi^2$ for both the parametric and  non-parametric \prospector{} SFH models, in contrast to the result with \bagpipes{}. It is possible that the nested sampling with {\tt dynesty} in \prospector{} provides a more robust constraint on the binned SFH than the nested sampling in \bagpipes{}, and may warrant further investigation.

\section{z = 7 Stellar Mass Function without `Little Red Dots'}
\label{sec:gsmf_no_agn}
As discussed in \autoref{sec:agn_discuss}, the `Little Red Dots' (LRDs) dominate the high mass end of our GSMF at $z = 7$. As the contribution of AGN to their photometry is still somewhat uncertain, and likely differs on an individual basis between galaxies, in the main results of this paper we do not remove LRDs from the GSMF estimates, or account for any possible AGN emission. In this Appendix we briefly present the alternative case, where we remove all objects which meet the color-color selection criteria of \cite{2024arXiv240109981K} and reconstruct the $z = 7$ GSMF. 

When we apply their `\textit{red2}' color selection, compactness criterion and SNR requirements to our robust sample, we find 34 galaxies which meet these cuts. There are 13 in the NEP-TDF, 17 in CEERS, 3 in the JADES DR1 field, and 1 in the NGDEEP field. The median redshift is 7.16, with all candidates falling between $z = 6.5$ (our redshift cut) and $z = 8.7$. The median fiducial \bagpipes{} stellar mass is $\log_{10} M_\star/M_{\odot} = 8.90$, with a maximum stellar mass of $\log_{10} M_\star/M_{\odot} = 10.70$. 

\begin{figure}
    \centering
    \includegraphics[width=0.5\columnwidth]{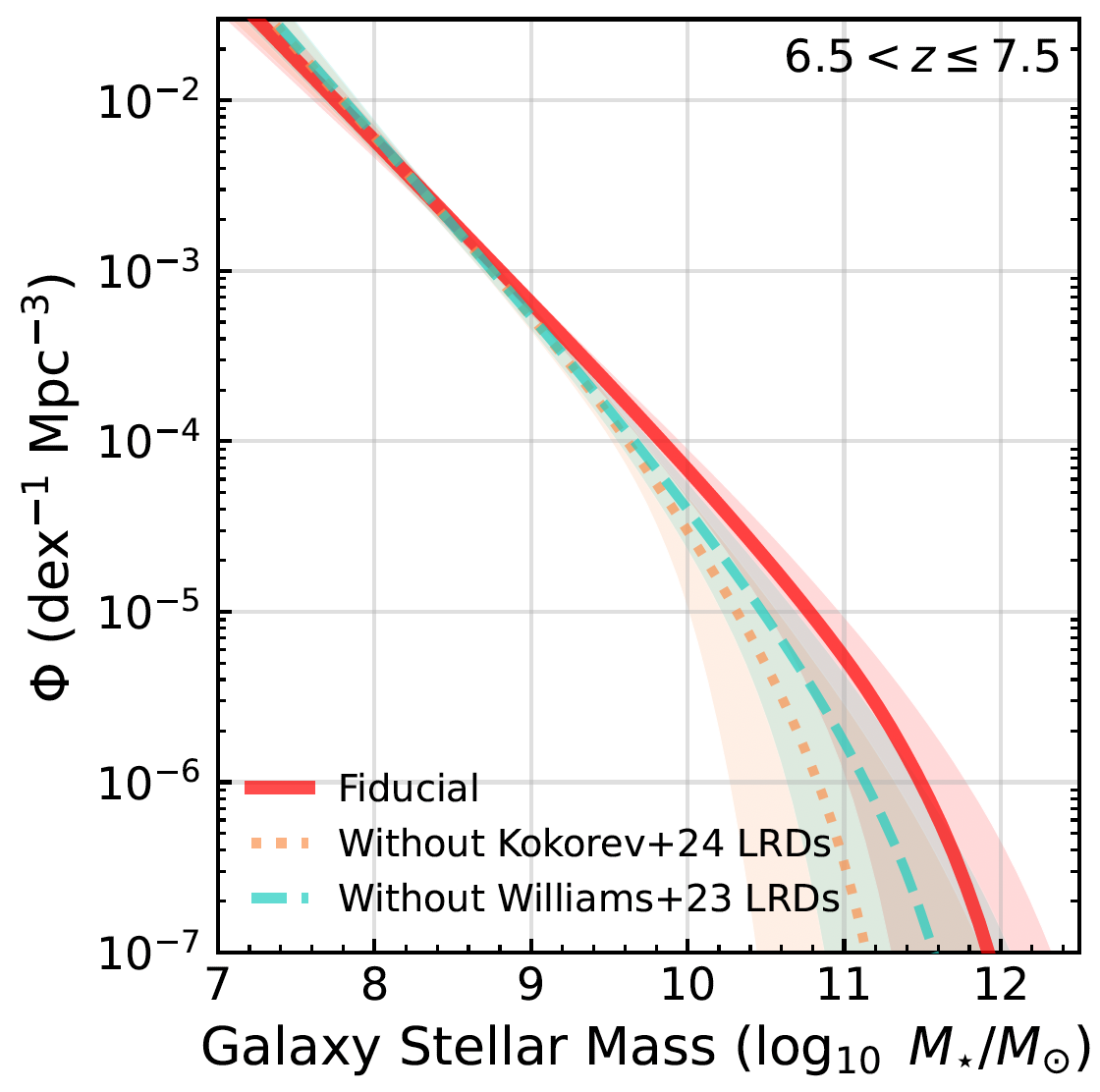}
    \caption{Galaxy Stellar Mass Function at $6.5 < z \leq 7.5$, excluding all `Little Red Dots', compared to our fiducial \bagpipes{} results. We differentiate between the selections of \cite{2024arXiv240109981K} and \cite{2024ApJ...968...34W}, which differ primarily by the strength of the F277W-F444W color required to characterise a galaxy as a Little Red Dot'.} 
    \label{fig:gsmf_noreddots}
\end{figure}

We exclude these 34 candidates from our sample and reconstruct the stellar mass function at $z=7$. No other changes are made to our GSMF construction or fitting procedures. \autoref{fig:gsmf_noreddots} shows the GSMF derived without including any `Little Red Dots', following the cut of \cite{2024arXiv240109981K}. Compared to the fiducial GSMF, this removes the two highest mass bins entirely, which demonstrates our reliance on these galaxies at the high mass regime. In terms of the derived Schechter parameters, the exponential mass cutoff M$^{\star}$, which is not well constrained, decreases from 11.57$^{+0.63}_{-0.85}$ to 10.64$^{+1.25}_{-0.98}$ when we exclude the LRDs. The median posterior $\phi^{\star}$ and $\alpha$ for the GSMF without LRDs are $-5.40^{+1.34}_{-1.43}$ and $-2.04^{+0.18}_{-0.13}$ respectively. 
We also test the more restrictive color cut of \cite{2024ApJ...968...34W}, which require a stronger F277W-F444W color criteria in order to provide an 80\% AGN purity \citep{greene2023uncover}. The majority of the LRDs we find above do not meet this criteria, with only 8/34 having F277W-F444W $>$ 1.6 mag, but these LRDs with the reddest colors also typically have the highest inferred stellar masses, so excluding just these 8 objects has a noticeable impact on the $z=7$  GSMF, as shown in \autoref{fig:gsmf_noreddots}.  

\section{Tabulated Schechter parameters for alternative GSMF estimates at z = 10.5}
\label{sec:alternative_tabulated}

We give the Schechter function parameters for our alternative GSMF fits at $z \sim 10.5$ in \autoref{tab:alternative_gmsf_params}. These are the Schechter parameters representing the fits shown in \autoref{fig:gsmf_z10.5_compar}. These GSMF estimates are equivalent to the fits given in \autoref{tab:params} for the fiducial GSMF, and are calculated using the same method, simply replacing the redshift and stellar mass PDF used in constructing the GSMF with those derived under the alternate SED fitting assumptions. 
\begin{table}[]
\caption{Schechter function parameters for the GSMF at $9.5 \leq z \leq 11.5$ for each of the alternative models shown in \autoref{fig:gsmf_z10.5_compar}. For $\alpha$, M$^{\star}$ and $\log_{10} \phi^{\star}$ we give both the median posterior and maximum likelihood values (in brackets). The details of the \bagpipes{} and \prospector{} configurations for each model are given in \autoref{sec:bagpipes} and \autoref{sec:prospector}. \label{tab:alternative_gmsf_params} } 
\begin{tabular}{cccccc}
\hline
SED Fitting Tool & SFH Model           & IMF         & $\alpha$ & M$^{\star}$ & $\log_{10} \phi^{\star}$ \\ \hline
Bagpipes         & ``continuity bursty" & \cite{kroupa2001variation} & $-1.93^{+0.21}_{-0.16} (-1.80) $ & $10.70^{+1.21}_{-1.06} (9.70)$ & $-5.94^{+1.33}_{-1.28} (-4.70)$ \\
Prospector       & ``continuity bursty" & \cite{kroupa2001variation} & $-1.72^{+0.26}_{-0.18} (-1.51) $ & $10.51^{+1.34}_{-0.99} (9.54)$ & $-5.57^{+1.13}_{-1.19} (-4.46)$ \\
Prospector       & delayed-$\tau$ & \cite{kroupa2001variation}  & $-1.98^{+0.22}_{-0.17} (-1.89) $ & $10.67^{+1.22}_{-1.07} (9.68)$ & $-6.22^{+1.37}_{-1.35} (-4.98)$ \\
Prospector       & ``continuity bursty" & HOT 45K  & $-1.93^{+0.25}_{-0.20} (-1.81) $ & $10.56^{+1.28}_{-1.10} (9.53)$ & $-6.09^{+1.39}_{-1.39} (-4.83)$ \\
Prospector       & delayed-$\tau$             & HOT 45K     & $-2.19^{+0.25}_{-0.22} (-2.18) $ & $10.54^{+1.14}_{-1.10} (9.55)$ & $-6.67^{+1.63}_{-1.45} (-5.31)$ \\
\hline
\end{tabular}
\end{table}

\section{z = 12.5 GSMF with no contamination limit}
\label{sec:z12.5_contam}
\begin{figure}
    \centering
    
    \includegraphics{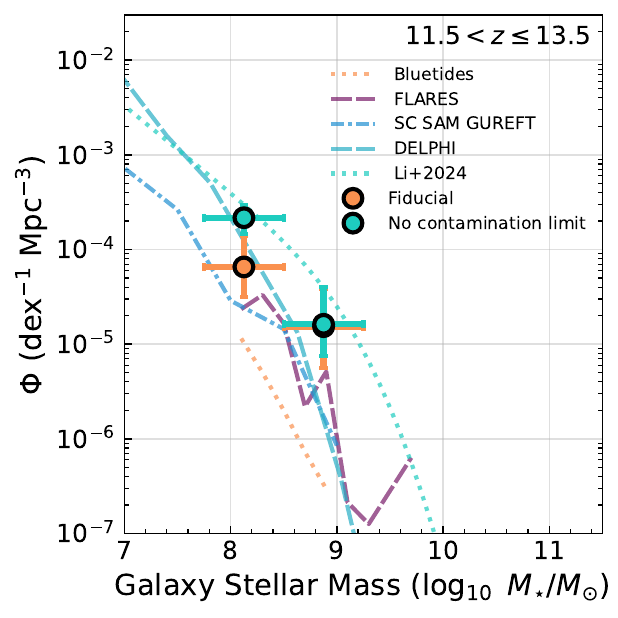}
    \caption{Galaxy Stellar Mass Function at $11.5 < z \leq 13.5$ without a 50\% contamination limit, compared to our fiducial \bagpipes{} results. }
    \label{fig:gsmf_contam}
\end{figure}

Our fiducial GSMF applies a 50\% contamination limit on all galaxies. Given the fields a galaxy is selected in and its stellar mass, our \jaguar{} contamination simulation computes a likelihood of contamination, based on simulated galaxies with the same stellar mass. The highest contamination is seen in the $z=12.5$ GSMF, for the $10^{7.5} < $ M$_{\star}$/M$_{\odot} \leq 10^{8.5}$ bin, and results in several galaxies being removed from our fiducial GSMF estimate. As the predictions of \jaguar{} are uncertain at these redshifts, it is hard to judge how accurate our contamination predictions are. In \autoref{fig:gsmf_contam} we have recomputed the $z = 12.5$ GSMF with no contamination limit, which boosts the lower stellar mass bin by $\sim$0.3~dex. This brings it closer to the predictions of \cite{li2023feedback}'s Feedback Free Model (FFB), which has higher star formation efficiency than most models. The FFB model shown is for $\epsilon_{\rm max} = 0.20$ specifically, and the models with higher SFE ($\epsilon_{\rm max} = 0.5 - 1$) overpredict the GSMF at this redshift compared to our observations. If our contamination is over-estimated in this redshift bin, then the FFB model of \cite{li2023feedback} or \cite{mauerhofer2023dust}'s DELPHI model provide the closest predictions, suggesting high but not extreme star formation efficiency is required to produce the observed GSMF at this redshift.

\clearpage

\bibliography{main.tex}{}
\bibliographystyle{aasjournal}



\end{document}